\documentclass[letterpaper,twocolumn,10pt]{article}
\usepackage{zhanggroup}

\usepackage{url}
\usepackage{tikz}
\usepackage{amsmath}
\usepackage{caption,subcaption}
\usepackage{picture,graphics}
\usepackage{amsthm}
\usepackage{amsmath}
\usepackage{amssymb}
\usepackage{xcolor}             
\usepackage{hyperref}
\usepackage{xspace}
\usepackage[absolute]{textpos}

\theoremstyle{definition}
\newtheorem{definition}{Definition}[section]

\newcommand{\refapp}[1]{\hyperref[#1]{Appendix~\ref*{#1}}}
\newcommand{\mypara}[1]{\smallskip\noindent\textbf{#1:}}

\begin{document}

\begin{textblock}{15}(3.2,1)
To Appear in Network and Distributed System Security (NDSS) Symposium 2025
\end{textblock}

\title{\Large \bf Understanding Data Importance in Machine Learning Attacks: \\Does Valuable Data Pose Greater Harm?} 

\date{}

\author{
{\rm Rui Wen}\ \ \
{\rm Michael Backes}\ \ \
{\rm Yang Zhang\thanks{Corresponding author}}
\\
\textit{CISPA Helmholtz Center for Information Security}\\
}

\maketitle

\begin{abstract}
Machine learning has revolutionized numerous domains, playing a crucial role in driving advancements and enabling data-centric processes.
The significance of data in training models and shaping their performance cannot be overstated. 
Recent research has highlighted the heterogeneous impact of individual data samples, particularly the presence of valuable data that significantly contributes to the utility and effectiveness of machine learning models. 
However, a critical question remains unanswered: are these valuable data samples more vulnerable to machine learning attacks? 
In this work, we investigate the relationship between data importance and machine learning attacks by analyzing five distinct attack types. 
Our findings reveal notable insights. 
For example, we observe that high importance data samples exhibit increased vulnerability in certain attacks, such as membership inference and model stealing. 
By analyzing the linkage between membership inference vulnerability and data importance, we demonstrate that sample characteristics can be integrated into membership metrics by introducing sample-specific criteria, therefore enhancing the membership inference performance. 
These findings emphasize the urgent need for innovative defense mechanisms that strike a balance between maximizing utility and safeguarding valuable data against potential exploitation.
\end{abstract}

\section{Introduction}
\label{sec:intro}

Machine learning has emerged as an indispensable tool across numerous domains, revolutionizing industries and empowering data-driven decision-making processes. 
Central to the essence of machine learning is the pivotal role played by data, serving as the bedrock for training models and exerting a profound influence on their performance and predictive accuracy. 
Concurrently, the crucial role of data in model training also exposes it as a noteworthy source of vulnerabilities.

Recent research has shed light on the heterogeneous impact of individual data samples, highlighting the presence of certain data that exhibit a heightened influence on the utility and overall effectiveness of machine learning models~\cite{JDWHGLZSS19,GZ19,KZ22,KL17,JWSXDKZLS21}. 
Understanding this variability is important for two main reasons. 
First, knowing how individual data samples affect model performance is key to improving machine learning explainability, offering new insights into model behavior, and enhancing interpretability~\cite{KL17,RSG16,LL17,PGILM23}. 
Second, this knowledge can guide data trading practices, where the importance of data is a significant factor~\cite{ADS19,GZ19}.

However, the influence of such diverse data on model leakage and security remains largely unexplored. 
Existing research predominantly concentrates on the models themselves; for example, studies~\cite{SZHBFB19,LWHSZBCFZ22} suggest that overfitted models are more prone to membership inference attacks. 
Nevertheless, even within the same model, distinct data samples exhibit varying vulnerabilities to attacks. 
This prompts a crucial question: do these valuable data samples also exhibit an increased vulnerability to a spectrum of machine learning attacks? 
Understanding the differential vulnerability of data samples has significant practical implications. 
In medical diagnostics, for example, patient records with rare but highly indicative symptoms are considered high importance samples. 
Assessing whether these records are more prone to attacks is crucial, as breaches could lead to discrimination, higher insurance premiums, or other serious consequences for individuals.

In this paper, our focus lies in investigating the relationship between data importance and machine learning attacks. 
Our primary objective is to thoroughly investigate whether valuable data samples, which contribute significantly to the utility of machine learning models, are exposed to an elevated risk of exploitation by malicious actors.

To achieve our objectives, we focus on five distinct types of attacks, encompassing both training-time and testing-time attacks. 
The training-time attack we consider is the backdoor attack~\cite{GDG17,LMALZWZ18,SWBMZ22}, while the testing-time attacks consist of membership inference attack~\cite{LZBZ22,SSSS17,SZHBFB19,CTCP21,LZ21}, model stealing attack~\cite{TMWP21,TZJRR16,SAB22}, attribute inference attack~\cite{MSCS19,SS20}, and data reconstruction attack~\cite{ZJPWLS20,YMALMHJK20,FJR15}. 
For each of these attacks, we thoroughly analyze the behavior and impact on both high importance and low importance data samples, aiming to uncover any discernible differences.

\mypara{Main Findings}
Our research has yielded significant findings that shed light on the heightened vulnerability of valuable data samples to privacy attacks. 
Specifically, our key findings are as follows:
\begin{itemize}
    \item Membership Inference Attack: High importance data samples exhibit a higher vulnerability compared to low importance samples, particularly in the low false-positive rate region. 
    For instance, in the CIFAR10 dataset, at a false positive rate (FPR) of $1\%$, the true positive rate (TPR) of high importance data is $10.2\times$ greater than that of low importance samples.
    \item Privacy Onion Effect: The concept of the privacy onion effect~\cite{CJZPTT22} can be extended to the distribution of data importance.
    Specifically, previously considered unimportant samples gain significance when the dataset removes the important samples.
    \item Model Stealing Attack: High importance samples demonstrate greater efficiency in stealing models when the target model is trained on the same distribution as the query distribution. 
    However, we empirically demonstrate that the importance does not transfer between different tasks.
    \item Backdoor Attack: Poisoning high importance data enhances the efficiency of the poisoning process, particularly when the size of the poison is small. 
    On the other hand, the influence on clean accuracy does not yield a definitive conclusion, poisoning either type of data has a limited impact on clean accuracy.
    \item Attribute Inference and Data Reconstruction Attacks: We observe no significant distinction between high and low importance data in these attacks.
\end{itemize}

Our research provides empirical evidence establishing a correlation between data importance and vulnerabilities across diverse attack scenarios. 
This introduces a novel perspective for analyzing sample-specific vulnerabilities, enriching our understanding of the security implications within the realm of machine learning. 

Beyond theoretical insights, our study showcases practical applications of these findings, illustrating how they can be utilized to devise more potent attacks. 
On one hand, these findings can be utilized in a \textit{passive} manner. 
For example, we empirically demonstrate that membership inference attacks can be improved by introducing sample-specific criteria based on sample importance. 
Additionally, adjusting the poisoning strategy according to sample importance proves to enhance the efficacy of backdoor attacks, particularly with a reduced poisoning rate.

More interestingly, we can \textit{actively} modify samples to alter their importance, which subsequently impacts both attack and defense performance. 
For example, recognizing that high importance samples are more vulnerable to membership inference attacks, attackers could increase the importance of targeted samples to heighten their vulnerability. 
This approach follows exactly the same idea as the attack accepted at CCS’22~\cite{TSJLJHC22}, effectively demonstrating how we can ``reinvent'' state-of-the-art attacks guided by findings in our work.

In summary, our work represents a pioneering step in systematically understanding the vulnerabilities of the machine learning ecosystem through the lens of data. 
These findings serve as a resounding call to action, urging researchers and practitioners to develop innovative defenses that strike a delicate balance between maximizing utility and safeguarding valuable data against malicious exploitation. 

\section{Background}

\subsection{Machine Learning Models}
\label{sec:background_ml}

Machine learning algorithms aim to construct models that effectively predict outputs based on given inputs. 
These models are typically represented by a parameterized function denoted as $f_{\theta}: \mathcal{X}\rightarrow\mathcal{Y}$, where $\mathcal{X}$ represents the input space and $\mathcal{Y}$ represents the output space encompassing all possible predictions. 
The process of determining optimal parameter values $\theta$ involves minimizing an objective function using gradient descent. 
Specifically, the objective is to minimize the classification loss 
\[
\mathop{\mathbb{E}}_{(x,y)}[\mathcal{L}(f_{\theta}(x),y)]
\]
where $(x,y)\in\mathcal{X}\times\mathcal{Y}$ denotes samples from the training dataset used to train the target model. 
This optimization process guides the model towards achieving optimal performance by iteratively adjusting the parameters.

\subsection{Data Importance}
\label{sec:background_importance}

The investigation of individual training sample importance in machine learning (ML) is a fundamental and intricate problem with broad implications, especially in data valuation. 
Understanding the importance of a single training sample within a learning task profoundly impacts data assessment, allocation of resources, and the quality of ML models. 

Leave-one-out (LOO) method has long been regarded as an intuitive approach for assessing the importance of data samples.  
Formally, let $D$ and $D_\text{val}$ represent the training set and the validation set, and $\mathcal{A}$ denote the learning algorithm. 
$U_{\mathcal{A}, D_\text{val}}$ denotes the validation accuracy of the model trained on $D$ using $\mathcal{A}$. 
The importance of a target sample $z$ can be quantified as the difference in utility before and after incorporating the target sample into the training set, expressed as:

\begin{small}
\begin{align*}
v_{loo}(z) \propto U_{\mathcal{A}, D_{val}}(D) - U_{\mathcal{A}, D_{val}}(D \backslash \{z\})
\end{align*}
\end{small}

Nevertheless, evaluating the importance of all $N$ samples in the training set necessitates retraining the model $N$ times, resulting in computational heaviness. 
To address this limitation, Koh and Liang~\cite{KL17} proposed influence functions as an approximation method, significantly reducing the computational cost from $O(Np^2+p^3)$ to $O(Np)$, where $p$ represents the number of model parameters.

Despite the effectiveness of LOO, Ghorbani and Zou~\cite{GZ19} have raised concerns about its ability to capture complex interactions between subsets of data. 
They argue that the Shapley value provides a more comprehensive framework for measuring data importance. 
The Shapley value, originally proposed by Shapley~\cite{S53}, assigns an importance value to each sample $z$ in the training set using the following formulation:

\begin{small}
\begin{align*}
\nu_\text{shap}(z) \propto \frac{1}{N} \sum_{S\subseteq D\setminus\{z\}} \frac{1}{{N-1 \choose |S|}}
\big[U_{\mathcal{A}, D_{val}}(S\cup \{z\})-U_{\mathcal{A}, D_{val}}(S)\big]
\end{align*}
\end{small}

To simplify the interpretation of the Shapley value assigned to each sample, one can conceptualize it as the contribution to accuracy in typical scenarios.

For instance, in a hypothetical scenario with 100 samples and a model achieving 90\% accuracy, a valuable sample may contribute 2\% accuracy, while a less valuable sample may only contribute 0.1\%. 
Consequently, the importance value assigned to a valuable sample is 0.02, whereas for a less valuable sample, it is 0.001. 
Samples with an importance of 0 signify no contribution to the model's accuracy, while values below 0 suggest a detrimental impact, possibly due to incorrect labels or samples lying outside the distribution.

The Shapley value takes into account the contributions of all possible subsets of the training set, offering a more holistic assessment of data importance. 
However, the accurate computation of the Shapley value based on the defined formula necessitates training $\mathcal{O}(2^N)$ machine learning models, rendering it impractical for complex datasets. 
As a result, existing methods employ approximate algorithms to estimate the Shapley value. 
For instance, Ghorbani and Zou~\cite{GZ19} introduced two Monte Carlo-based approaches for Shapley value approximation. 
To expedite evaluation time and enable analysis of large datasets, Jia et al.~\cite{JDWHGLZSS19} utilized the $K$-nearest neighbors ($K$NN) algorithm to approximate the target learning algorithm, reducing the time complexity to $\mathcal{O}(N\log N)$.

\section{Evaluation Setup}
\label{sec:eval_setup}

In this work, we deploy $K$NN-Shapley~\cite{JDWHGLZSS19}  to assess the importance of samples in the training set, which takes a dataset as input and assigns an importance value to each sample in the dataset. 
This decision is justified by two main considerations.

Firstly, from the perspective of utility, traditional data attribution methods struggle to account for the complex interactions within data subsets. 
Previous research, as discussed by Gupta and Zou~\cite{GZ19}, highlights this limitation. 
Consequently, we adopt Shapley value-based approaches for a more accurate assessment of data importance. 
We further examine the efficacy of two non-Shapley-based measurement techniques: Leave-one-out (LOO) and the advanced data attribution method, Trak\footnote{Trak quantifies the influence of each sample on specific test samples within a dataset. 
To adhere to the established definition of data importance, we calculate the average influence exerted by each sample across the entire test dataset as the importance of each sample.}~\cite{PGILM23}. 
As evidenced in~\autoref{figure:append_othermeasure}, when comparing the performance of models trained with 5000 samples of varying significance, the accuracy discrepancy is less than 7\% for these methods. 
In contrast, the $K$NN-Shapley method identifies samples that yield an accuracy difference exceeding 20\%, thereby demonstrating its superior capability in accurately quantifying importance, we defer the details to~\refapp{append:effective_knn}.

Secondly, regarding scalability, most measurement methods are highly computationally inefficient. 
For instance, employing Leave-one-out to calculate importance values for CIFAR10 necessitates over 80 hours on 8$\times$A100 GPUs. 
Shapley value-based methods are generally \textit{more demanding}. 
In Jia et al.'s work~\cite{JWSXDKZLS21}, the authors provide a runtime demonstration (\autoref{fig:runtime} for ease reference) showing that existing measurement methods, except for $K$NN-Shapley, do not scale efficiently to large datasets, even as CIFAR10.

Furthermore, the comprehensive evaluations conducted by prior studies~\cite{JWSXDKZLS21} consistently underscore the effectiveness and accuracy of $K$NN-Shapley. 
Therefore, considering both utility and scalability, $K$NN-Shapley emerges as the sole feasible method for conducting experiments.

\mypara{Datasets}
Our evaluation encompasses three widely-used benchmark datasets, namely CIFAR10~\cite{CIFAR}, CelebA~\cite{LLWT15},
and TinyImageNet~\cite{TinyImageNet}.
CIFAR10 comprises a collection of 60,000 colored images evenly distributed across ten classes, representing common objects encountered in everyday life, including airplanes, birds, and dogs.
CelebA is a large-scale face dataset that encompasses over 40 annotated binary attributes. 
To ensure balance in our analysis, we follow previous works~\cite{NT21,SWBMZ22,ZQZ22,LWHSZBCFZ22} that select the three most balanced attributes (Heavy Makeup, Mouth Slightly Open, and Smiling) to create an 8-class ($2^3$) classification task. 
Note that our findings are not dependent on this specific attribute selection, as validated in~\refapp{append:attributeselection}.
Additionally, our evaluation incorporates TinyImageNet, which constitutes a subset of the ImageNet dataset. 
It encompasses 200 distinct object classes, each with 500 training images.
We further validate the generalizability of our conclusions across different modalities, with detailed information deferred to~\autoref{sec:transferability}.

\begin{figure*}[!t]
\centering
\begin{subfigure}{0.49\columnwidth}
\includegraphics[width=\columnwidth]{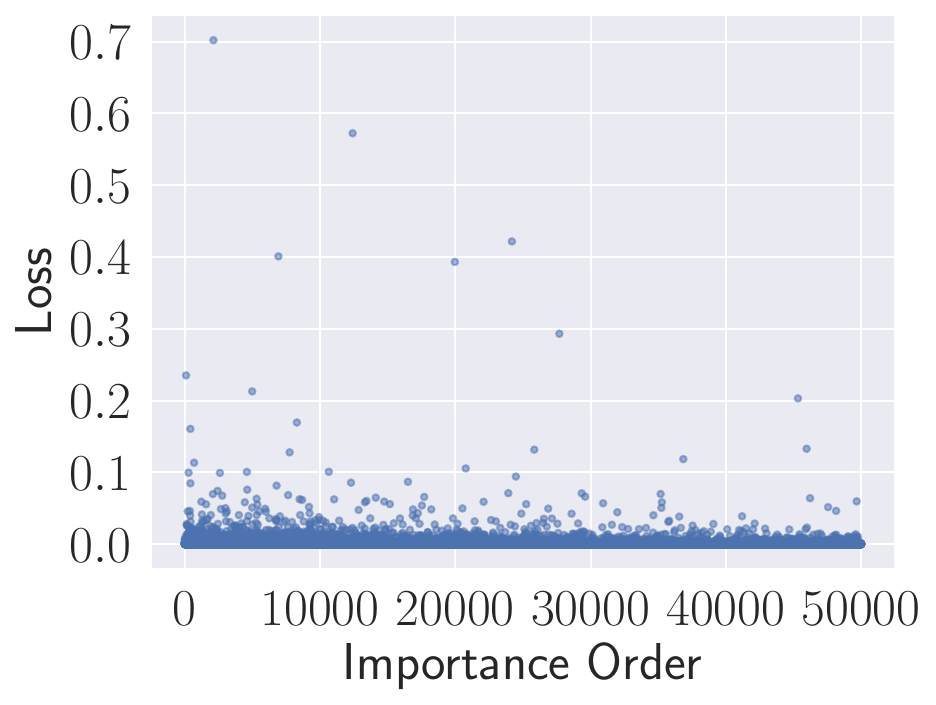}
\caption{Loss Distribution}
\label{fig:loss_overview}
\end{subfigure}
\begin{subfigure}{0.49\columnwidth}
\includegraphics[width=\columnwidth]
{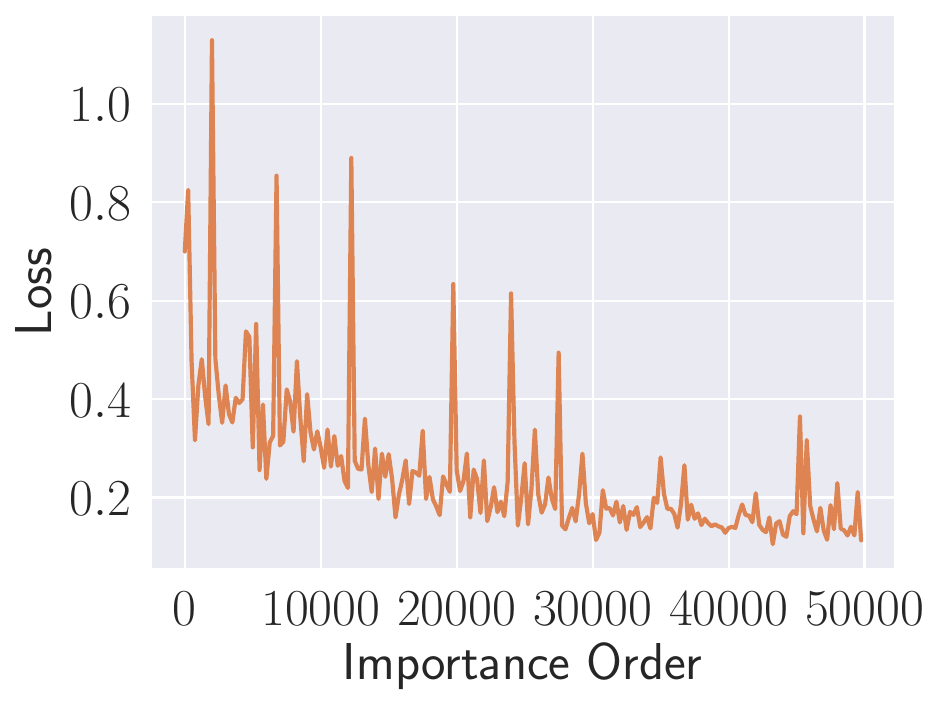}
\caption{CIFAR10}
\label{fig:loss_cifar}
\end{subfigure}
\begin{subfigure}{0.49\columnwidth}
\includegraphics[width=\columnwidth]{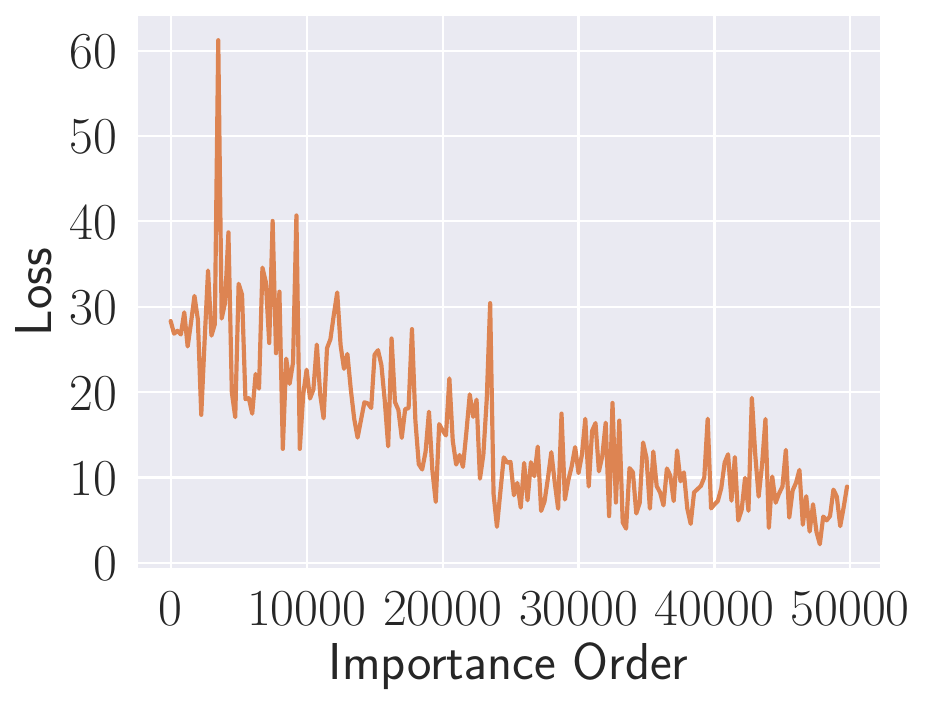}
\caption{CelebA}
\label{fig:loss_celeba}
\end{subfigure}
\begin{subfigure}{0.49\columnwidth}
\includegraphics[width=\columnwidth]{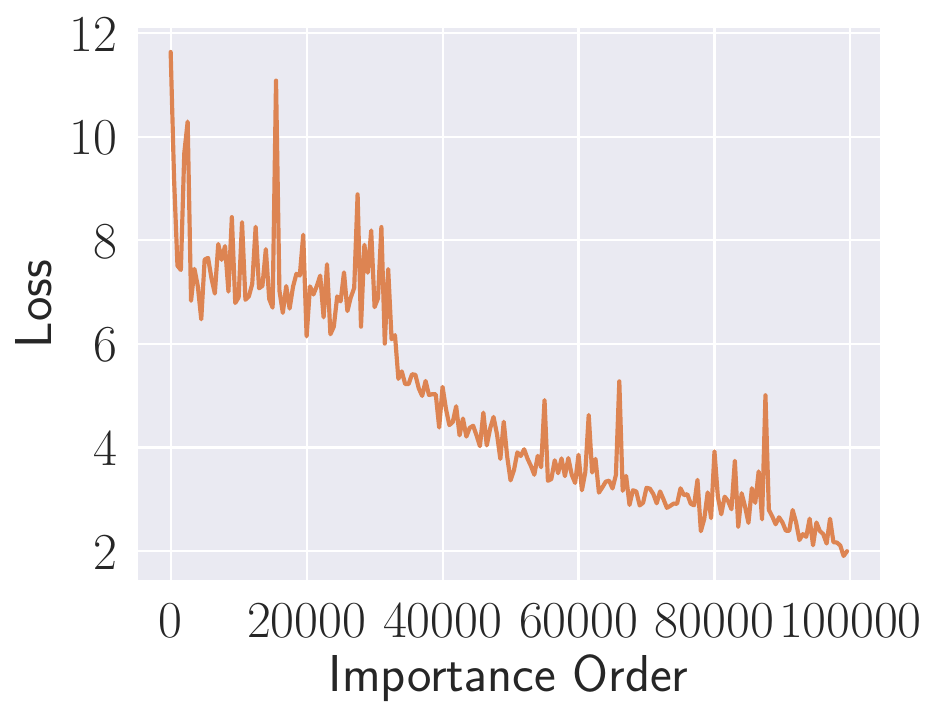}
\caption{TinyImageNet}
\label{fig:loss_tinyimagenet}
\end{subfigure}
\caption{Relationship between loss and importance value. 
Low importance samples statistically have higher losses.}
\label{figure:loss}
\end{figure*}

\subsection{Learning Characteristic}
\label{sec:learn_charac}

In order to gain a deep understanding of the disparities between high and low importance data, we delve into the learning characteristics, such as loss, associated with these samples. 
To the best of our knowledge, our study represents the first endeavor to investigate the learning characteristics of samples with varying degrees of importance, diverging from the conventional focus solely on their contribution to the final performance. 

To quantify these learning characteristics, we initially train a model using the complete dataset, comprising both high and low importance data. 
Subsequently, we compute the loss for each individual data point and explore the correlation between the loss and its corresponding importance value.

In~\autoref{fig:loss_overview}, we present a visual representation of the relationship between loss and importance value. 
The x-axis represents the importance order of a sample in the dataset, with 1 denoting the lowest importance and 50000 representing the most valuable data. 
Initially, it may seem that there is no discernible pattern between loss and importance value, as both low and high importance samples can exhibit either low or high loss. 
However, upon further analysis, we statistically observe that higher importance samples tend to demonstrate lower loss, as depicted in~\autoref{fig:loss_cifar}, \autoref{fig:loss_celeba}, and \autoref{fig:loss_tinyimagenet}.

To arrive at this conclusion, we divide the samples into 200 bins based on their importance value. 
For instance, the lowest 1 to 250 samples are categorized into bin one, 251 to 500 are allocated to bin two, and so forth. 
For each bin, we calculate the sum of the losses and plot these 200 data points to generate the final curve.
Despite some fluctuations observed in the curve, it is evident that valuable samples tend to exhibit lower loss. 
This finding aligns with our expectations, as lower loss signifies greater representativeness, thereby facilitating easier learning and enhancing their contribution to the overall utility of the model.

Having established the effectiveness of importance assignment and gained preliminary insights into the learning characteristics, we proceed to conduct representative machine learning attacks to investigate the impact of data importance in such attacks. 
Our experimental investigations are carried out using the ResNet18 architecture, and in~\autoref{sec:transferability}, we demonstrate the generalizability of our conclusions to different architectures.

\begin{figure*}[!t]
\centering
\begin{subfigure}{0.49\columnwidth}
\includegraphics[width=\columnwidth]{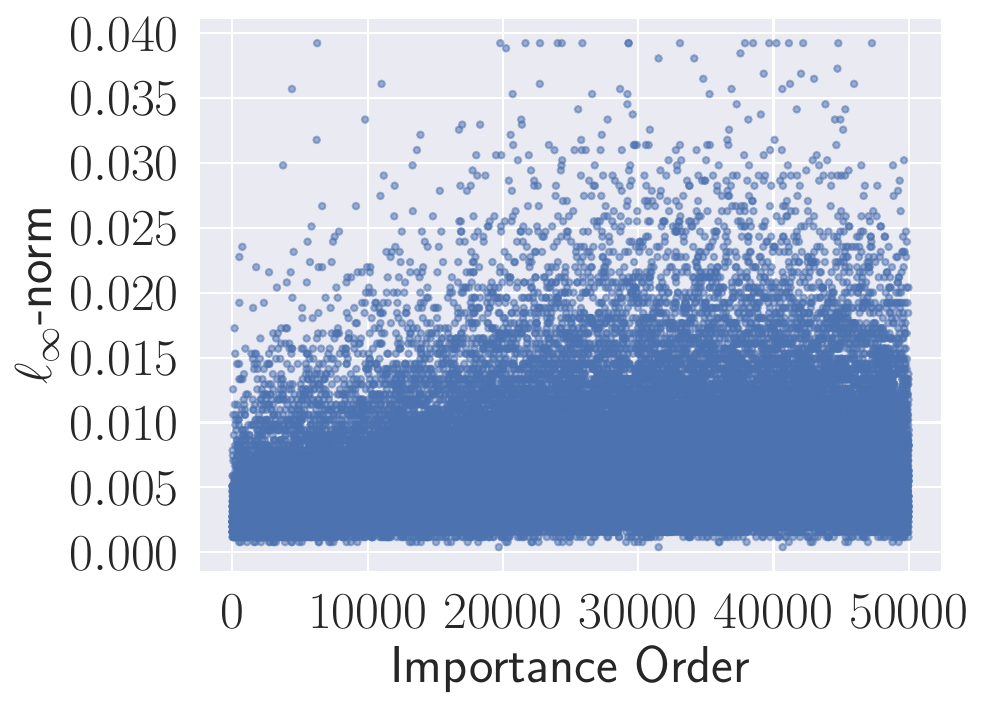}
\caption{Distance Distribution}
\label{fig:distance_overview}
\end{subfigure}
\begin{subfigure}{0.49\columnwidth}
\includegraphics[width=\columnwidth]{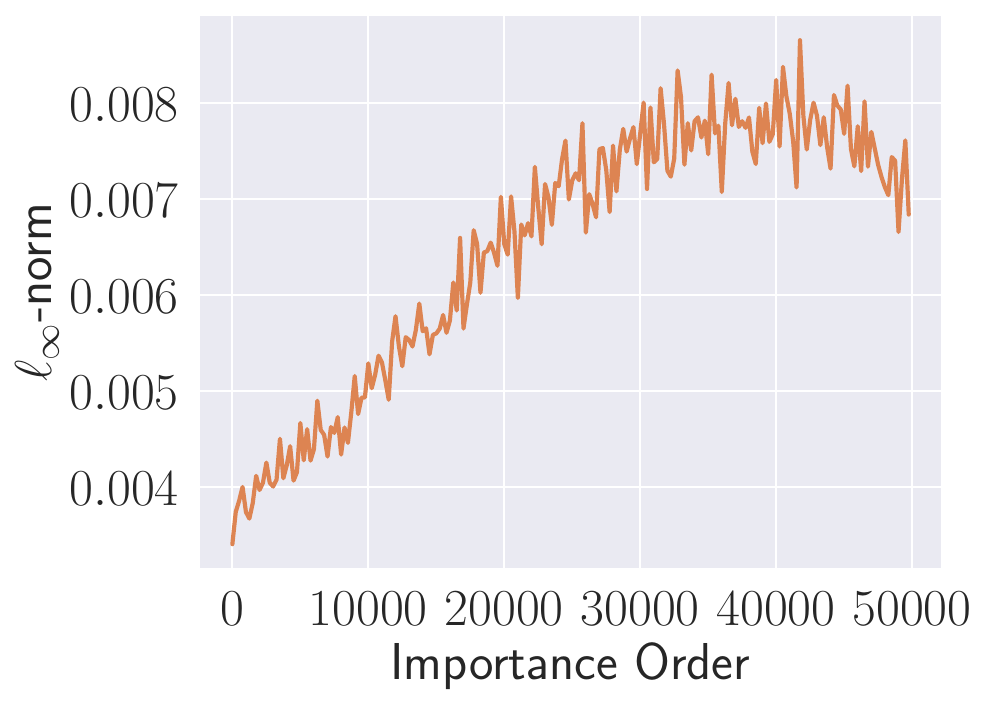}
\caption{CIFAR10}
\label{fig:distance_cifar}
\end{subfigure}
\begin{subfigure}{0.51\columnwidth}
\includegraphics[width=\columnwidth]{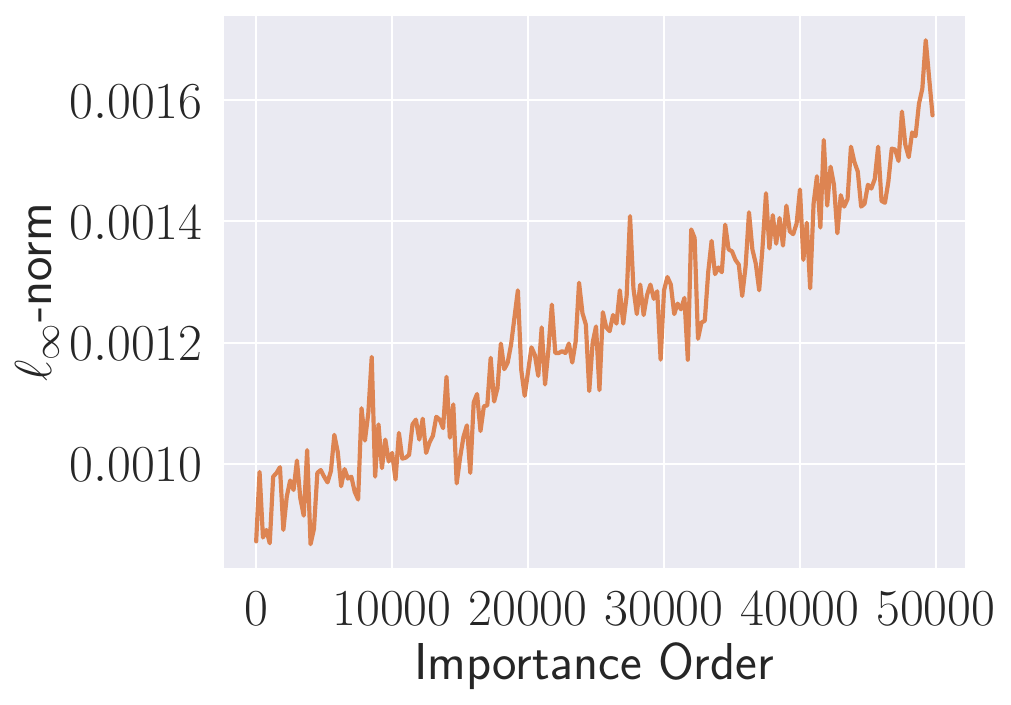}
\caption{CelebA}
\label{fig:distance_celeba}
\end{subfigure}
\begin{subfigure}{0.53\columnwidth}
\includegraphics[width=\columnwidth]{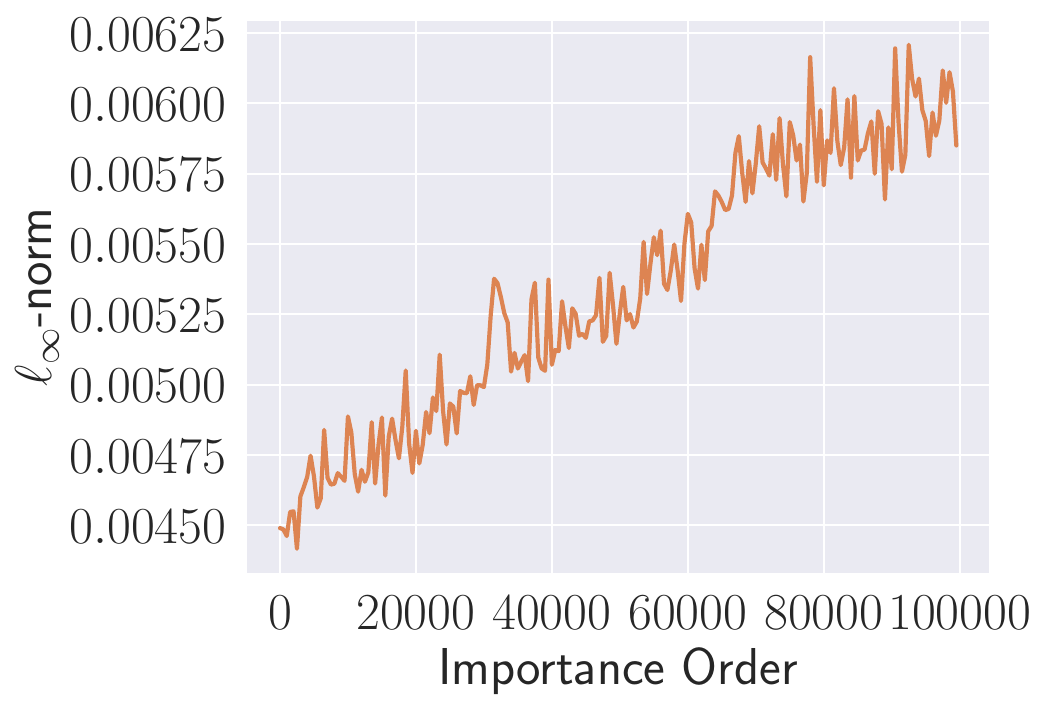}
\caption{TinyImageNet}
\label{fig:distance_tinyimagenet}
\end{subfigure}
\caption{Relationship between distance to the decision boundary and importance value. Low importance samples are
statistically closer to the decision boundary. 
The distance measured with different norms can be found in~\refapp{appendix_mia}.}
\label{figure:distance}
\end{figure*}

\section{Membership Inference Attack}
\label{sec:mia}

Membership Inference Attack (MIA)~\cite{LZBZ22,SSSS17,SZHBFB19,CTCP21,LZ21,YMMBS22} is a prominent privacy attack utilized to determine whether a specific data sample belongs to a training dataset. 
This attack is widely employed to assess the privacy of training data due to its simplicity and broad applicability.

In the attack scenario, the adversary $\mathcal{A}$ is granted access to a target model and is tasked with determining the membership status of a given data sample $(x, y)$. 
Formally, the membership inference attack can be defined as a security game, referred to as Membership Inference Security Game, which is described as follows:

\begin{definition}[Membership Inference Security Game~\cite{CCNSTT22}]
\label{def:game}
The game proceeds between a challenger $\mathcal{C}$ and an adversary $\mathcal{A}$:
\begin{enumerate}
    \item The challenger samples a training dataset $D \gets \mathbb{D}$ and trains a model $f_\theta \gets \mathcal{T}(D)$ on the dataset $D$.
    \item The challenger flips a bit $b$, and if $b=0$, samples a fresh challenge point from the distribution $(x, y) \gets \mathbb{D}$ {(such that $(x, y) \notin D$)}. 
    Otherwise, the challenger selects a point from the training set $(x, y) \overset{\$}{\gets} D$.
    \item The challenger sends $(x, y)$ to the adversary.
    \item The adversary gets query access to the distribution $\mathbb{D}$, and to the model $f_\theta$, and outputs a bit $\hat{b} \gets \mathcal{A}^{\mathbb{D}, f}(x, y)$.
    \item Output 1 if $\hat{b}=b$, and 0 otherwise.
\end{enumerate}
\end{definition}

The adversary $\mathcal{A}$ is provided with auxiliary information about the data distribution $\mathbb{D}$. 
This allows the adversary to sample a shadow dataset from the same or a similar distribution, which is a common assumption in the existing literature.

The attack accuracy for the adversary is defined as follows:
\[\emph{Acc} = \Pr_{x,y,f,b}[\mathcal{A}^{\mathbb{D},f}(x, y) = b]. \]
To assess the privacy leakage caused by membership inference attacks (MIAs), we employ two metrics commonly used in prior research, focusing on both worst-case and average-case performance:
\begin{enumerate}
    \item \textbf{(Log-scale) ROC Analysis}~\cite{CCNSTT22}, which focuses on the true-positive rate at low false-positive rates, effectively capturing the worst-case privacy vulnerabilities of machine learning models.
    \item \textbf{Membership Advantage}~\cite{YGFJ18,SSM19}, defined as 
    \[\emph{Adv}=2\times(\emph{Acc}-0.5).\]
    This metric represents the advantage over random guessing, multiplied by 2, providing an average-case measure to gain an overview of the attack's efficacy.
\end{enumerate}

In this work, we investigate four specific membership inference attacks. 
For the CIFAR10 and CelebA tasks, a training set of 50,000 samples is employed, while for the TinyImageNet task, we utilize a training set of 100,000 samples to construct the target model.

To assess the membership status of samples, we first adopt a methodology based on previous research~\cite{CTCP21,LZ21} that considers the distance to the decision boundary as a reflection of membership status. 
Specifically, they claim that samples located near the decision boundary are more likely to be non-members, whereas samples positioned in the central region of the decision area are more likely to be members. 

We calculate the distance to the decision boundary for all samples in the training dataset.
Specifically, for each sample, we iteratively perturb it using Projected Gradient Descent (PGD) with a small step size until it is classified into a different class. 
Subsequently, we compute the distance between the perturbed sample and its original counterpart. 
In this analysis, the distance is measured using the $\ell_\infty$ norm, and we find consistent results across different norms such as $\ell_1$ and $\ell_2$, as evidenced by the corresponding findings presented in~\refapp{appendix_mia}.

\begin{figure*}[!t]
\centering
\begin{subfigure}{0.56\columnwidth}
\includegraphics[width=\columnwidth]{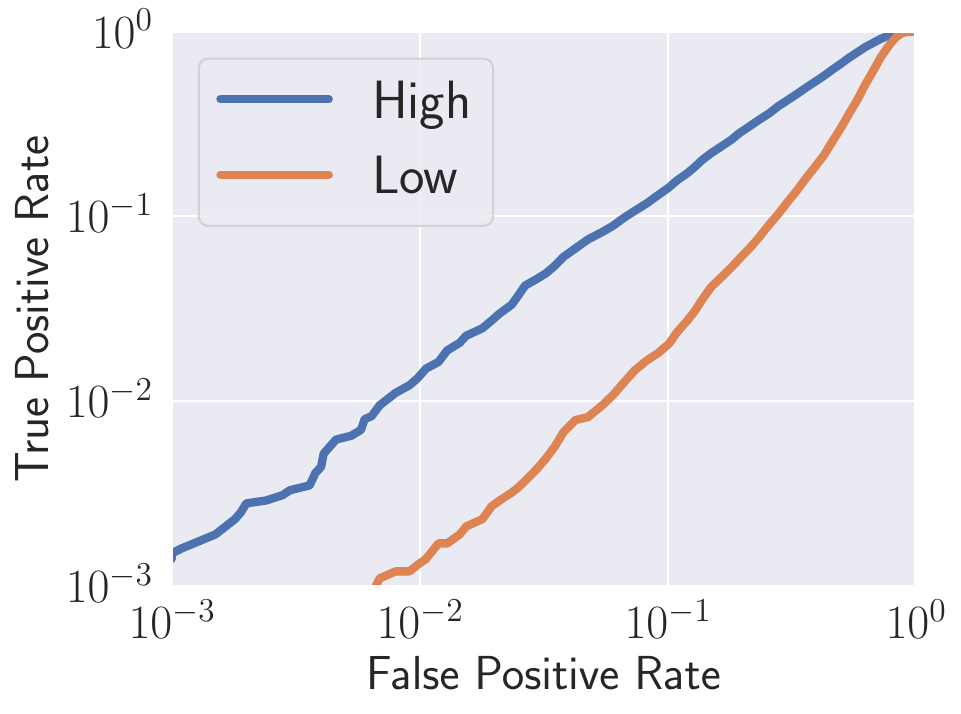}
\caption{CIFAR10}
\label{fig:distance_mia_cifar}
\end{subfigure}
\begin{subfigure}{0.56\columnwidth}
\includegraphics[width=\columnwidth]{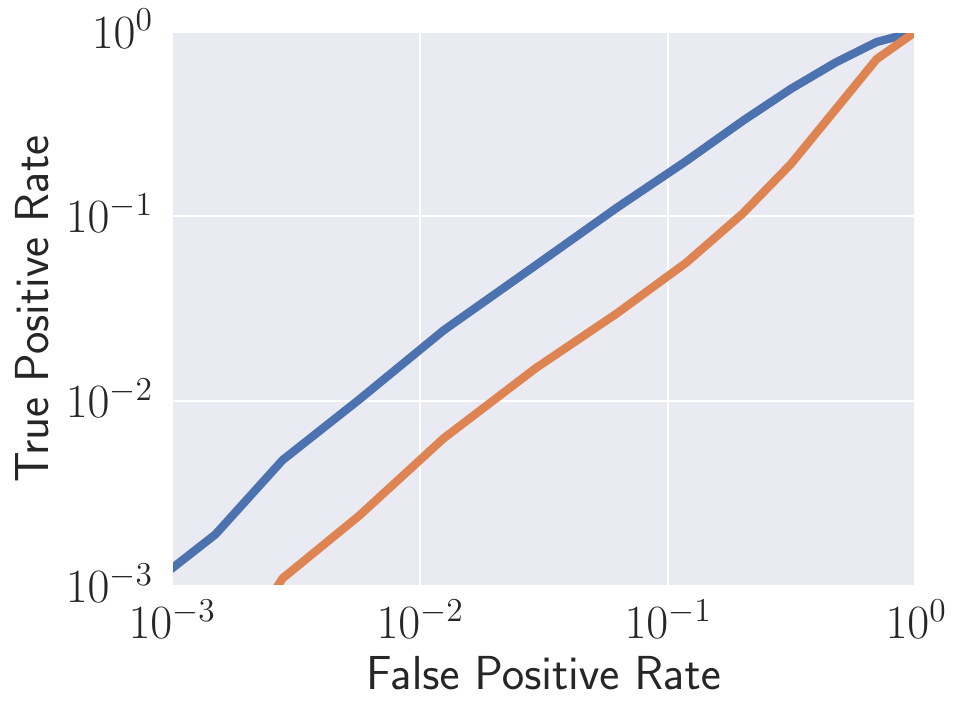}
\caption{CelebA}
\label{fig:distance_mia_celeba}
\end{subfigure}
\begin{subfigure}{0.56\columnwidth}
\includegraphics[width=\columnwidth]{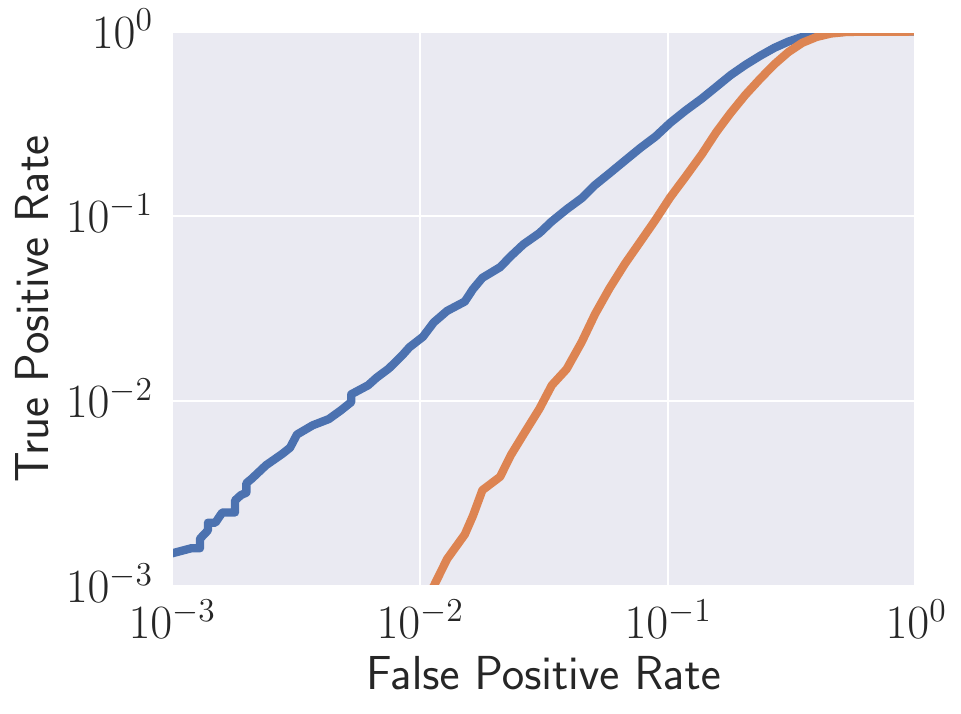}
\caption{TinyImageNet}
\label{fig:distance_mia_tinyimagenet}
\end{subfigure}
\caption{Log-scale ROC curve: membership inference attack based on the distance to the decision boundary. 
High importance samples exhibited substantially higher true-positive rates, particularly in the
low false-positive rate region. 
Results with different norms can be found in~\refapp{appendix_mia}.}
\label{figure:mia}
\end{figure*}

Our initial visualization focuses on examining the distances of samples with different importance values. 
Similar to the observation made regarding the distribution of loss values in~\autoref{sec:learn_charac}, no direct relationship is discernible between the importance value and the distance to the decision boundary. 
Notably, samples with similar importance values may exhibit substantial differences in their distances to the decision boundary.
In contrast, we further analyze the statistical characteristics of these samples, as performed in~\autoref{sec:learn_charac} and present the ``group distance'' in~\autoref{fig:distance_cifar}, ~\autoref{fig:distance_celeba}, and \autoref{fig:distance_tinyimagenet}.
The results reveal that low importance samples are statistically closer to the decision boundary, which aligns with the previous conclusion that low importance samples tend to have higher loss compared to high importance samples.

We follow the same procedure to derive the distance for samples in the testing dataset and launch the membership inference attack based on the distance.
We identify 10,000 samples with the highest importance value as the high group, and an equivalent number of samples with the lowest value as the low group. 
The resulting ROC curves, depicted in~\autoref{figure:mia}, are presented on a logarithmic scale to compare the performance between these two groups.

From the figures, we observe significant differences in the behavior of high importance samples and low importance samples, particularly in the low false-positive rate area. 
Specifically, for the CIFAR10 dataset, high importance samples demonstrate a true-positive rate (TPR) $10.2\times$ higher than low importance samples at a low false-positive rate of $1\%$. 
For the TinyImageNet dataset, the difference is even more pronounced, with high importance samples exhibiting a TPR that is $27.9\times$ higher than that of low importance samples at the same false-positive rate.

These observations provide compelling and empirical evidence supporting the notion that high importance samples are considerably more vulnerable to membership inference attacks, which satisfies our expectation as the importance of data samples can be regarded as the proxy of memorization~\cite{DSA21}. 
These findings thus pose a significant and tangible threat to the safeguarding of high importance data privacy. 
On the other hand, these findings may also prompt researchers to consider adopting strategic sampling methods for more effective privacy auditing~\cite{NSTPC21,JUO20}.

We further validate the generalizability of this finding across various attack methodologies by conducting experiments with three additional metric-based attacks: \textit{prediction confidence-based} attack~\cite{YGFJ18,SSM19}, \textit{entropy-based} attack~\cite{SSSS17,SZHBFB19}, and \textit{modified prediction entropy-based} attack~\cite{SM21}. 
The first two attacks were enhanced by introducing class-dependent thresholds, as demonstrated by Song and Mittal~\cite{SM21}.

\begin{figure*}[!t]
\centering
\begin{subfigure}{0.56\columnwidth}
\includegraphics[width=\columnwidth]{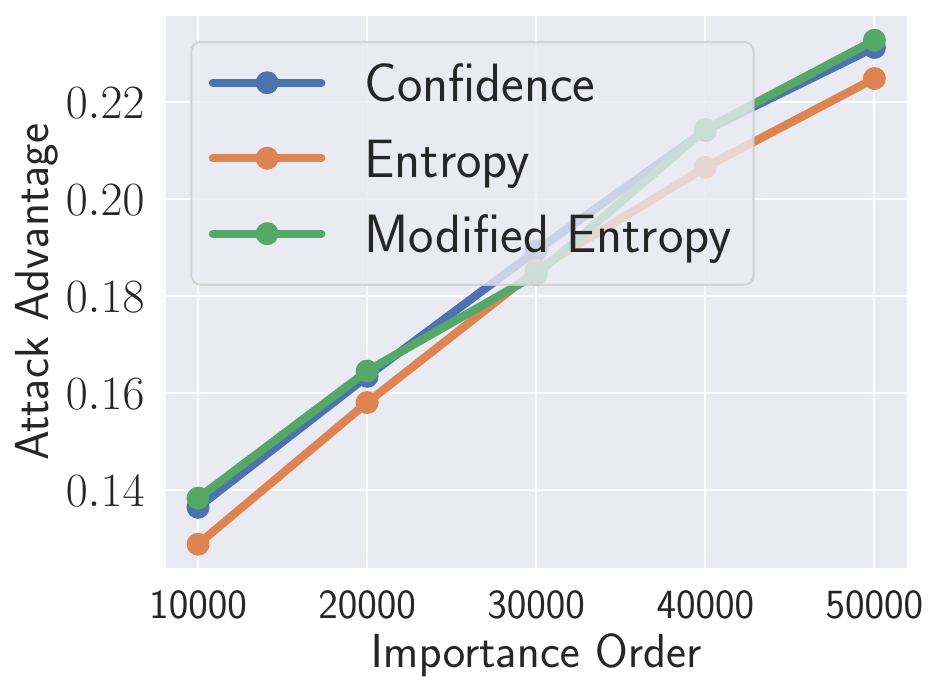}
\caption{CIFAR10}
\label{fig:mia_cifar}
\end{subfigure}
\begin{subfigure}{0.56\columnwidth}
\includegraphics[width=\columnwidth]{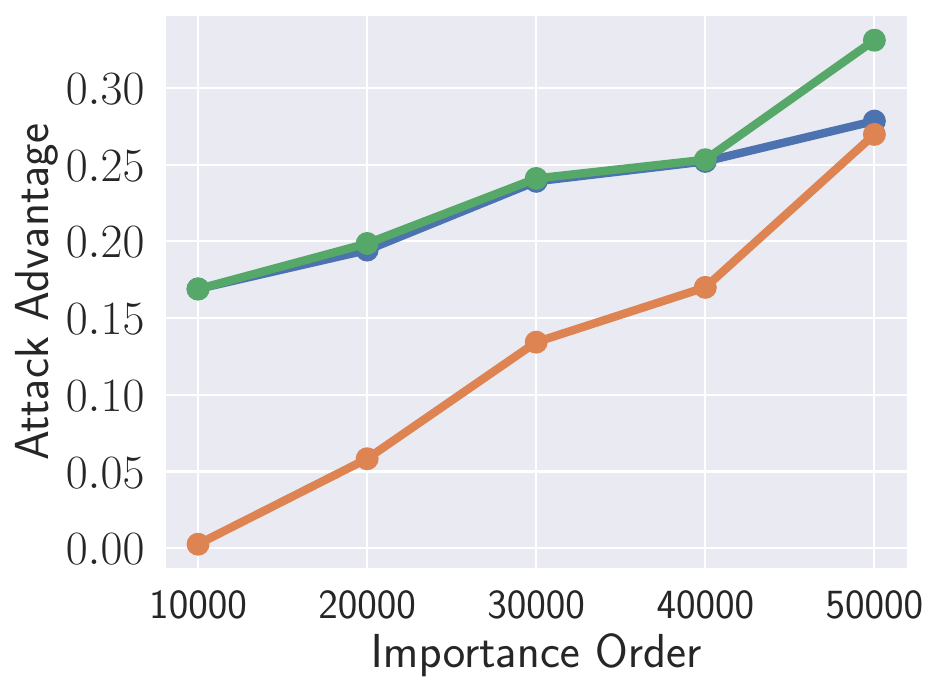}
\caption{CelebA}
\label{fig:mia_celeba}
\end{subfigure}
\begin{subfigure}{0.56\columnwidth}
\includegraphics[width=\columnwidth]{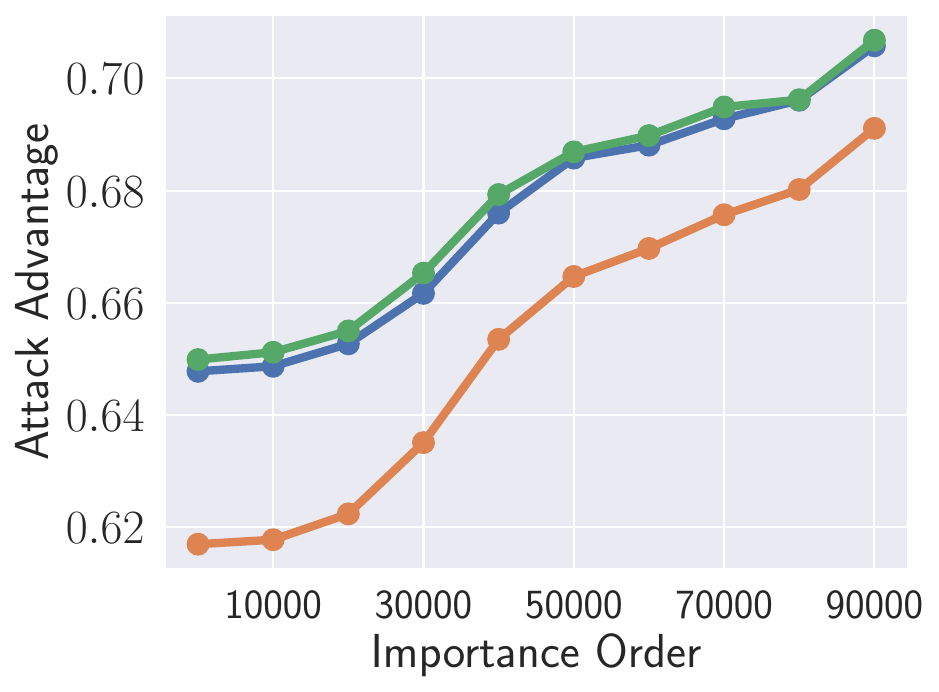}
\caption{TinyImageNet}
\label{fig:mia_tinyimagenet}
\end{subfigure}
\caption{Membership advantage: membership inference attack based on three metrics. 
Attack advantage steadily escalates as the importance value of the samples increases.}
\label{figure:mia_metricbase}
\end{figure*}

\begin{figure}[!t]
\centering
\includegraphics[width=0.68\columnwidth]{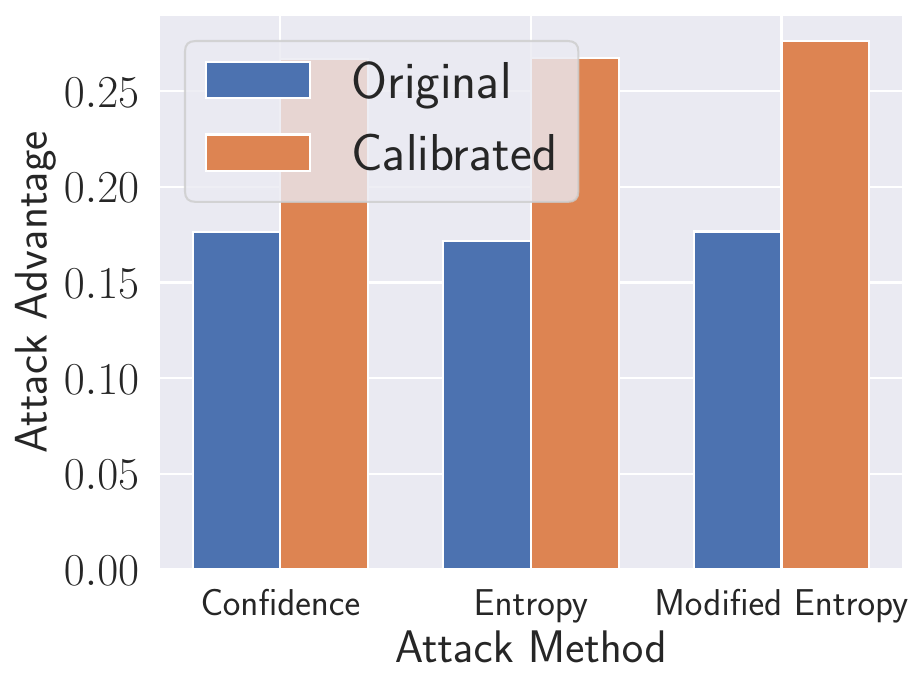}
\caption{Incorporation of importance values in calibrating membership inference metrics improves the attack performance, demonstrating the strength of employing sample-specific membership criteria.}
\label{figure:mia_cali}
\end{figure}

By grouping the samples based on their importance values in intervals of 10,000 samples (equivalent to the size of the testing dataset), we conducted the aforementioned attacks on these subsets. 
The membership advantage achieved for each subset is illustrated in~\autoref{figure:mia_metricbase}. 
Notably, a clear monotonic increase in attack advantage is observed as the importance value increases, establishing a positive correlation between the importance value and the susceptibility of membership inference.

\begin{figure*}[!t]
\centering
\begin{subfigure}{0.49\columnwidth}
\includegraphics[width=\columnwidth]{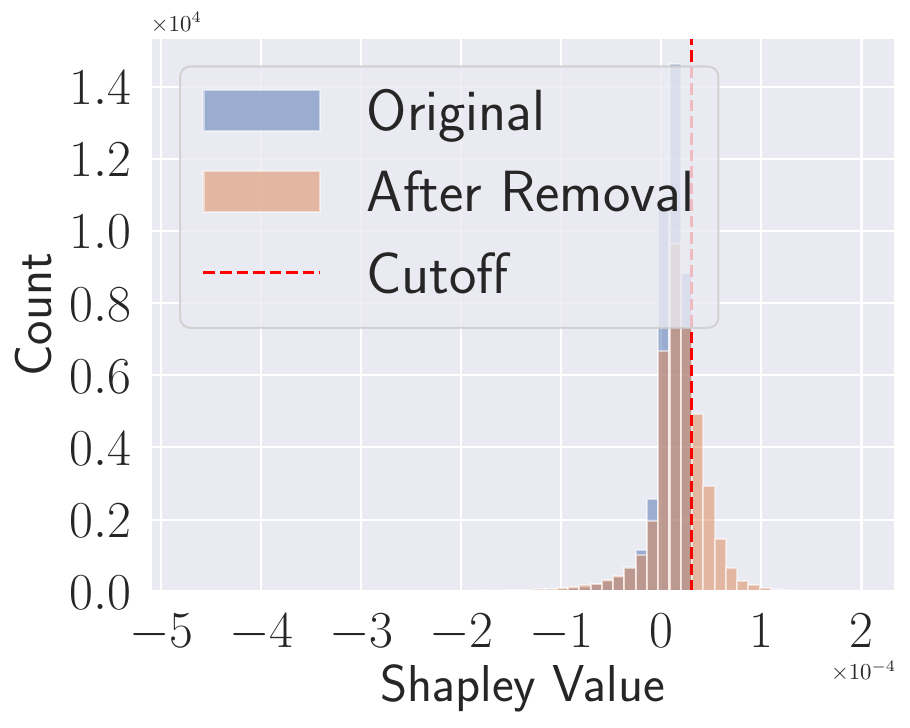}
\caption{Distribution Shift}
\label{fig:onion_shift}
\end{subfigure}
\begin{subfigure}{0.479\columnwidth}
\includegraphics[width=\columnwidth]{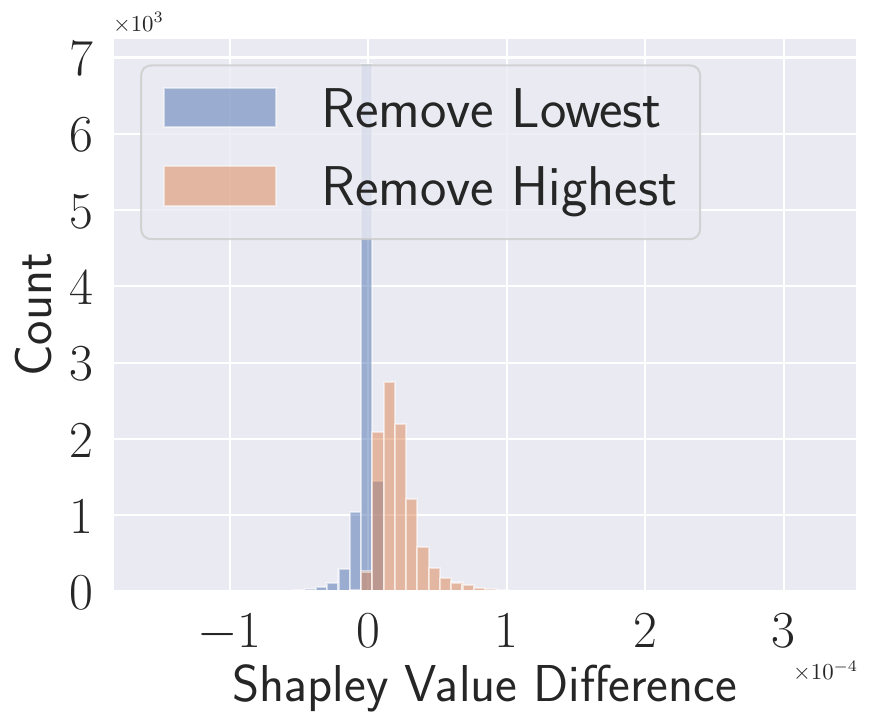}
\caption{CIFAR10}
\label{fig:onion_cifar}
\end{subfigure}
\begin{subfigure}{0.49\columnwidth}
\includegraphics[width=\columnwidth]{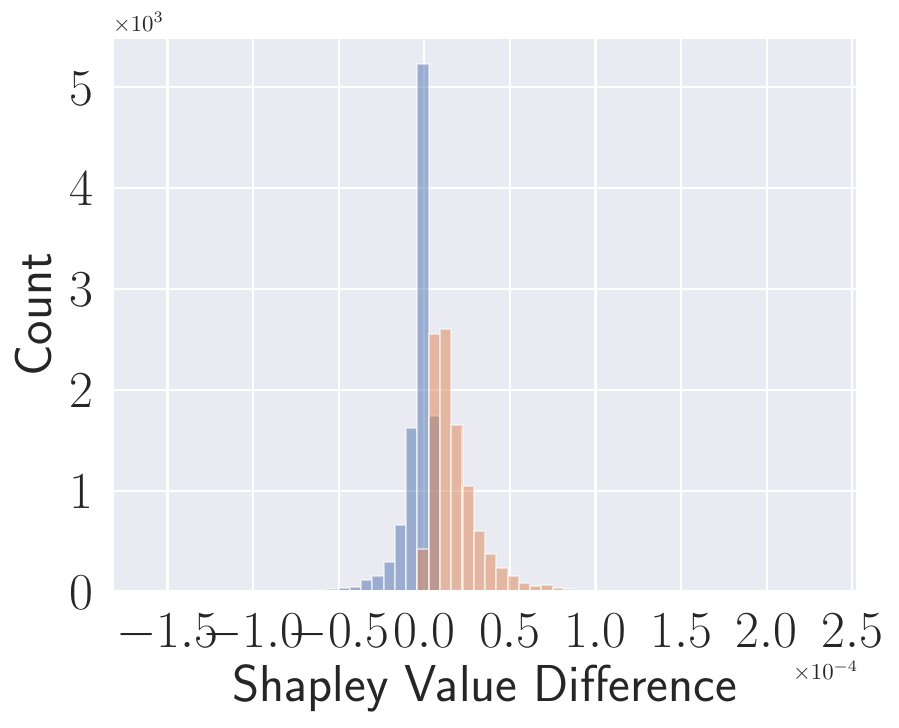}
\caption{Celeba}
\label{fig:onion_celeba}
\end{subfigure}
\begin{subfigure}{0.49\columnwidth}
\includegraphics[width=\columnwidth]{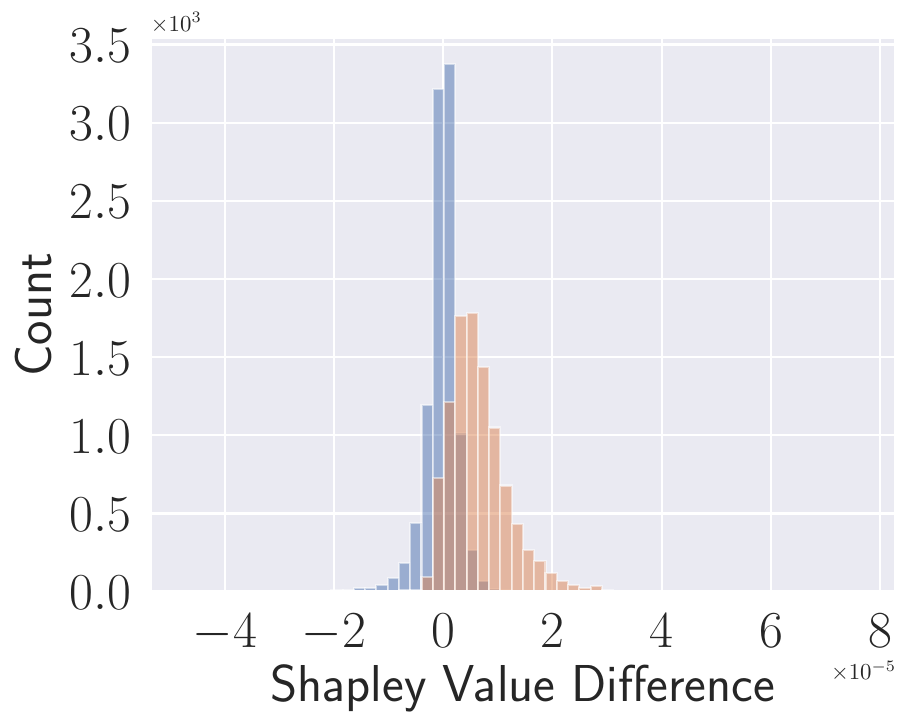}
\caption{TinyImageNet}
\label{fig:onion_tinyimagenet}
\end{subfigure}
\caption{The privacy onion effect can be extended to the importance distribution.
For~\autoref{fig:onion_shift}, we remove all examples with importance values larger than the red line. 
Samples that remain after removal have their importance value increase past the red line. 
Subsequent figures confirm that removing highly important samples elevates the significance of the remaining set, while the removal of less significant samples has no such effect, ruling out dataset size influence.}
\label{figure:mia_onion}
\end{figure*}

This empirical trend aligns with our expectations. 
As evidenced by the metrics in~\autoref{figure:loss} and~\autoref{figure:distance}, samples with lower importance inherently present greater learning challenges compared to their higher-importance counterparts. 
Even post-learning, these samples exhibit worse membership metrics compared to those of higher importance. 
This circumstance renders them challenging to distinguish from non-member samples, especially when certain non-member samples manifest a lower learning difficulty and consequently exhibit better metrics compared to the more challenging member samples.

Drawing from this insight, one potential strategy to enhance the efficiency of membership inference attacks is to compare each sample with others of comparable difficulty. 
A pragmatic approach to actualize this entails introducing sample-specific criteria. 
Rather than employing a uniform threshold across the entire testing dataset, such criteria should intricately correlate with the sample's characteristics, with their designated importance level serving as a robust quantitative index to reflect this alignment.

In this study, we initiate an exploration into the feasibility of such an approach.
To seamlessly integrate our method into the existing metric-based framework, we introduce a sample-specific threshold in a consistent manner: while maintaining a uniform threshold for the dataset, we modify the membership metrics by incorporating an importance-related term:
\[
\mathtt{CaliMem}(x) = \mathtt{OriMem}(x) + k\times\mathtt{Shapley}(x)
\]
Here, $\mathtt{OriMem}(x)$ denotes the conventional membership metric, including elements such as confidence, entropy, and modified entropy. 
The term $\mathtt{Shapley}(x)$ signifies the importance value attributed to the specific sample $x$, and $\mathtt{CaliMem}(x)$ represents the recalibrated membership metric.

As a proof of concept, we empirically determine the hyperparameter $k$ in an exploratory manner, adjusting its magnitude until optimal performance is attained. 
The experimental outcomes, depicted in \autoref{figure:mia_cali}, illustrate that the incorporation of importance calibration notably enhances the efficacy of metric-based attacks. 
Nevertheless, it is pertinent to acknowledge that identifying the optimal hyperparameter and devising more refined methods for integrating importance values warrant further investigation.

We also emphasize that this improvement does not necessitate additional requirements compared to standard attacks; specifically, the adversary does not need full access to the training dataset to obtain importance values. 
Although importance values cannot be calculated for single samples, most membership inference attacks assume access to a shadow dataset. 
We validate the feasibility of our approach using a shadow dataset. 
Specifically, we randomly select 1,000 samples from the CIFAR10 dataset to calculate their importance value, we first calculate the importance value for all samples using the whole CIFAR10 dataset as the ground truth. 
Then we assume the adversary can only access a shadow dataset containing 10,000 samples, and calculate the importance value for each sample using only the shadow dataset.
We found a correlation coefficient of 0.957 between these values, indicating that using the shadow dataset could provide a good approximation.

\subsection{Privacy Onion Effect}
\label{sec:onion}

Carlini et al.~\cite{CJZPTT22} have identified the onion effect of memorization, which refers to the phenomenon wherein ``\emph{removing the layer of outlier points that are most vulnerable to a privacy attack exposes a new layer of previously-safe points to the same attack.}'' 
Their research demonstrates this effect by removing samples that are at the highest risk of being compromised through membership inference, resulting in formerly safe samples becoming vulnerable to the attack.

Building upon the insights from the preceding section, our empirical findings confirm a positive correlation between membership inference vulnerability and data importance. 
This prompts an intriguing question: Does this effect reflect in the importance values assigned to the data? 
Put differently, when high importance samples are removed from a dataset, do previously designated low importance samples gain significance?

To avoid ambiguity, it is imperative to understand that the term ``importance'' in this context is not subjective or relative. 
The removal of high importance data points does not inherently increase the importance of those initially deemed low importance. 
Furthermore, it is conceivable for a dataset to exclusively consist of low importance samples. 
This clarification is indispensable; otherwise, the studied question may seem trivial.

Our results indicate that removing important samples indeed makes samples previously considered unimportant gain importance.
Specifically, upon removing 10,000 data points with the highest importance scores, we recalculated the importance for the remaining samples. 
As depicted in~\autoref{fig:onion_shift}, this removal led to a noticeable redistribution in data point importance, with previously low important data points now being assigned greater significance.

However, a note of caution is warranted in interpreting this result. 
The removal of a substantial number of samples (10,000 in this context) might introduce a baseline drift. 
Therefore, attributing the observed importance augmentation solely to the exclusion of important samples might be premature. 
To further validate our findings, we executed controlled experiments wherein we systematically excluded either the most or least significant data points. 
This approach mitigates potential biases stemming from dataset size discrepancies. 
To emphasize the impact of these exclusions, we quantified the importance value discrepancies for data points ranked between 10,000 and 20,000 in descending order of importance in the original dataset, as they remained for both removal procedures.

Our results, as visualized in~\autoref{fig:onion_cifar}, ~\autoref{fig:onion_celeba}, and~\autoref{fig:onion_tinyimagenet}, underscore the pronounced disparities in how the exclusion of high importance versus low importance data samples influences the remaining dataset's importance distribution. 
Using CIFAR10 as an illustrative case, removing the most significant data points caused 99.14\% of the remaining data points to be reevaluated as more important. 
In contrast, removing the least significant data points led to a 45.88\% decrease in importance for the affected data points. 
These findings robustly support our hypothesis that data points previously deemed of lesser importance assume greater significance when high importance data points are excluded, and such a conclusion cannot be attributed to dataset size variation.

\begin{figure}[!t]
\centering
\includegraphics[width=0.68\columnwidth]{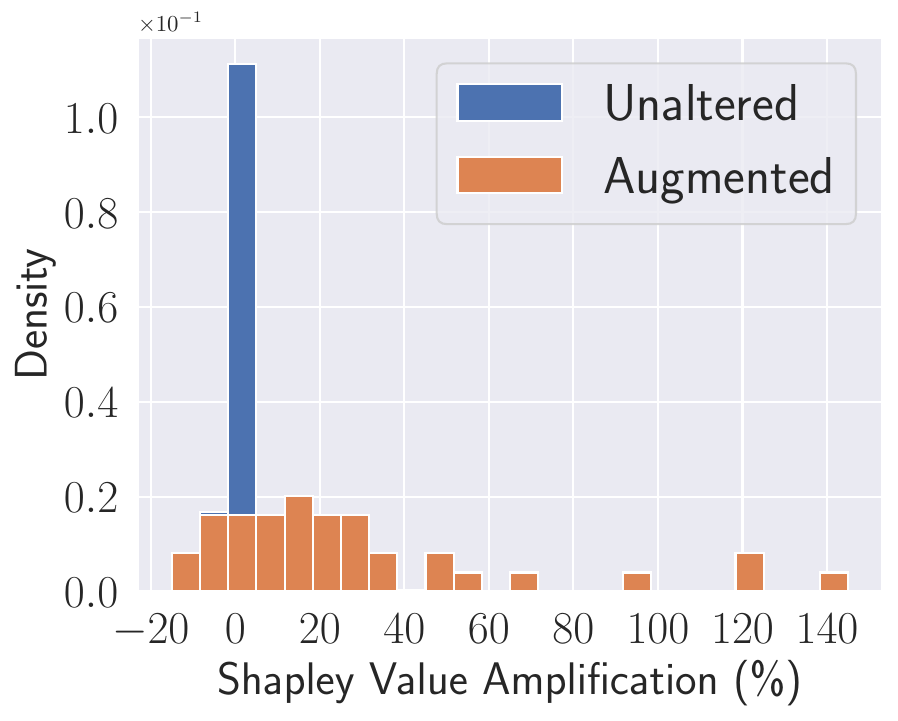}
\caption{Impact of sample duplication on importance values. 
Following the duplication process (16 duplications per sample), 45 out of 50 target samples showed a marked increase in importance, averaging a 53.02\% rise, while the importance of control samples (unduplicated) remained relatively stable. }
\label{figure:mia_active}
\end{figure}

\subsection{Actively Modify Sample Importance}
\label{sec:active_mia}

\begin{figure*}[h]
\centering
\begin{subfigure}{0.49\columnwidth}
\includegraphics[width=\columnwidth]{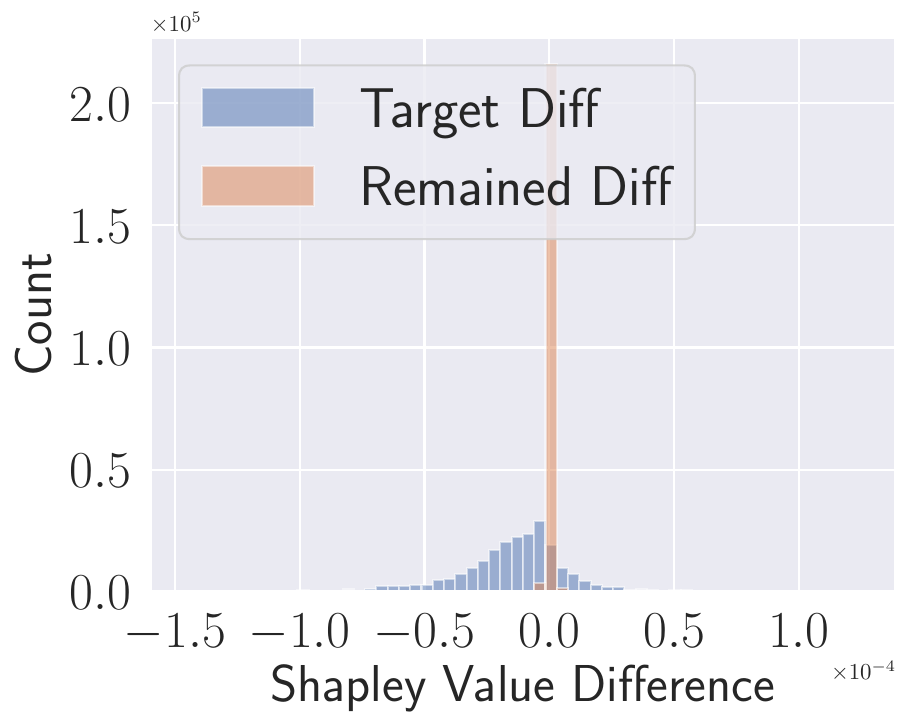}
\caption{ColorJitter}
\label{fig:replace_colorjitter}
\end{subfigure}
\begin{subfigure}{0.49\columnwidth}
\includegraphics[width=\columnwidth]{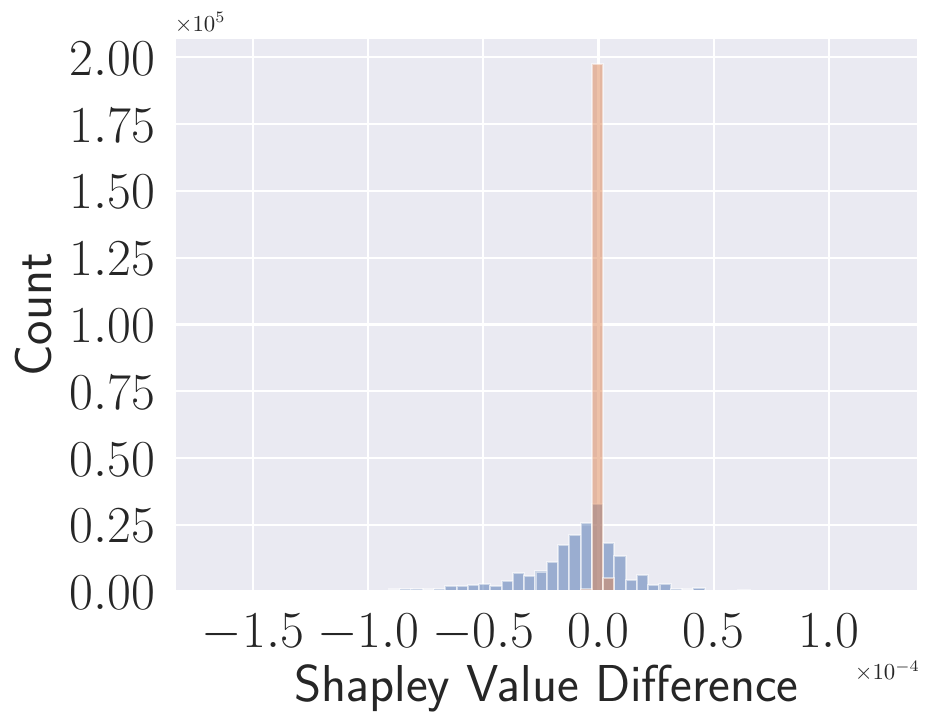}
\caption{GrayScale}
\label{fig:replace_grayscale}
\end{subfigure}
\begin{subfigure}{0.49\columnwidth}
\includegraphics[width=\columnwidth]{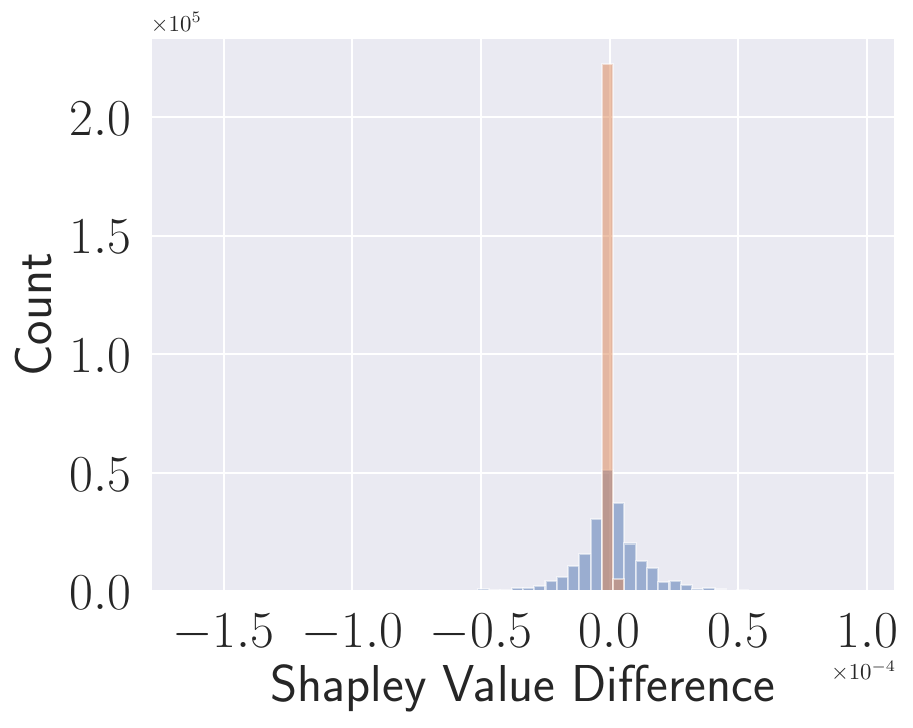}
\caption{HorizontalFlip}
\label{fig:replace_horizontalflip}
\end{subfigure}
\begin{subfigure}{0.49\columnwidth}
\includegraphics[width=\columnwidth]{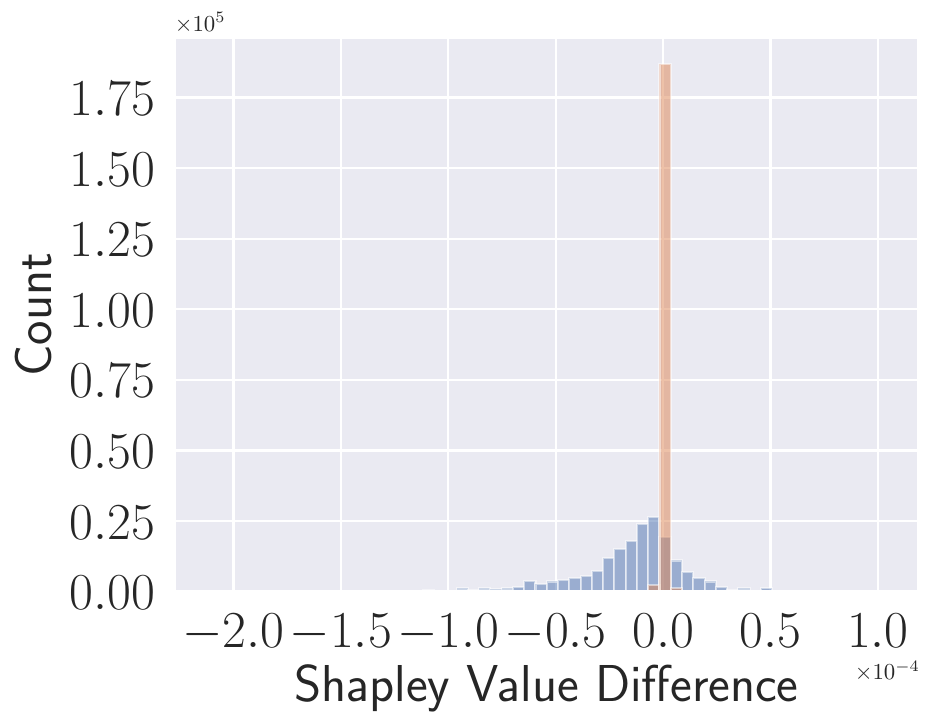}
\caption{VerticalFlip}
\label{fig:replace_verticalflip}
\end{subfigure}
\caption{Comparison of importance values before and after replacing 1,000 samples with their augmented versions using ColorJitter, Grayscale, HorizontalFlip, and VerticalFlip techniques. 
The plot illustrates that augmentation caused variable changes in importance, with some samples gaining and others losing importance.}
\label{figure:replace_influnce}
\end{figure*}

\begin{figure*}[h]
\centering
\begin{subfigure}{0.49\columnwidth}
\includegraphics[width=\columnwidth]{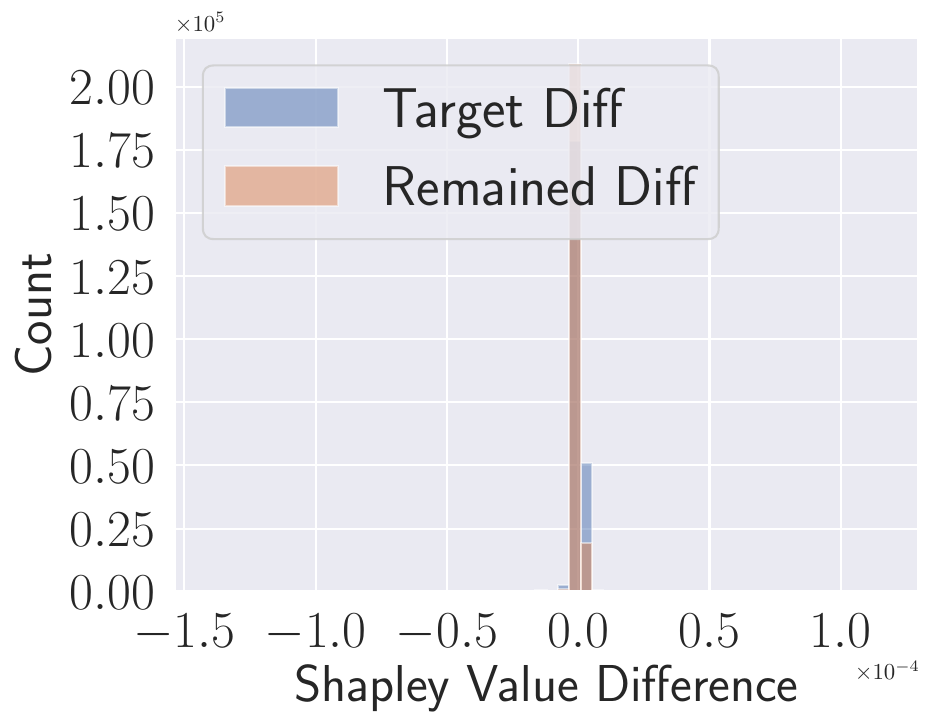}
\caption{ColorJitter}
\label{fig:add_colorjitter}
\end{subfigure}
\begin{subfigure}{0.49\columnwidth}
\includegraphics[width=\columnwidth]{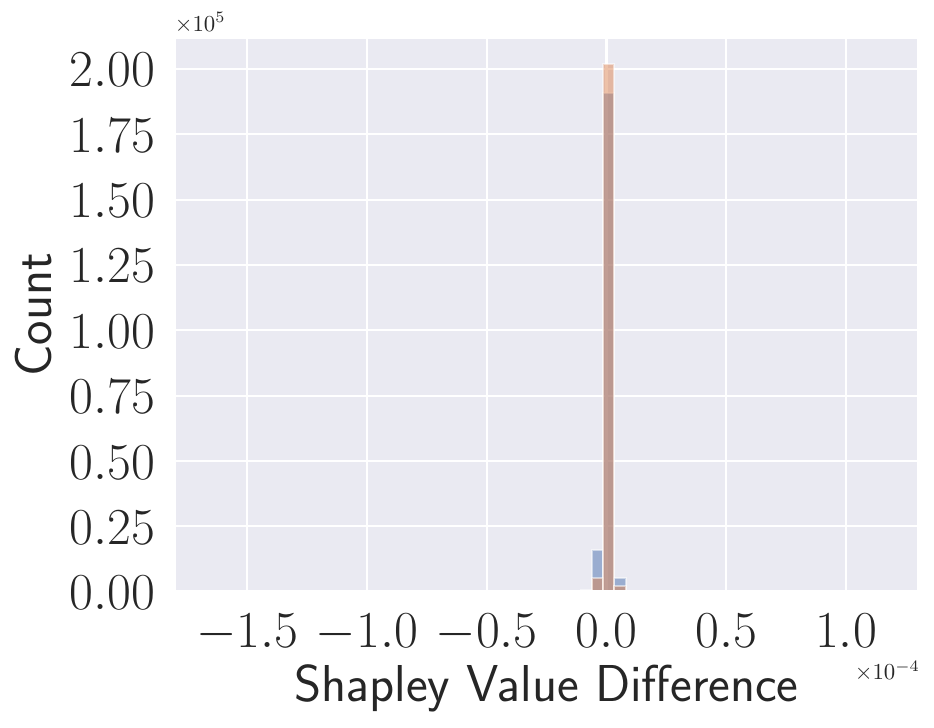}
\caption{GrayScale}
\label{fig:add_grayscale}
\end{subfigure}
\begin{subfigure}{0.49\columnwidth}
\includegraphics[width=\columnwidth]{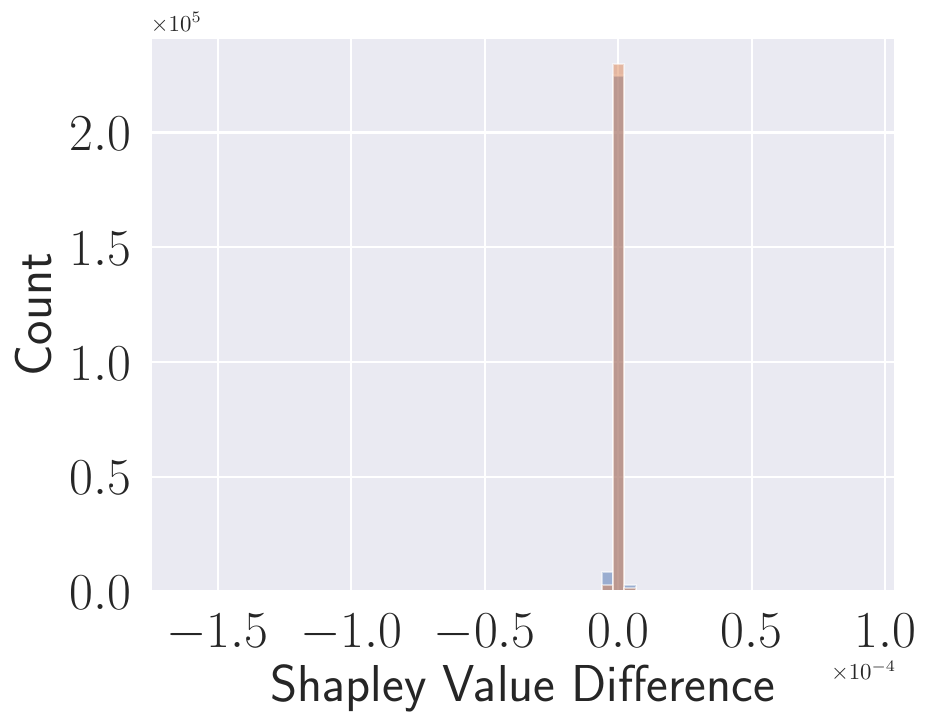}
\caption{HorizontalFlip}
\label{fig:add_horizontalflip}
\end{subfigure}
\begin{subfigure}{0.49\columnwidth}
\includegraphics[width=\columnwidth]{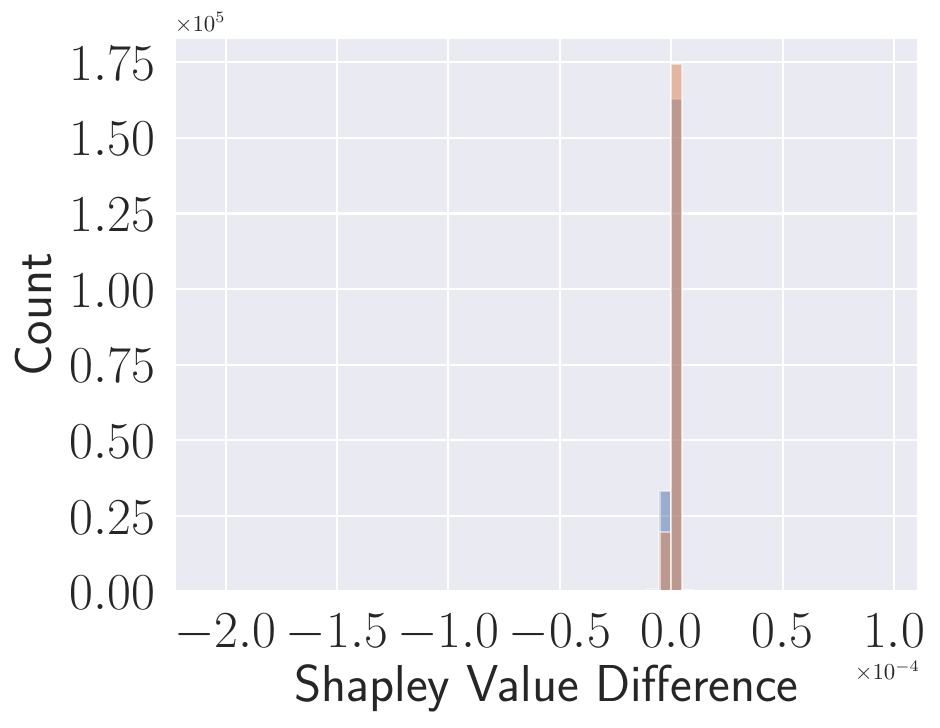}
\caption{VerticalFlip}
\label{fig:add_verticalflip}
\end{subfigure}
\caption{Analysis of the impact on importance values when 1,000 augmented samples are added to the original dataset, compared to a baseline scenario where the same 1,000 samples were duplicated. 
The figure demonstrates that the presence of both original and augmented samples had minimal effect on the importance of the original samples, showing no significant differences from simple duplication.}
\label{figure:add_influnce}
\end{figure*}

As discussed in the previous section, altering the dataset can influence the importance of samples. 
Given the observed linkage between membership vulnerability and importance value, an interesting question arises: can we actively use our findings to design more advanced attacks by modifying the importance of target samples?

However, directly altering sample importance is challenging due to the absence of a standardized method or framework. 
In this section, we explore an ad-hoc approach aimed at increasing the importance of target samples. 
Specifically, we select a set of target samples and duplicate each one multiple times with consistently incorrect labels. 
This strategy intuitively heightens the influence of these samples by causing the model to consider them as ``outliers'' due to the prevalence of incorrect duplicates. 
We tested this idea by duplicating 50 target samples 16 times and reassessing their importance values. 
As shown in ~\autoref{figure:mia_active}, the duplicated samples generally exhibited an increase in importance. 
Specifically, 45 of the 50 target samples experienced an increase in importance, with an average increase of 53.02\%, while the importance of the unaltered samples remained nearly constant.

This approach is exactly the membership poisoning attack proposed by Tramèr et al.~\cite{TSJLJHC22}, where they comprehensively demonstrated the method's efficiency. 
This indicates that increasing the importance of samples can be a practical approach to enhance their vulnerability. 
By revisiting their technique from a data importance perspective, we highlight that actively modifying sample importance could be a promising strategy for developing sophisticated attack techniques or formulating robust defenses. 

\subsection{Data Augmentation}
\label{sec:augmentation}

In previous discussions, it has been demonstrated that manipulating data samples can alter their importance values and, consequently, their susceptibility to attacks. 
Given that data augmentation is the most widely used method for data manipulation, it would be interesting to ask: does data augmentation affect the importance of a sample?

In our study, we examined the impact of four data augmentation techniques—ColorJitter, Grayscale, HorizontalFlip, and VerticalFlip—under two specific scenarios. 
In both scenarios, we selected 1,000 samples for augmentation while leaving the rest of the dataset unaltered:

\mypara{Augmented Versions Only} 
In this scenario, we aimed to investigate how data augmentation impacts the importance of the augmented samples and the unaltered samples in the dataset. 
Specifically, we \textit{replaced} 1,000 samples with their augmented versions, recalculated their importance values, and compared these values to the original. 
As illustrated in~\autoref{figure:replace_influnce}, we found that the augmented versions had variable effects on importance: some samples gained higher importance, while others lost it. 
Overall, a slight majority of the samples experienced a decrease in importance following augmentation. 
However, the augmented samples had a negligible impact on the remaining non-augmented samples.

\mypara{Original and Augmented Versions} 
In this scenario, we examined the effect of having both augmented and original versions of certain samples in the dataset. 
We \textit{added} 1,000 augmented samples to the original dataset and recalculated the importance values for the original samples. 
To control for dataset size, we also considered a baseline case where the same 1,000 samples were duplicated. 
As shown in~\autoref{figure:add_influnce}, the presence of both original and augmented samples had minimal impact on the importance of the original samples, with no significant differences compared to simple duplication.

We acknowledge that more complex augmentation techniques, such as those using generative models, may have different effects. 
The exploration of these complex augmentation techniques remains an avenue for future research.

\mypara{Takeaways}
Our findings highlight the vulnerability of high importance samples to membership inference attacks. 
Significant differences were observed in the behavior of high importance and low importance samples, particularly in the low false-positive rate region, where high importance samples exhibited substantially higher true-positive rates. 
This emphasizes the necessity of addressing the privacy risks associated with high importance samples and implementing effective safeguards. 
Simultaneously, it encourages researchers to explore strategic sampling methods to enhance the effectiveness of privacy audits.

The observation also suggests a potential enhancement to membership inference attacks through the introduction of sample-specific criteria. 
We empirically validate the practicality of using importance values to calibrate membership metrics, thereby enhancing attack efficiency. 

Moreover, our findings reveal the ``privacy onion effect'' within the sample importance distribution, where previously overlooked samples gain importance when key samples are removed. 
Furthermore, by revisiting an advanced membership poisoning attack from the perspective of data importance, we suggest that actively manipulating sample importance can be a potent strategy for developing sophisticated cybersecurity measures, both offensive and defensive, but finding general manipulating methods needs further investigation.

\begin{figure}[!b]
\centering
\includegraphics[width=0.88\columnwidth]{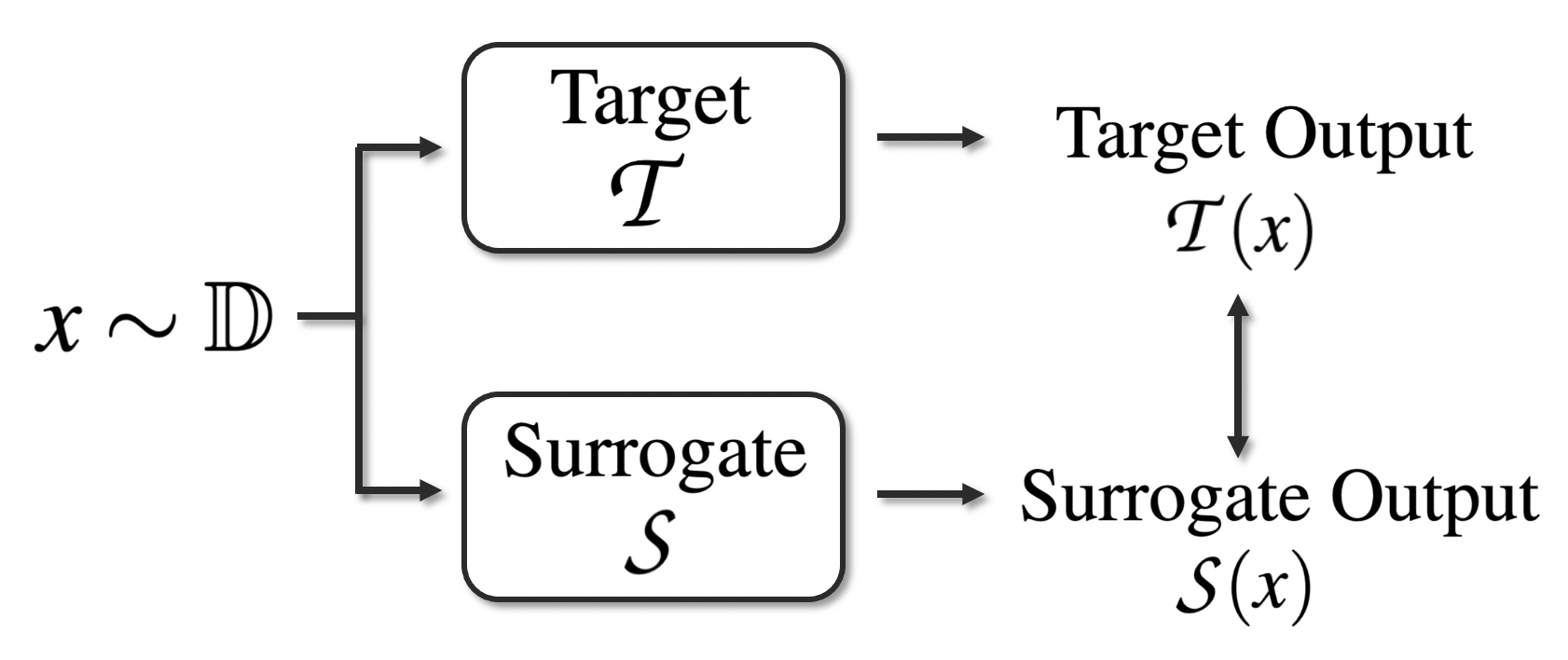}
\caption{The workflow of model stealing attack, the adversary leverages the target model to guide the surrogate model.}
\label{figure:modelsteal_workflow}
\end{figure}

\begin{figure*}[!t]
\centering
\begin{subfigure}{0.53\columnwidth}
\includegraphics[width=\columnwidth]{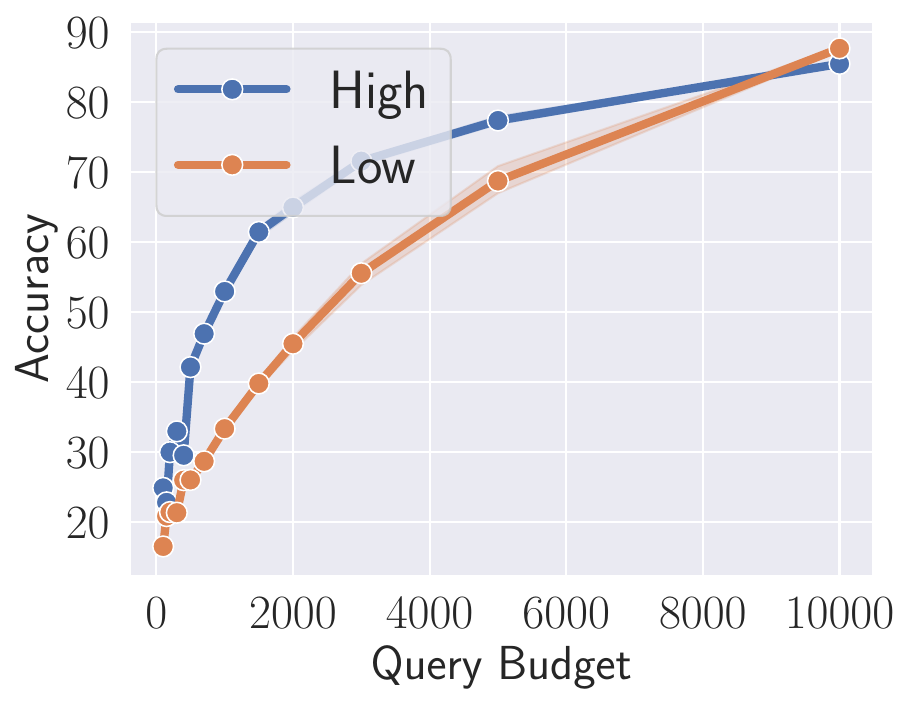}
\caption{CIFAR10}
\label{fig:steal_cifar}
\end{subfigure}
\begin{subfigure}{0.53\columnwidth}
\includegraphics[width=\columnwidth]{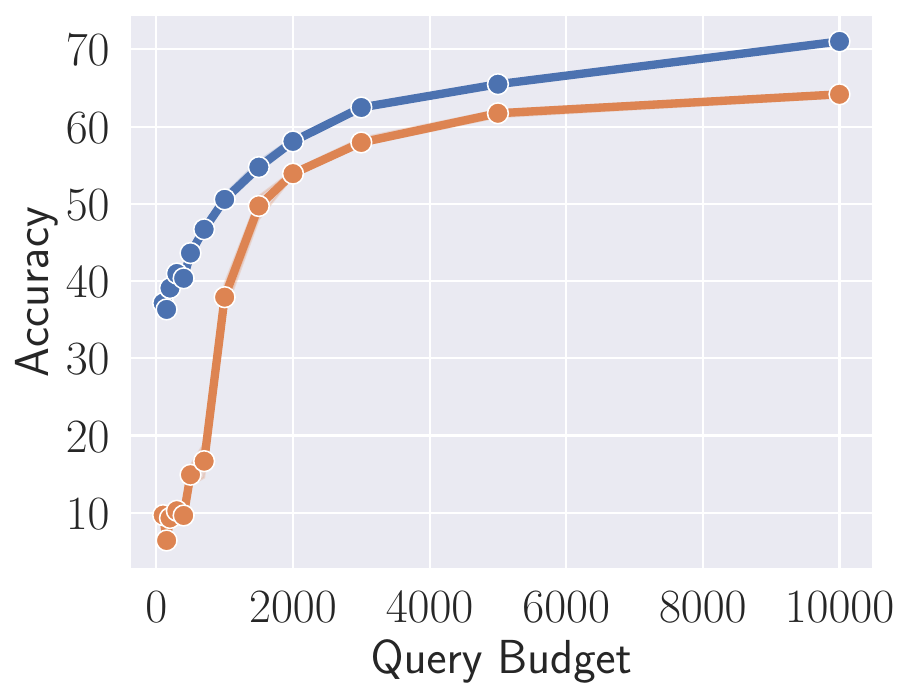}
\caption{CelebA}
\label{fig:steal_celeba}
\end{subfigure}
\begin{subfigure}{0.53\columnwidth}
\includegraphics[width=\columnwidth]{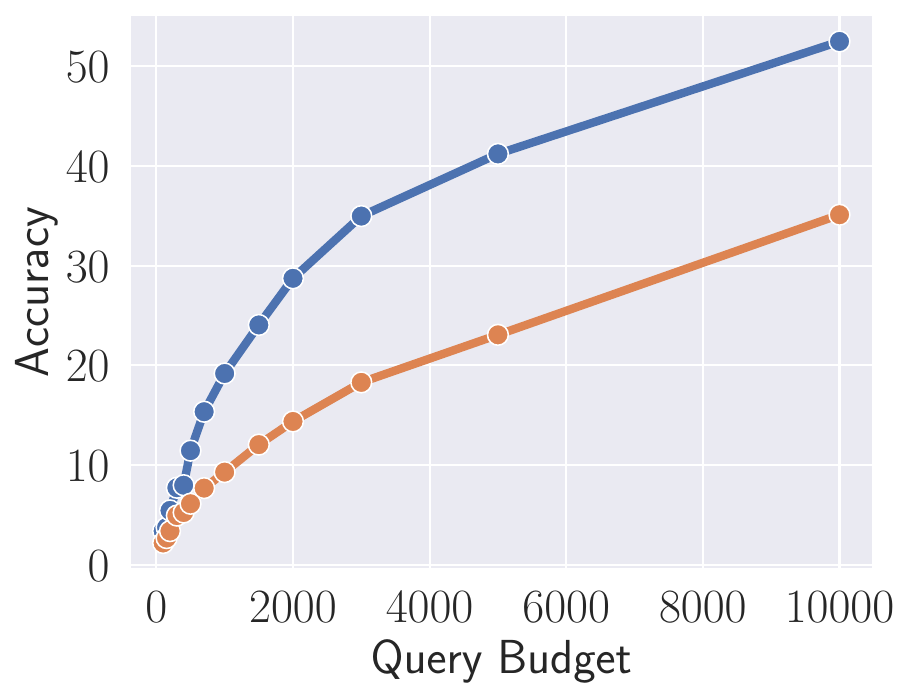}
\caption{TinyImageNet}
\label{fig:steal_tinyimagenet}
\end{subfigure}
\caption{Model stealing attack that queried with data from the same distribution. 
High importance samples exhibit greater efficiency in stealing models when the target model is trained on the same distribution as the query
distribution. }
\label{figure:modelsteal}
\end{figure*}

\section{Model Stealing}
\label{sec:ms}

Model stealing attack~\cite{TMWP21,TZJRR16,SAB22,JCBKP20,CJM20} differs from membership inference attack as it aims to compromise the confidentiality of the model itself rather than exploiting privacy information about training samples. 
This type of attack does not have information about the target model's architecture or parameters but seeks to create a surrogate model that emulates the functionality of the target model. 
Such attacks can be employed by adversaries for various purposes, including monetary gains or as a preliminary step for subsequent attacks~\cite{PMGJCS17}.

The workflow of a model stealing attack is visualized in~\autoref{figure:modelsteal_workflow}. 
The adversary samples data from a specific distribution $\mathbb{D}$ and simultaneously queries the target and surrogate models. 
To ensure similarity between the surrogate and target models, the adversary optimizes the surrogate model to produce similar outputs $\mathcal{S}(x)$ as the target outputs $\mathcal{T}(x)$. 
While the attack approach is straightforward, selecting an appropriate query data distribution poses a challenge, as it directly impacts the stolen accuracy and query efficiency. 
Recent research has explored efficient and data-free methods for launching these attacks~\cite{OSF19,KPQ21,SAB22}, yet the question of selecting high-quality samples when the target task is known remains intriguing.

In this work, we focus on the primary scenario where the adversary can query the target model to obtain corresponding posteriors, while having knowledge of the target task. 
Specifically, the adversary could query the model with a dataset from the same or a similar distribution. 
This scenario has practical applications, such as creating a surrogate model to facilitate further attacks or to save on labeling costs. 
We limit our discussion to this primary scenario and do not delve into more advanced model stealing techniques that focus on reducing the dataset assumption, as our interest lies in understanding how different data interact with the model stealing process.

Our goal is to investigate whether query samples with different importance values exhibit varying efficiency in stealing models. 
We explore two settings in our experiments. 
First, we launch the attack using query data from the same distribution as the target model trained on. 
For example, if the target model is trained on CIFAR10, we employ CIFAR10 data to query the model. 
The second scenario involves using data from different distributions, specifically CelebA and TinyImageNet, to query the CIFAR10 model.
We choose accuracy and query budget as the metrics to evaluate the success of the attack, using less query budget to achieve higher accuracy denotes better attack performance. 

\subsection{Same Distribution Query}
\label{sec:ms_same}

Three target models were trained using a standard training procedure, resulting in testing accuracies of $95.15\%$ for CIFAR10, $79.05\%$ for CelebA, and $65.01\%$ for TinyImageNet. 
After training the target models, our attack solely interacts with the target models through their outputs without accessing or reading their parameters.

To initiate the attack, we establish a query budget ranging from 100 to 10,000. 
Once the query budget is determined, we prioritize collecting high importance data until the budget is exhausted, and the same principle applies to the collection of low importance data.

The attack results are illustrated in~\autoref{figure:modelsteal}, highlighting the superior efficiency of high importance samples in the model stealing process. 
For instance, when the query budget is set to $1000$, high importance data steal a CIFAR10 model with $53.77\%$ accuracy, which is $1.6\times$ higher than the model stolen by low importance data ($33.29\%$).
This trend holds true for the other two datasets as well. 
Taking TinyImageNet as an example, when the query budget is 1000, high importance data yield a model accuracy of $19.25\%$, whereas low importance data only result in a model accuracy of $9.25\%$, exhibiting a notable $2.1$-fold disparity.

One plausible explanation for this difference may arise from variations in class balance, given that the query sets are chosen based on sample importance. 
It is conceivable that the low importance query set may lack samples from certain classes, thereby resulting in suboptimal performance. 
Prior research has suggested that a more balanced data distribution could potentially improve model stealing performance \cite{SAB22,LWBZ24}. 
However, upon examining the data distribution for both high and low significance samples, we observed no significant disparities. 
For example, in the case of CIFAR10, the entropy values for the top-10,000 high importance and low importance distributions were 3.282 and 3.245, respectively. 
Even when considering 1000 samples, the corresponding entropy values were 3.161 (high importance) and 3.229 (low importance). 
For context, a perfectly uniform distribution has an entropy of 3.322. 
This indicates that both the high and low importance subsets closely approximate a uniform distribution. 
Such findings reinforce our assertion that high importance data can indeed augment model stealing performance, mitigating concerns related to distributional biases.

\begin{figure}[!t]
\centering
\begin{subfigure}{0.47\columnwidth}
\includegraphics[width=\columnwidth]{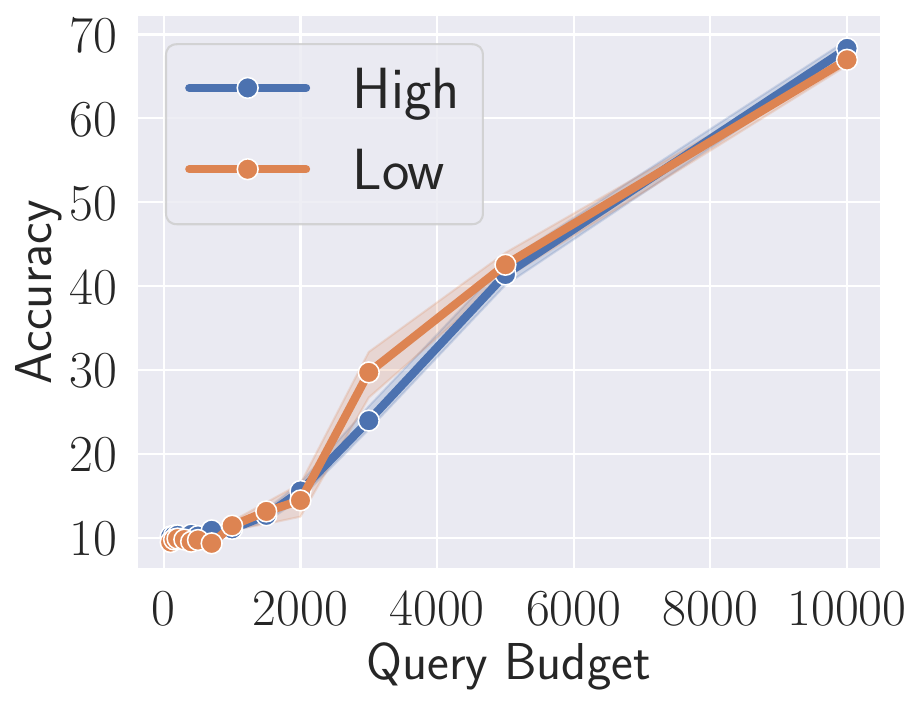}
\caption{CelebA}
\label{fig:celeba_steal_cifar}
\end{subfigure}
\begin{subfigure}{0.47\columnwidth}
\includegraphics[width=\columnwidth]{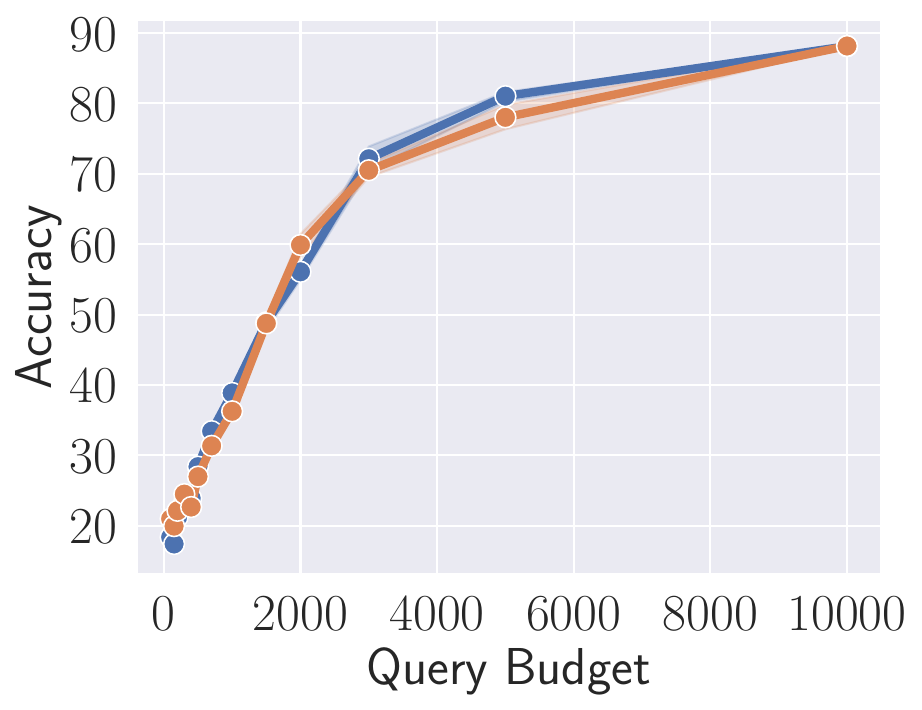}
\caption{TinyImageNet}
\label{tiny_steal_cifar}
\end{subfigure}
\caption{Model stealing attack that queried with data from different distributions. 
The target model is trained on the CIFAR10 task. 
Results show that importance does not transfer between different tasks.}
\label{figure:modelsteal_diff_steal}
\end{figure}

\begin{figure*}[!t]
\centering
\begin{subfigure}{0.53\columnwidth}
\includegraphics[width=\columnwidth]{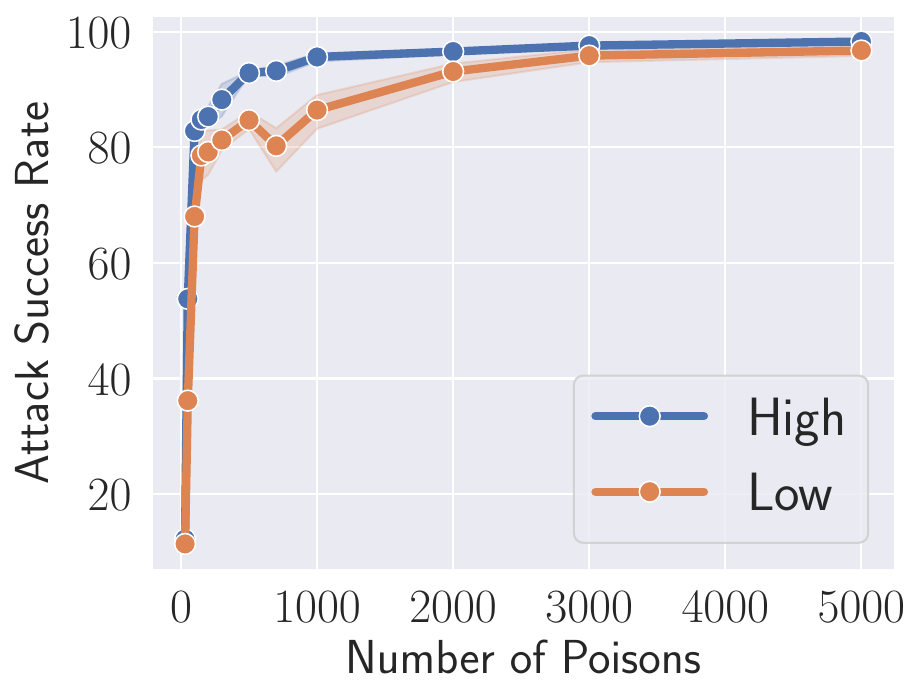}
\caption{CIFAR10}
\label{fig:backdoor_asr_cifar}
\end{subfigure}
\begin{subfigure}{0.53\columnwidth}
\includegraphics[width=\columnwidth]{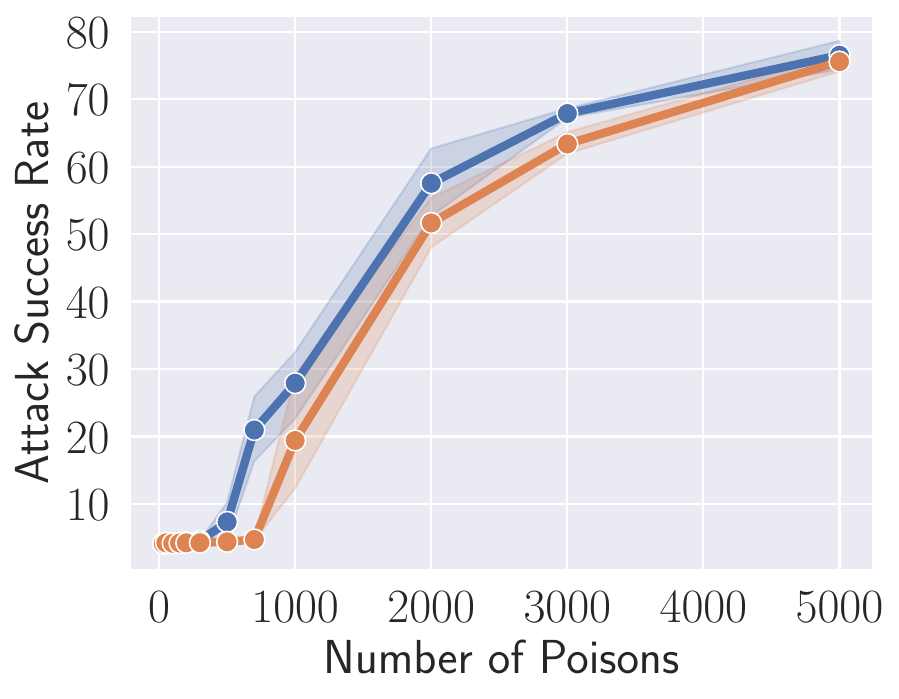}
\caption{CelebA}
\label{fig:backdoor_asr_celeba}
\end{subfigure}
\begin{subfigure}{0.53\columnwidth}
\includegraphics[width=\columnwidth]{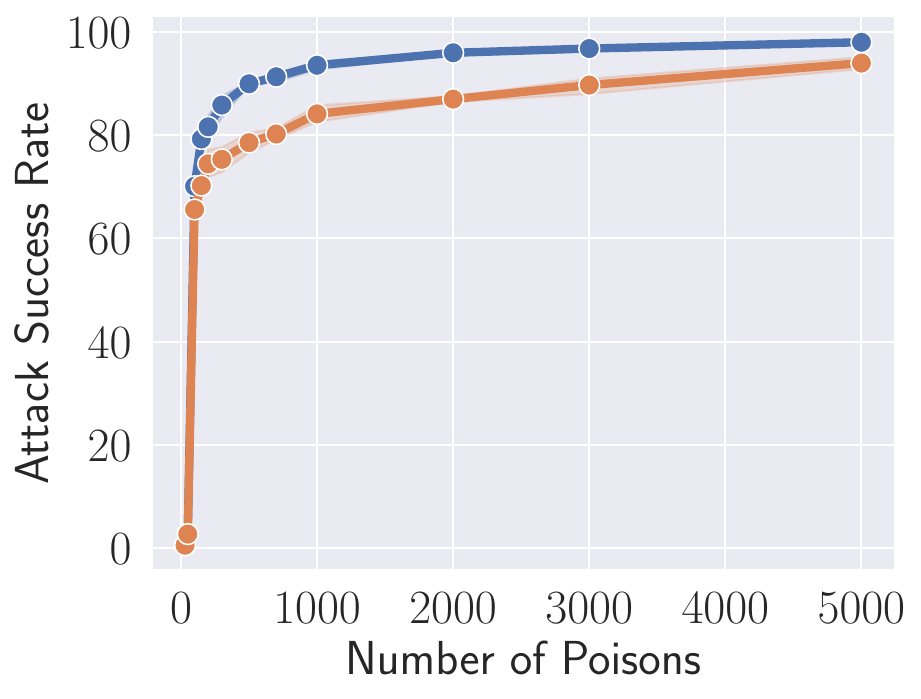}
\caption{TinyImageNet}
\label{fig:backdoor_asr_tinyimagenet}
\end{subfigure}
\caption{Relationship between attack success rate and the poisoning rate, high importance samples enhance the efficiency of the poisoning process, particularly when the poisoning rate is small.}
\label{figure:backdoor_asr}
\end{figure*}

\subsection{Different Distribution Query}
\label{sec:ms_diff}

The previous section highlighted the enhanced efficiency of high importance samples in stealing models trained on the same task. 
However, it remains uncertain whether this efficiency persists when the target model is trained using a different dataset or task, and whether importance values can be transferred across tasks. 

To investigate, we conducted experiments involving query data that differed from the distribution used to train the target model. 
Specifically, we employed a CIFAR10 model as the target model and queried it with the CelebA and TinyImageNet datasets.

Interestingly, as depicted in~\autoref{figure:modelsteal_diff_steal}, we observed that the advantage of high importance samples disappeared in this cross-task scenario. 
When the query budget was consistent, the stolen accuracy for both high and low importance samples was comparable. 
This suggests that samples deemed important for one task may not transfer effectively to arbitrary tasks.

\mypara{Takeaways}
Our findings demonstrate that high importance samples exhibit greater efficiency in stealing models when the target model is trained on the same distribution as the query distribution. 
Importantly, this enhanced efficiency cannot be solely attributed to distribution bias.
This suggests that adversaries, when aware of the target task, can employ high importance samples to optimize attack performance with a reduced query budget. 
However, this conclusion does not hold when the target task differs from the query distribution.
Consequently, this implies that selecting a group of high importance samples as a ``universal'' query set for efficient model stealing attacks, regardless of the target task, is not feasible.

\section{Backdoor Attack}
\label{sec:backdoor}

Backdoor attack~\cite{GDG17,LMALZWZ18,SWBMZ22} is a training-time attack that involves actively interfering with the training process to manipulate the resulting model. 
Its primary objective is to introduce malicious behavior into the model, making it behave like a benign model for normal inputs. 
However, when a specific trigger is detected, the backdoored model intentionally misclassifies the input to a predetermined class. 
This type of attack can have severe consequences, such as compromising the integrity and reliability of the model, leading to potential security breaches, data manipulation, or unauthorized access to sensitive information. 

Despite the severe consequences that a backdoor attack may cause, the attack itself is relatively easy to achieve by poisoning the training dataset, thereby posing an even stronger threat. 
For instance, a straightforward attack approach called BadNets~\cite{GDG17} adds a fixed trigger to a portion of the training dataset, resulting in a perfect attack where almost all triggered samples are misclassified into the target class, while the accuracy on the original task remains largely unaffected.

In the context of backdoor attacks, the poison rate plays a critical role as it directly influences the effectiveness and concealment of the attack. 
A higher poison rate can lead to an increased attack success rate, but it also raises the risk of detection since a large number of samples need to be modified. 
Conversely, a lower poison rate may offer better concealment, but it may not achieve optimal attack performance. 
Additionally, there are situations where the adversary can only control a small set of samples, making it impossible to poison a large number of samples to achieve the attack. 
Consequently, the problem of backdooring a model with a limited poison rate becomes an interesting and challenging research question.

In this section, we conduct empirical investigations to explore whether poisoning samples with different importance levels influences the attack performance under the same poison rate. 
We utilize two metrics to evaluate the attack performance:
\begin{enumerate}
    \item \textbf{Accuracy}. 
    This metric assesses the deviation of the backdoored model from the clean model. 
    We measure the performance of the backdoored model on the clean dataset, and a successful attack should result in accuracy close to that of the clean model, making it difficult to detect.
    \item \textbf{Attack Success Rate (ASR)}. 
    This metric evaluates the functionality of the backdoored model and is measured on the triggered dataset. 
    A desirable backdoored model should exhibit a high ASR, indicating its ability to misclassify all triggered samples into the target label.
\end{enumerate}
By analyzing these metrics, we aim to gain insights into the influence of data importance on the attack performance and further understand the trade-offs between attack effectiveness and concealment in the context of backdoor attacks.

In this part, we adopt the same approach as BadNets to backdoor the model, with hyperparameter details provided in~\refapp{append:bd_hyperparameter}. 
Additionally, we validate the generalizability of our conclusion across five other backdoor attacks—Blend~\cite{CLLLS17}, SSBA~\cite{LLWLHL21}, LF~\cite{ZPMJ21}, SIG~\cite{BKT19}, and CTRL~\cite{LPXDJYW23}—which utilize various trigger patterns or target different learning paradigms, as discussed in~\refapp{append:more_backdoor}.

We present the visualized attack success rate in~\autoref{figure:backdoor_asr}. 
As depicted in the figure, there is a noticeable increase in the attack success rate as the number of poisons increases. 
Concurrently, we observe significant differences between poisoning high importance samples and low importance samples. 
Specifically, poisoning an equal number of high importance samples proves to be more effective in increasing the attack success rate compared to poisoning low importance samples. 
This phenomenon becomes more pronounced when the poisoning rate is small. 
For instance, in the case of CIFAR10, with a poisoning size of 50, poisoning high importance data results in a model with an ASR of $54.42\%$, whereas poisoning low importance data only achieves $37.74\%$, indicating a $1.44\times$ advantage. 
Similar trends can be observed across the other two datasets.

However, we also find that when the poisoning rate is large, the difference is not significant. 
We believe this is due to the trade-off between the importance advantage and the attack upper bound. 
As the number of poisons increases, the advantage of using high importance data becomes more evident. 
However, achieving the optimal ASR for the backdoor attack does not require a large amount of data. 
Generally, poisoning approximately 10\% of the dataset is sufficient. 
Therefore, as the number of poisons increases, the gap between high importance and low importance samples is reduced. 
Nevertheless, it is still observed that poisoning high importance samples requires poisoning fewer samples to achieve its optimal ASR.

\begin{figure*}[!t]
\centering
\includegraphics[width=1.66\columnwidth]{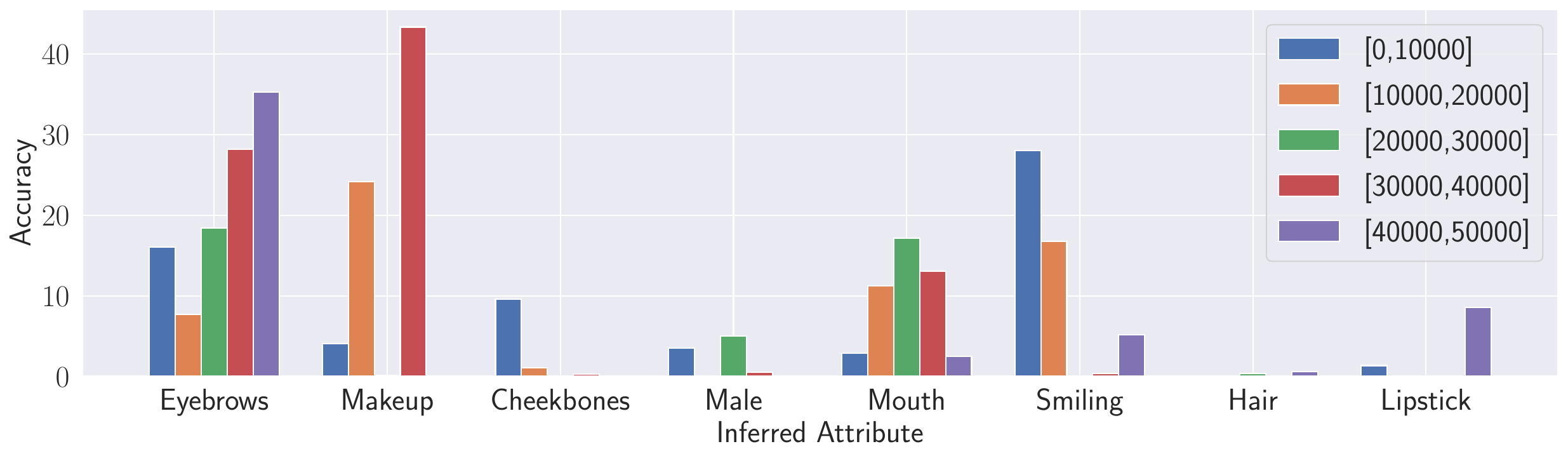}
\caption{Attribute inference attack performance on different attributes, no significant correlation between attribute inference attacks and data importance is observed. 
These confirm our hypothesis that the importance of data samples is context-dependent and can vary based on the specific task at hand.}
\label{figure:attrinf}
\end{figure*}

In scenarios where adversaries have limited access to data, determining the true importance of samples can be challenging, which impacts the feasibility of selectively poisoning high importance samples. 
In this case, we empirically demonstrate that calculating the importance value using just a fraction of the training set can provide a good approximation of the true importance. 
For example, with just 2\% of the CIFAR10 data available, the computed importance values correlate strongly with those derived from the entire dataset, achieving a correlation coefficient of \(0.811 \pm 0.016\). 
The accuracy of these approximations improves with more data: with 5\% of the data, the correlation coefficient rises to \(0.899 \pm 0.006\), and with 20\% of the data, it exceeds $0.96$. 
These results demonstrate that even with limited data access, it is feasible to closely estimate the importance of samples, facilitating effective attack planning under realistic constraints.

Additionally, our investigation into the impact on clean accuracy reveals no significant trends suggesting that poisoning samples of differing importance levels affects clean accuracy. 
In both scenarios, the influence on clean accuracy remains below $2\%$, indicating the concealment of the backdoor attack. 
Due to space constraints, detailed results are deferred to~\refapp{append:bd_acc}.

\mypara{Takeaways}
Our experimental results demonstrate that poisoning high importance samples enhances the efficiency of the poisoning process, particularly when the poisoning rate is small. 
This insight offers valuable guidance for developing attack strategies aimed at compromising models with restricted data accessibility. 
Beyond refining trigger patterns for effective injections, prioritizing the poisoning of high importance samples emerges as a promising approach. 
On the other hand, the influence on clean accuracy does not yield a definitive conclusion, as poisoning either type of data has a limited impact on clean accuracy.

\section{Attribute Inference Attack}
\label{sec:attrinf}

Attribute inference attack is a privacy attack that aims to infer sensitive attributes that are not directly related to the original task of a machine learning model.  
For instance, a model trained to predict age from profile photos may unintentionally learn to predict race as well~\cite{SR20,MSCS19,SS20}.
This type of attack has significant implications for privacy and fairness, as the inadvertent leakage of sensitive attributes can have far-reaching consequences, including the violation of privacy rights, potential discrimination, and the undermining of trust in machine learning systems.

\begin{figure}[!b]
\centering
\includegraphics[width=0.99\columnwidth]{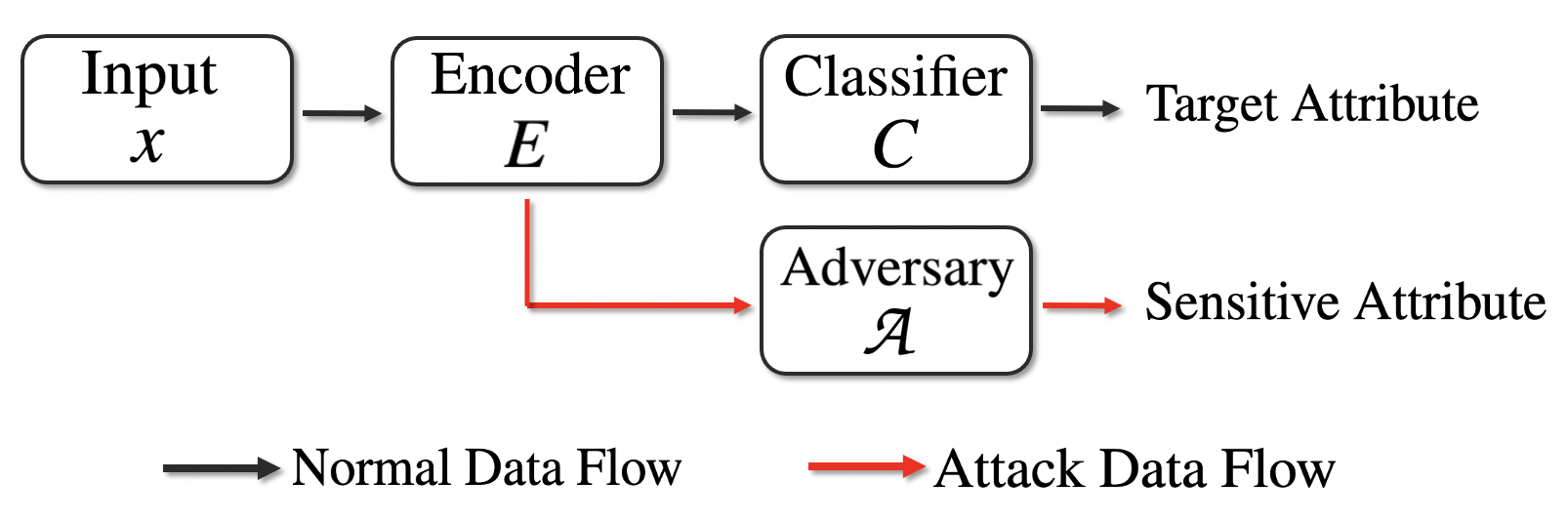}
\caption{The attack scenario for attribute inference attack.
The adversary can get the embeddings and aims to infer sensitive attributes based on the information encoded in embeddings.}
\label{figure:attr_workflow}
\end{figure}

In this work, we focus on a commonly considered attack scenario as depicted in~\autoref{figure:attr_workflow}, where the adversary exploits the embeddings of a target sample obtained from the target model to predict its sensitive attributes~\footnote{We acknowledge that there exists a separate line of research on attribute inference attacks targeting \textit{tabular data}~\cite{MDKLB22,JE22,FLJLPR14}, which primarily aims to \textit{reconstruct} missing attribute values in original records. 
Given that these attacks employ different technical methodologies and pursue distinct objectives,  our conclusions may not necessarily apply to such work.}
To perform attribute inference, the adversary assumes auxiliary information about the training dataset and collects a shadow dataset from similar distributions. 
They train a shadow model to mimic the behavior of the target model and use the embeddings and sensitive attributes to train an attack classifier.

In this section, we investigate the impact of data importance on the CelebA dataset, which contains several attributes that can be inferred. 
We categorize the samples into five groups, each comprising 10,000 samples, based on their importance values ranging from low to high. 
Following this categorization, we train five models using these groups as target models.

To perform the attack, we utilize 10,000 samples, disjoint from the 50,000 training samples, to train a shadow model. 
This shadow model is employed to generate datasets for training the attack model, where the inputs are embeddings, and the associated sensitive attributes serve as labels. 
We train a two-layer fully connected network as the attack model, which is then utilized to infer the sensitive attribute from the embeddings.

\begin{figure*}[!t]
\centering
\begin{subfigure}{0.53\columnwidth}
\includegraphics[width=\columnwidth]{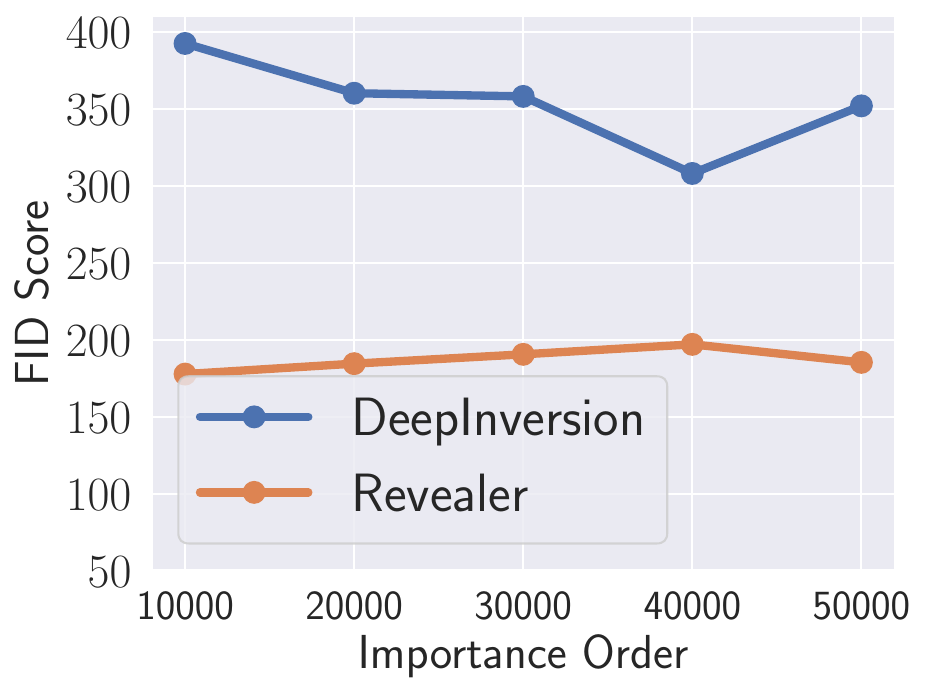}
\caption{CIFAR10}
\label{fig:recon_cifar}
\end{subfigure}
\begin{subfigure}{0.53\columnwidth}
\includegraphics[width=\columnwidth]{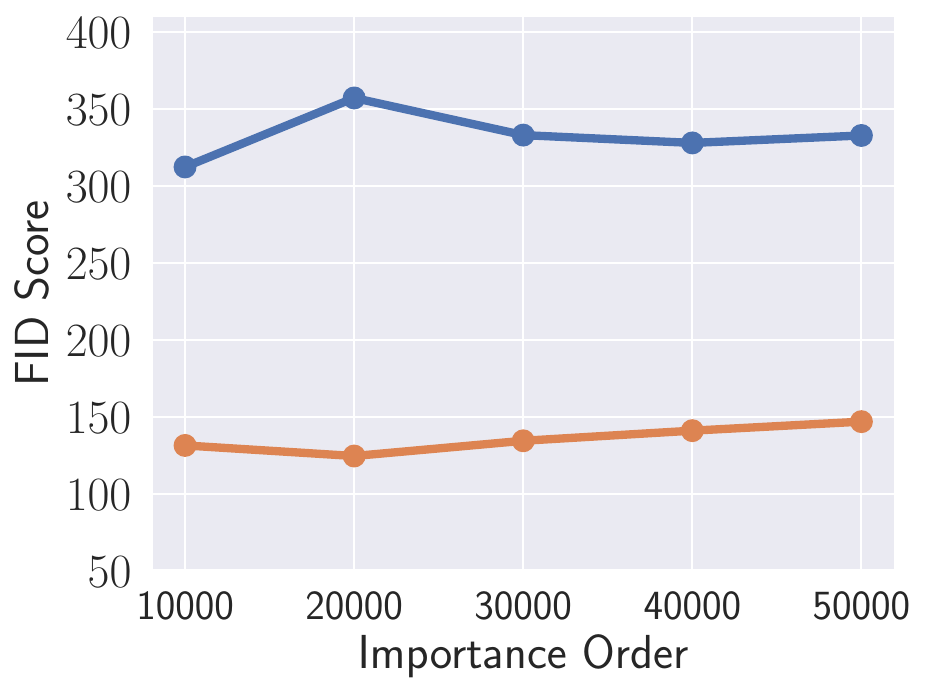}
\caption{CelebA}
\label{fig:recon_celeba}
\end{subfigure}
\begin{subfigure}{0.53\columnwidth}
\includegraphics[width=\columnwidth]{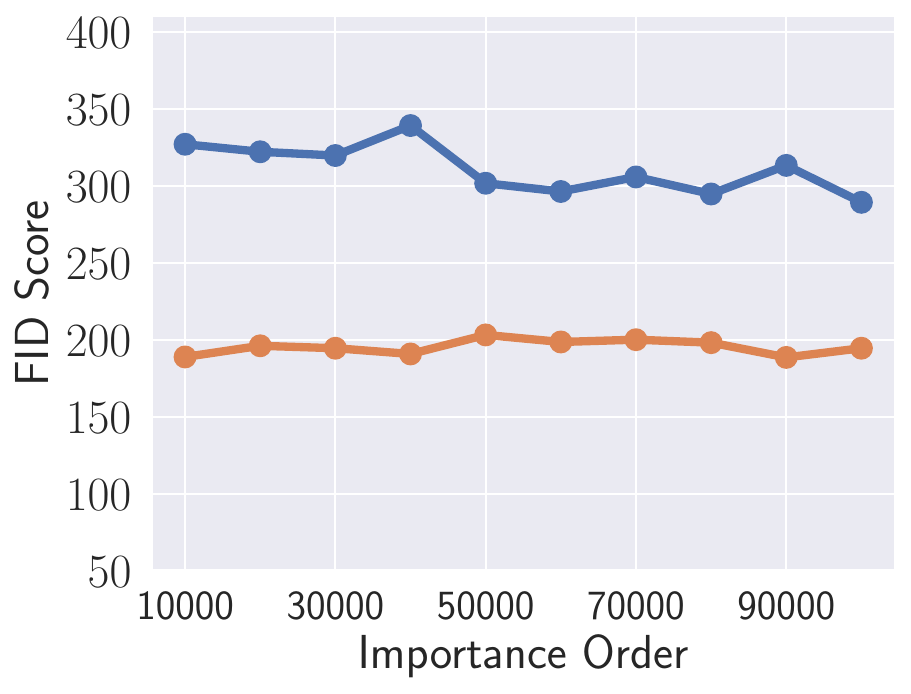}
\caption{TinyImageNet}
\label{fig:recon_tinyimagenet}
\end{subfigure}
\caption{Relationship between reconstruction quality and data importance, reconstruction performance remains steady regardless of the importance level of the data samples. }
\label{figure:reconstruct}
\end{figure*}

To evaluate the attack performance, we utilize relative accuracy as the metric, comparing the accuracy against a random guessing baseline that varies for different attributes due to the uneven distribution of the CelebA dataset. 

The experimental results, as shown in~\autoref{figure:attrinf}, reveal no significant connection between data importance and the success of attribute inference attacks.
For instance, the ``Arched Eyebrows'' attribute is easily inferred for high importance samples, while only low importance samples can be inferred for the ``High Cheekbones'' attribute. 
Furthermore, the vulnerability to attribute inference for the ``Mouth Slightly Open'' attribute is most prominent among samples with middle importance values. 
These results demonstrate that there is no significant correlation between attribute inference attacks and data importance.

One possible explanation for these results is that the importance value of data samples may vary depending on the prediction task. 
In other words, the significance of certain features or attributes may differ across different prediction tasks. 
For example, while whiskers may be an important feature for predicting gender, it may hold less importance when predicting income.
We further validate our conjecture by visualizing the correlation among importance values assigned to different attributes in~\refapp{app:attr_corr}. 
It indicates that a sample's elevated importance on one attribute may not align with its importance on another attribute.

\begin{figure}[!b]
\centering
\includegraphics[width=0.82\columnwidth]{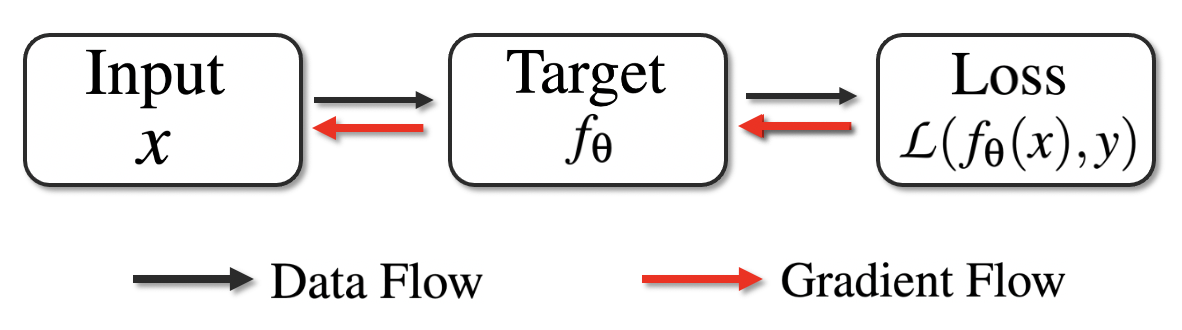}
\caption{The workflow of a basic data reconstruction attack, the adversary optimizes the input to maximize the likelihood of the target class.}
\label{figure:ds_workflow}
\end{figure}

\mypara{Takeaways}
Our findings indicate that there is not a straightforward correlation between the significance of data samples and the performance of an attack, aligning with our initial hypothesis. 
A pivotal insight from this section demonstrates that the importance of data samples is context-dependent and can vary based on the specific task at hand, which resonates with our earlier discovery in~\autoref{sec:ms_diff}.

\section{Data Reconstruction Attack}
\label{sec:rec}

Data reconstruction attack~\cite{ZJPWLS20,YMALMHJK20,FJR15,YZCL19,SBBFZ20} refers to recovering the target dataset with limited access to the target model, with the aid of additional knowledge possessed by the adversary.
While data reconstruction attack shares similarities with membership inference attack, there are significant differences that make data reconstruction a stronger attack. 
Specifically, membership inference operates at the sample level, determining the membership status of individual samples. 
In contrast, data reconstruction is a dataset-level attack aimed at extracting the entire training dataset. 
This distinction necessitates different technical approaches for data reconstruction.

In this work, we employ two data reconstruction attacks, namely DeepInversion~\cite{YMALMHJK20} and Revealer~\cite{ZJPWLS20}, to investigate the influence of data importance on the reconstruction process. 
These attacks are based on the optimization of input samples, as illustrated in~\autoref{figure:ds_workflow}. 
Specifically, given a target class y, both methods initialize a sample x and iteratively update it to maximize the likelihood or probability of belonging to that class while keeping the model parameters fixed. 
This optimization process is guided by the following loss function:
\[
\min_{x}\mathcal{L}(f_{\theta}(x),y)
\]
DeepInversion leverages statistical information encoded in the batch normalization layer to enhance the quality of reconstructions, while Revealer employs a Generative Adversarial Network (GAN) to generate high-quality reconstructions.

To investigate the impact of data importance on the performance of data reconstruction attacks, we partition the samples into groups of 10,000 based on their importance values, ranging from low importance to high importance. 
Subsequently, we train one target model for each sample group, resulting in a total of five models for CIFAR10 and CelebA datasets, and ten models for TinyImageNet. 
We then apply two reconstruction attacks on each of them.
We leverage Fr\'echet Inception Distance (FID) to measure the similarity between reconstructed samples and the training samples, given its established utility in evaluating the quality of generated distributions~\cite{ZJPWLS20,YMALMHJK20,SHCAK22}. 
A smaller FID denotes better reconstruction quality.
For each target model, we generate 10,000 reconstructions, matching the size of the training dataset. 
Subsequently, we calculate the FID score, quantifying the discrepancy between the reconstructions and the corresponding training dataset.

The findings presented in~\autoref{figure:reconstruct} suggest that there is no significant distinction between high and low importance data samples in terms of data reconstruction. 
Taking CIFAR10 as an example, DeepInversion exhibits a maximum deviation of only 13.02\% compared to the mean value, indicating a consistent performance. 
Similar results are observed with Revealer, where the maximum deviation is merely 5.35\% compared to the mean value.
Moreover, this consistent performance extends to more complex datasets. 
For instance, in the case of CelebA dataset, the maximum deviation is less than 8.27\%, while for TinyImageNet, the deviation is less than 4.01\%. 
These findings suggest that the reconstruction performance remains steady regardless of the importance level of the data samples.

\section{Transferability Study}
\label{sec:transferability}

\begin{figure*}[!t]
\centering
\begin{subfigure}{0.49\columnwidth}
\includegraphics[width=\columnwidth]{mia/cifar10_resnet18.pdf}
\caption{ResNet18 (CIFAR10)}
\label{fig:trans_mia_res18}
\end{subfigure}
\begin{subfigure}{0.49\columnwidth}
\includegraphics[width=\columnwidth]{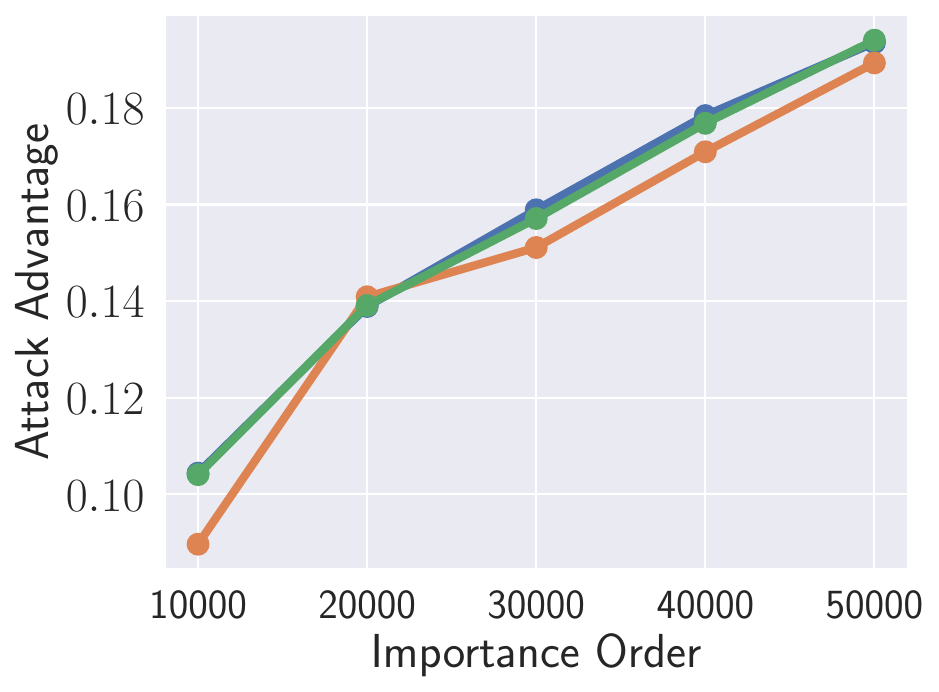}
\caption{MobileNetV2 (CIFAR10)}
\label{fig:trans_mia_mobilev2}
\end{subfigure}
\begin{subfigure}{0.49\columnwidth}
\includegraphics[width=\columnwidth]{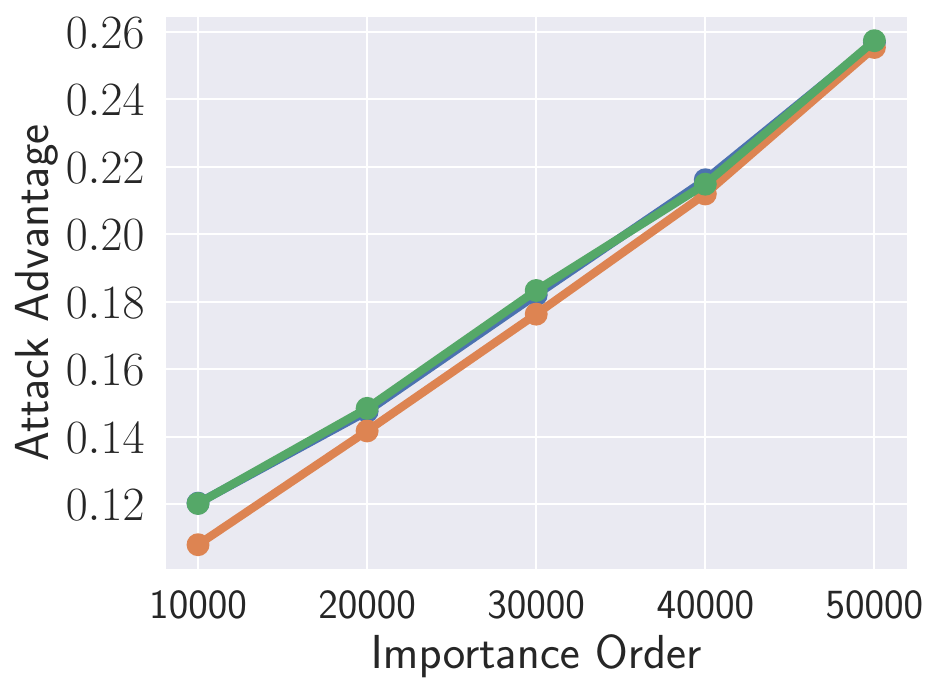}
\caption{ResNet50 (CIFAR10)}
\label{fig:trans_mia_res50}
\end{subfigure}
\begin{subfigure}{0.49\columnwidth}
\includegraphics[width=\columnwidth]{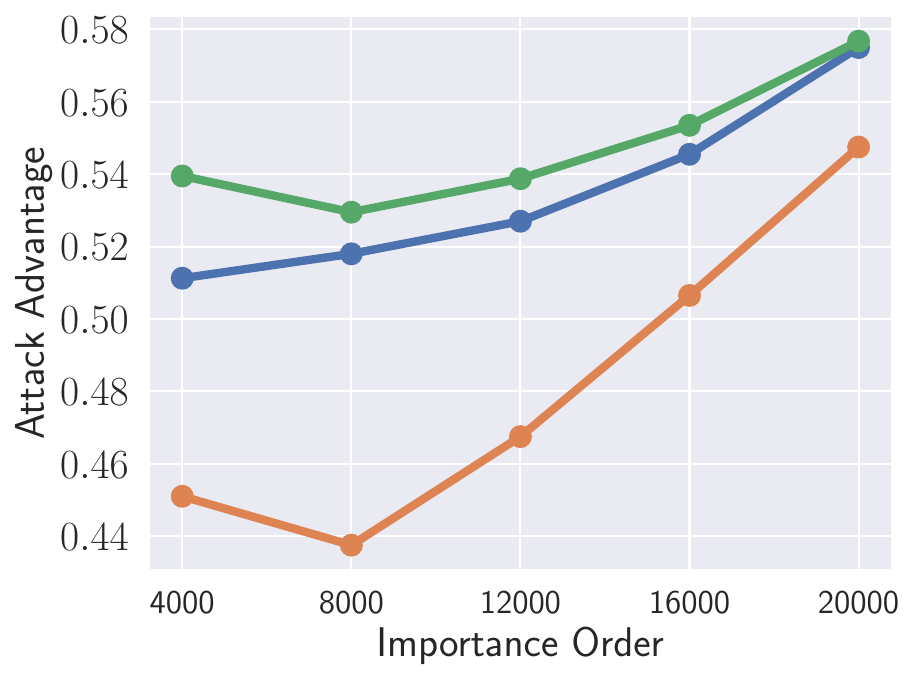}
\caption{MLP (Purchase)}
\label{fig:trans_mia_purchase}
\end{subfigure}
\caption{Relationship between membership inference attack advantage and data importance. 
Results show that our conclusion can be generalized to different model architectures and data modalities.}
\label{figure:trans_mia}
\end{figure*}

In order to fortify the generalizability of our conclusions, this section investigates the transferability of our findings across various model architectures and data modalities. 
For the vision modality, we conducted experiments employing two distinct model architectures, namely MobileNetV2~\cite{SHZZC18} and ResNet50~\cite{HZRS16}. 

\looseness=-1
To evaluate the transferability to diverse data modalities, we introduced the tabular dataset Purchase-100~\cite{Purchases}, consisting of 600 binary features for classifying 100 classes. 
We focus on the tabular modality considering that existing Shapley methods predominantly support vision and tabular modalities. 
We utilize a Multilayer Perceptron (MLP) to process the Purchase task, aligning with established practices in prior research~\cite{SSSS17,SZHBFB19}. 

Our experimental results consistently support our conclusions, irrespective of the model architecture and data modality. 
For example, in~\autoref{figure:trans_mia}, the three left figures depict a consistent relationship between the advantage of membership inference attacks and the importance of data across all three model architectures. 
This trend is also evident when performing the attack based on the distance to the boundary (see results in~\refapp{append:transfer}).
Additionally, ~\autoref{fig:trans_mia_purchase} illustrates that this conclusion holds for the tabular modality. 
Although slight fluctuations are observed in the low importance area, the overall picture demonstrates a consistent relationship between importance and membership vulnerability, aligning with the conclusions drawn from the vision modality.

Furthermore, this conclusion extends to other attack types, such as model stealing and backdoor attacks. 
Due to space constraints, we defer the results to~\refapp{append:transfer}. 
These findings reaffirm the consistent impact of data importance across different attack scenarios, underscoring the generalizability of our observations.

\section{Limitations and Future Work}
\label{sec:limitation}

While our research provides valuable insights into the relationship between data importance and vulnerability to specific attacks, several limitations exist that warrant further investigation. 
Our study focuses on a specific set of attacks. 
Although these are important, they may not cover the entire spectrum of potential threats. 
Other types of attacks could exhibit different relationships between data importance and vulnerability, and understanding how these various attacks interact with data importance remains an open area for exploration.

Extending our findings to Large Language Models (LLMs) presents substantial challenges despite their promising advancements. 
The primary obstacle is the computational cost associated with calculating importance values. 
To manage this burden, current methods often resort to computationally lighter algorithms like $K$NN for classification tasks. 
However, it is unclear whether similar computationally efficient approaches can be adapted to approximate auto-regression models, especially since LLMs exhibit unique emergent characteristics when scaled beyond certain thresholds. 
Additionally, the considerably larger datasets typical of LLMs further complicate the feasibility of extending these methods.

Furthermore, our research does not examine more complex augmentation techniques, such as those utilizing generative models. 
Future work should investigate whether these advanced techniques affect data importance and vulnerability differently. 
Additionally, exploring whether there exists a generalizable method to manipulate data importance across various augmentation techniques would be invaluable. 
To foster further research and collaboration, we have open-sourced our evaluation framework, available at \url{https://github.com/TrustAIRLab/importance-in-mlattacks}. This will enable other researchers to examine whether the observed data discrepancies hold for new types of attacks, thereby benefiting the broader community.

\section{Conclusion}
\label{sec:conclusion}

In this paper, our research systematically studies the vulnerability of heterogeneous data when confronted with machine learning attacks. 
Our findings underscore a heightened susceptibility of high importance data samples to privacy attacks, including membership inference attacks and model stealing attacks. 
Our findings also carry practical implications, inspiring researchers to design
more efficient attacks. 
For example, we empirically showcase the potential enhancement of membership inference attacks through the incorporation of sample-specific criteria based on importance values. 
Additionally, we demonstrate that our findings can be strategically employed to guide the creation of more advanced attacks through the active manipulation of sample importance.

\section*{Acknowledgements}

We thank all anonymous reviewers for their constructive comments.
This work is partially funded by the European Health and Digital Executive Agency (HADEA) within the project ``Understanding the individual host response against Hepatitis D Virus to develop a personalized approach for the management of hepatitis D'' (DSolve, grant agreement number 101057917) and the BMBF with the project ``Repräsentative, synthetische Gesundheitsdaten mit starken Privatsphärengarantien'' (PriSyn, 16KISAO29K).

\bibliographystyle{plain}
\bibliography{normal_generated_py3}

\begin{thebibliography}{10}

\bibitem{CIFAR}
\url{https://www.cs.toronto.edu/~kriz/cifar.html}.

\bibitem{TinyImageNet}
\url{https://www.kaggle.com/c/tiny-imagenet}.

\bibitem{Purchases}
\url{https://www.kaggle.com/c/acquire-valued-shoppers-challenge/data}.

\bibitem{ADS19}
Anish Agarwal, Munther~A. Dahleh, and Tuhin Sarkar.
\newblock {A Marketplace for Data: An Algorithmic Solution}.
\newblock In {\em {Entertainment Computing (EC)}}, pages 701--726. ACM, 2019.

\bibitem{BS21}
Eugene Bagdasaryan and Vitaly Shmatikov.
\newblock {Blind Backdoors in Deep Learning Models}.
\newblock In {\em {USENIX Security Symposium (USENIX Security)}}, pages 1505--1521. USENIX, 2021.

\bibitem{BKT19}
Mauro Barni, Kassem Kallas, and Benedetta Tondi.
\newblock {A New Backdoor Attack in {CNNS} by Training Set Corruption Without Label Poisoning}.
\newblock In {\em {IEEE International Conference on Image Processing (ICIP)}}, pages 101--105. IEEE, 2019.

\bibitem{CCNSTT22}
Nicholas Carlini, Steve Chien, Milad Nasr, Shuang Song, Andreas Terzis, and Florian Tram{\`{e}}r.
\newblock {Membership Inference Attacks From First Principles}.
\newblock In {\em {IEEE Symposium on Security and Privacy (S\&P)}}, pages 1897--1914. IEEE, 2022.

\bibitem{CJM20}
Nicholas Carlini, Matthew Jagielski, and Ilya Mironov.
\newblock {Cryptanalytic Extraction of Neural Network Models}.
\newblock In {\em {Annual International Cryptology Conference (CRYPTO)}}, pages 189--218. Springer, 2020.

\bibitem{CJZPTT22}
Nicholas Carlini, Matthew Jagielski, Chiyuan Zhang, Nicolas Papernot, Andreas Terzis, and Florian Tram{\`{e}}r.
\newblock {The Privacy Onion Effect: Memorization is Relative}.
\newblock In {\em {Annual Conference on Neural Information Processing Systems (NeurIPS)}}. NeurIPS, 2022.

\bibitem{CKJQ21}
Si~Chen, Mostafa Kahla, Ruoxi Jia, and Guo{-}Jun Qi.
\newblock {Knowledge-Enriched Distributional Model Inversion Attacks}.
\newblock In {\em {IEEE International Conference on Computer Vision (ICCV)}}, pages 16158--16167. IEEE, 2021.

\bibitem{CLLLS17}
Xinyun Chen, Chang Liu, Bo~Li, Kimberly Lu, and Dawn Song.
\newblock {Targeted Backdoor Attacks on Deep Learning Systems Using Data Poisoning}.
\newblock {\em {CoRR abs/1712.05526}}, 2017.

\bibitem{CTCP21}
Christopher A.~Choquette Choo, Florian Tram{\`e}r, Nicholas Carlini, and Nicolas Papernot.
\newblock {Label-Only Membership Inference Attacks}.
\newblock In {\em {International Conference on Machine Learning (ICML)}}, pages 1964--1974. PMLR, 2021.

\bibitem{DSA21}
Vasisht Duddu, Sebastian Szyller, and N.~Asokan.
\newblock {SHAPr: An Efficient and Versatile Membership Privacy Risk Metric for Machine Learning}.
\newblock {\em {CoRR abs/2112.02230}}, 2021.

\bibitem{FJR15}
Matt Fredrikson, Somesh Jha, and Thomas Ristenpart.
\newblock {Model Inversion Attacks that Exploit Confidence Information and Basic Countermeasures}.
\newblock In {\em {ACM SIGSAC Conference on Computer and Communications Security (CCS)}}, pages 1322--1333. ACM, 2015.

\bibitem{FLJLPR14}
Matt Fredrikson, Eric Lantz, Somesh Jha, Simon Lin, David Page, and Thomas Ristenpart.
\newblock {Privacy in Pharmacogenetics: An End-to-End Case Study of Personalized Warfarin Dosing}.
\newblock In {\em {USENIX Security Symposium (USENIX Security)}}, pages 17--32. USENIX, 2014.

\bibitem{GZ19}
Amirata Ghorbani and James~Y. Zou.
\newblock {Data Shapley: Equitable Valuation of Data for Machine Learning}.
\newblock In {\em {International Conference on Machine Learning (ICML)}}, pages 2242--2251. PMLR, 2019.

\bibitem{GDG17}
Tianyu Gu, Brendan Dolan-Gavitt, and Siddharth Grag.
\newblock {Badnets: Identifying Vulnerabilities in the Machine Learning Model Supply Chain}.
\newblock {\em {CoRR abs/1708.06733}}, 2017.

\bibitem{HVYSI22}
Niv Haim, Gal Vardi, Gilad Yehudai, Ohad Shamir, and Michal Irani.
\newblock {Reconstructing Training Data from Trained Neural Networks}.
\newblock In {\em {Annual Conference on Neural Information Processing Systems (NeurIPS)}}. NeurIPS, 2022.

\bibitem{HZRS16}
Kaiming He, Xiangyu Zhang, Shaoqing Ren, and Jian Sun.
\newblock {Deep Residual Learning for Image Recognition}.
\newblock In {\em {IEEE Conference on Computer Vision and Pattern Recognition (CVPR)}}, pages 770--778. IEEE, 2016.

\bibitem{HWWBSZ21}
Xinlei He, Rui Wen, Yixin Wu, Michael Backes, Yun Shen, and Yang Zhang.
\newblock {Node-Level Membership Inference Attacks Against Graph Neural Networks}.
\newblock {\em {CoRR abs/2102.05429}}, 2021.

\bibitem{JCBKP20}
Matthew Jagielski, Nicholas Carlini, David Berthelot, Alex Kurakin, and Nicolas Papernot.
\newblock {High Accuracy and High Fidelity Extraction of Neural Networks}.
\newblock In {\em {USENIX Security Symposium (USENIX Security)}}, pages 1345--1362. USENIX, 2020.

\bibitem{JUO20}
Matthew Jagielski, Jonathan Ullman, and Alina Oprea.
\newblock {Auditing Differentially Private Machine Learning: How Private is Private SGD?}
\newblock In {\em {Annual Conference on Neural Information Processing Systems (NeurIPS)}}. NeurIPS, 2020.

\bibitem{JE22}
Bargav Jayaraman and David Evans.
\newblock {Are Attribute Inference Attacks Just Imputation?}
\newblock In {\em {ACM SIGSAC Conference on Computer and Communications Security (CCS)}}, pages 1569--1582. ACM, 2022.

\bibitem{JDWHGLZSS19}
Ruoxi Jia, David Dao, Boxin Wang, Frances~Ann Hubis, Nezihe~Merve G{\"{u}}rel, Bo~Li, Ce~Zhang, Costas~J. Spanos, and Dawn Song.
\newblock {Efficient Task-Specific Data Valuation for Nearest Neighbor Algorithms}.
\newblock {\em {Proceedings of the VLDB Endowment}}, 2019.

\bibitem{JWSXDKZLS21}
Ruoxi Jia, Fan Wu, Xuehui Sun, Jiacen Xu, David Dao, Bhavya Kailkhura, Ce~Zhang, Bo~Li, and Dawn Song.
\newblock {Scalability vs. Utility: Do We Have To Sacrifice One for the Other in Data Importance Quantification?}
\newblock In {\em {IEEE Conference on Computer Vision and Pattern Recognition (CVPR)}}, pages 8239--8247. IEEE, 2021.

\bibitem{KPQ21}
Sanjay Kariyappa, Atul Prakash, and Moinuddin~K. Qureshi.
\newblock {{MAZE:} Data-Free Model Stealing Attack Using Zeroth-Order Gradient Estimation}.
\newblock In {\em {IEEE Conference on Computer Vision and Pattern Recognition (CVPR)}}, pages 13814--13823. IEEE, 2021.

\bibitem{KL17}
Pang~Wei Koh and Percy Liang.
\newblock {Understanding Black-box Predictions via Influence Functions}.
\newblock In {\em {International Conference on Machine Learning (ICML)}}, pages 1885--1894. PMLR, 2017.

\bibitem{KZ22}
Yongchan Kwon and James Zou.
\newblock {Beta Shapley: a Unified and Noise-reduced Data Valuation Framework for Machine Learning}.
\newblock In {\em {International Conference on Artificial Intelligence and Statistics (AISTATS)}}, pages 8780--8802. JMLR, 2022.

\bibitem{LPXDJYW23}
Changjiang Li, Ren Pang, Zhaohan Xi, Tianyu Du, Shouling Ji, Yuan Yao, and Ting Wang.
\newblock {An Embarrassingly Simple Backdoor Attack on Self-supervised Learning}.
\newblock In {\em {IEEE International Conference on Computer Vision (ICCV)}}, pages 4344--4355. IEEE, 2023.

\bibitem{LLWLHL21}
Yuezun Li, Yiming Li, Baoyuan Wu, Longkang Li, Ran He, and Siwei Lyu.
\newblock {Invisible Backdoor Attack with Sample-Specific Triggers}.
\newblock In {\em {IEEE International Conference on Computer Vision (ICCV)}}, pages 16443--16452. IEEE, 2021.

\bibitem{LZ21}
Zheng Li and Yang Zhang.
\newblock {Membership Leakage in Label-Only Exposures}.
\newblock In {\em {ACM SIGSAC Conference on Computer and Communications Security (CCS)}}, pages 880--895. ACM, 2021.

\bibitem{LMALZWZ18}
Yingqi Liu, Shiqing Ma, Yousra Aafer, Wen-Chuan Lee, Juan Zhai, Weihang Wang, and Xiangyu Zhang.
\newblock {Trojaning Attack on Neural Networks}.
\newblock In {\em {Network and Distributed System Security Symposium (NDSS)}}. Internet Society, 2018.

\bibitem{LWBZ24}
Yiyong Liu, Rui Wen, Michael Backes, and Yang Zhang.
\newblock {Efficient Data-Free Model Stealing with Label Diversity}.
\newblock {\em {CoRR abs/2404.00108}}, 2024.

\bibitem{LZBZ22}
Yiyong Liu, Zhengyu Zhao, Michael Backes, and Yang Zhang.
\newblock {Membership Inference Attacks by Exploiting Loss Trajectory}.
\newblock In {\em {ACM SIGSAC Conference on Computer and Communications Security (CCS)}}, pages 2085--2098. ACM, 2022.

\bibitem{LWHSZBCFZ22}
Yugeng Liu, Rui Wen, Xinlei He, Ahmed Salem, Zhikun Zhang, Michael Backes, Emiliano~De Cristofaro, Mario Fritz, and Yang Zhang.
\newblock {ML-Doctor: Holistic Risk Assessment of Inference Attacks Against Machine Learning Models}.
\newblock In {\em {USENIX Security Symposium (USENIX Security)}}, pages 4525--4542. USENIX, 2022.

\bibitem{LLWT15}
Ziwei Liu, Ping Luo, Xiaogang Wang, and Xiaoou Tang.
\newblock {Deep Learning Face Attributes in the Wild}.
\newblock In {\em {IEEE International Conference on Computer Vision (ICCV)}}, pages 3730--3738. IEEE, 2015.

\bibitem{LL17}
Scott~M. Lundberg and Su{-}In Lee.
\newblock {A Unified Approach to Interpreting Model Predictions}.
\newblock In {\em {Annual Conference on Neural Information Processing Systems (NeurIPS)}}, pages 4765--4774. NIPS, 2017.

\bibitem{MDKLB22}
Shagufta Mehnaz, Sayanton~V. Dibbo, Ehsanul Kabir, Ninghui Li, and Elisa Bertino.
\newblock {Are Your Sensitive Attributes Private? Novel Model Inversion Attribute Inference Attacks on Classification Models}.
\newblock In {\em {USENIX Security Symposium (USENIX Security)}}, pages 4579--4596. USENIX, 2022.

\bibitem{MSCS19}
Luca Melis, Congzheng Song, Emiliano~De Cristofaro, and Vitaly Shmatikov.
\newblock {Exploiting Unintended Feature Leakage in Collaborative Learning}.
\newblock In {\em {IEEE Symposium on Security and Privacy (S\&P)}}, pages 497--512. IEEE, 2019.

\bibitem{MBDPWLR17}
Luis Mu{\~{n}}oz{-}Gonz{\'{a}}lez, Battista Biggio, Ambra Demontis, Andrea Paudice, Vasin Wongrassamee, Emil~C. Lupu, and Fabio Roli.
\newblock {Towards Poisoning of Deep Learning Algorithms with Back-gradient Optimization}.
\newblock In {\em {Workshop on Security and Artificial Intelligence (AISec)}}, pages 27--38. ACM, 2017.

\bibitem{NSTPC21}
Milad Nasr, Shuang Song, Abhradeep Thakurta, Nicolas Papernot, and Nicholas Carlini.
\newblock {Adversary Instantiation: Lower Bounds for Differentially Private Machine Learning}.
\newblock In {\em {IEEE Symposium on Security and Privacy (S\&P)}}. IEEE, 2021.

\bibitem{NT20}
Tuan~Anh Nguyen and Anh Tran.
\newblock {Input-Aware Dynamic Backdoor Attack}.
\newblock In {\em {Annual Conference on Neural Information Processing Systems (NeurIPS)}}. NeurIPS, 2020.

\bibitem{NT21}
Tuan~Anh Nguyen and Anh~Tuan Tran.
\newblock {WaNet - Imperceptible Warping-based Backdoor Attack}.
\newblock In {\em {International Conference on Learning Representations (ICLR)}}, 2021.

\bibitem{OSF19}
Tribhuvanesh Orekondy, Bernt Schiele, and Mario Fritz.
\newblock {Knockoff Nets: Stealing Functionality of Black-Box Models}.
\newblock In {\em {IEEE Conference on Computer Vision and Pattern Recognition (CVPR)}}, pages 4954--4963. IEEE, 2019.

\bibitem{PMGJCS17}
Nicolas Papernot, Patrick~D. McDaniel, Ian Goodfellow, Somesh Jha, Z.~Berkay Celik, and Ananthram Swami.
\newblock {Practical Black-Box Attacks Against Machine Learning}.
\newblock In {\em {ACM Asia Conference on Computer and Communications Security (ASIACCS)}}, pages 506--519. ACM, 2017.

\bibitem{PGILM23}
Sung~Min Park, Kristian Georgiev, Andrew Ilyas, Guillaume Leclerc, and Aleksander Madry.
\newblock {TRAK: Attributing Model Behavior at Scale}.
\newblock In {\em {International Conference on Machine Learning (ICML)}}, pages 27074--27113. PMLR, 2023.

\bibitem{RSG16}
Marco~T{\'{u}}lio Ribeiro, Sameer Singh, and Carlos Guestrin.
\newblock {Why Should I Trust You?: Explaining the Predictions of Any Classifier}.
\newblock In {\em {ACM Conference on Knowledge Discovery and Data Mining (KDD)}}, pages 1135--1144. ACM, 2016.

\bibitem{SBBFZ20}
Ahmed Salem, Apratim Bhattacharya, Michael Backes, Mario Fritz, and Yang Zhang.
\newblock {Updates-Leak: Data Set Inference and Reconstruction Attacks in Online Learning}.
\newblock In {\em {USENIX Security Symposium (USENIX Security)}}, pages 1291--1308. USENIX, 2020.

\bibitem{SWBMZ22}
Ahmed Salem, Rui Wen, Michael Backes, Shiqing Ma, and Yang Zhang.
\newblock {Dynamic Backdoor Attacks Against Machine Learning Models}.
\newblock In {\em {IEEE European Symposium on Security and Privacy (Euro S\&P)}}, pages 703--718. IEEE, 2022.

\bibitem{SZHBFB19}
Ahmed Salem, Yang Zhang, Mathias Humbert, Pascal Berrang, Mario Fritz, and Michael Backes.
\newblock {ML-Leaks: Model and Data Independent Membership Inference Attacks and Defenses on Machine Learning Models}.
\newblock In {\em {Network and Distributed System Security Symposium (NDSS)}}. Internet Society, 2019.

\bibitem{SHZZC18}
Mark Sandler, Andrew~G. Howard, Menglong Zhu, Andrey Zhmoginov, and Liang{-}Chieh Chen.
\newblock {MobileNetV2: Inverted Residuals and Linear Bottlenecks}.
\newblock In {\em {IEEE Conference on Computer Vision and Pattern Recognition (CVPR)}}, pages 4510--4520. IEEE, 2018.

\bibitem{SAB22}
Sunandini Sanyal, Sravanti Addepalli, and R.~Venkatesh Babu.
\newblock {Towards Data-Free Model Stealing in a Hard Label Setting}.
\newblock {\em {CoRR abs/2204.11022}}, 2022.

\bibitem{SHNSSDG18}
Ali Shafahi, W~Ronny Huang, Mahyar Najibi, Octavian Suciu, Christoph Studer, Tudor Dumitras, and Tom Goldstein.
\newblock {Poison Frogs! Targeted Clean-Label Poisoning Attacks on Neural Networks}.
\newblock In {\em {Annual Conference on Neural Information Processing Systems (NeurIPS)}}, pages 6103--6113. NeurIPS, 2018.

\bibitem{S53}
Lloyd~S Shapley.
\newblock {A Value for n-person Games}.
\newblock {\em {Contributions to the Theory of Games}}, 1953.

\bibitem{SSSS17}
Reza Shokri, Marco Stronati, Congzheng Song, and Vitaly Shmatikov.
\newblock {Membership Inference Attacks Against Machine Learning Models}.
\newblock In {\em {IEEE Symposium on Security and Privacy (S\&P)}}, pages 3--18. IEEE, 2017.

\bibitem{SR20}
Congzheng Song and Ananth Raghunathan.
\newblock {Information Leakage in Embedding Models}.
\newblock In {\em {ACM SIGSAC Conference on Computer and Communications Security (CCS)}}, pages 377--390. ACM, 2020.

\bibitem{SS20}
Congzheng Song and Vitaly Shmatikov.
\newblock {Overlearning Reveals Sensitive Attributes}.
\newblock In {\em {International Conference on Learning Representations (ICLR)}}, 2020.

\bibitem{SM21}
Liwei Song and Prateek Mittal.
\newblock {Systematic Evaluation of Privacy Risks of Machine Learning Models}.
\newblock In {\em {USENIX Security Symposium (USENIX Security)}}. USENIX, 2021.

\bibitem{SSM19}
Liwei Song, Reza Shokri, and Prateek Mittal.
\newblock {Privacy Risks of Securing Machine Learning Models against Adversarial Examples}.
\newblock In {\em {ACM SIGSAC Conference on Computer and Communications Security (CCS)}}, pages 241--257. ACM, 2019.

\bibitem{SHCAK22}
Lukas Struppek, Dominik Hintersdorf, Antonio De~Almeida Correia, Antonia Adler, and Kristian Kersting.
\newblock {Plug {\&} Play Attacks: Towards Robust and Flexible Model Inversion Attacks}.
\newblock In {\em {International Conference on Machine Learning (ICML)}}, pages 20522--20545. PMLR, 2022.

\bibitem{TSJLJHC22}
Florian Tram{\`e}r, Reza Shokri, Ayrton~San Joaquin, Hoang Le, Matthew Jagielski, Sanghyun Hong, and Nicholas Carlini.
\newblock {Truth Serum: Poisoning Machine Learning Models to Reveal Their Secrets}.
\newblock In {\em {ACM SIGSAC Conference on Computer and Communications Security (CCS)}}. ACM, 2022.

\bibitem{TZJRR16}
Florian Tram{\`e}r, Fan Zhang, Ari Juels, Michael~K. Reiter, and Thomas Ristenpart.
\newblock {Stealing Machine Learning Models via Prediction APIs}.
\newblock In {\em {USENIX Security Symposium (USENIX Security)}}, pages 601--618. USENIX, 2016.

\bibitem{TMWP21}
Jean{-}Baptiste Truong, Pratyush Maini, Robert~J. Walls, and Nicolas Papernot.
\newblock {Data-Free Model Extraction}.
\newblock In {\em {IEEE Conference on Computer Vision and Pattern Recognition (CVPR)}}, pages 4771--4780. IEEE, 2021.

\bibitem{WFLKZM21}
Kuan{-}Chieh Wang, Yan Fu, Ke~Li, Ashish Khisti, Richard~S. Zemel, and Alireza Makhzani.
\newblock {Variational Model Inversion Attacks}.
\newblock In {\em {Annual Conference on Neural Information Processing Systems (NeurIPS)}}, pages 9706--9719. NeurIPS, 2021.

\bibitem{WLBZ24}
Rui Wen, Zheng Li, Michael Backes, and Yang Zhang.
\newblock {Membership Inference Attacks Against In-Context Learning}.
\newblock In {\em {ACM SIGSAC Conference on Computer and Communications Security (CCS)}}. ACM, 2024.

\bibitem{WWBZS23}
Rui Wen, Tianhao Wang, Michael Backes, Yang Zhang, and Ahmed Salem.
\newblock {Last One Standing: A Comparative Analysis of Security and Privacy of Soft Prompt Tuning, LoRA, and In-Context Learning}.
\newblock {\em {CoRR abs/2310.11397}}, 2023.

\bibitem{WZLBWZ23}
Rui Wen, Zhengyu Zhao, Zhuoran Liu, Michael Backes, Tianhao Wang, and Yang Zhang.
\newblock {Is Adversarial Training Really a Silver Bullet for Mitigating Data Poisoning?}
\newblock In {\em {International Conference on Learning Representations (ICLR)}}, 2023.

\bibitem{WCZZWYS22}
Baoyuan Wu, Hongrui Chen, Mingda Zhang, Zihao Zhu, Shaokui Wei, Danni Yuan, and Chao Shen.
\newblock {BackdoorBench: A Comprehensive Benchmark of Backdoor Learning}.
\newblock In {\em {Annual Conference on Neural Information Processing Systems (NeurIPS)}}. NeurIPS, 2022.

\bibitem{YZCL19}
Ziqi Yang, Jiyi Zhang, Ee-Chien Chang, and Zhenkai Liang.
\newblock {Neural Network Inversion in Adversarial Setting via Background Knowledge Alignment}.
\newblock In {\em {ACM SIGSAC Conference on Computer and Communications Security (CCS)}}, page 225–240. ACM, 2019.

\bibitem{YMMBS22}
Jiayuan Ye, Aadyaa Maddi, Sasi~Kumar Murakonda, Vincent Bindschaedler, and Reza Shokri.
\newblock {Enhanced Membership Inference Attacks against Machine Learning Models}.
\newblock In {\em {ACM SIGSAC Conference on Computer and Communications Security (CCS)}}, pages 3093--3106. ACM, 2022.

\bibitem{YGFJ18}
Samuel Yeom, Irene Giacomelli, Matt Fredrikson, and Somesh Jha.
\newblock {Privacy Risk in Machine Learning: Analyzing the Connection to Overfitting}.
\newblock In {\em {IEEE Computer Security Foundations Symposium (CSF)}}, pages 268--282. IEEE, 2018.

\bibitem{YMALMHJK20}
Hongxu Yin, Pavlo Molchanov, Jose~M. Alvarez, Zhizhong Li, Arun Mallya, Derek Hoiem, Niraj~K. Jha, and Jan Kautz.
\newblock {Dreaming to Distill: Data-Free Knowledge Transfer via DeepInversion}.
\newblock In {\em {IEEE Conference on Computer Vision and Pattern Recognition (CVPR)}}, pages 8712--8721. IEEE, 2020.

\bibitem{ZPMJ21}
Yi~Zeng, Won Park, Z.~Morley Mao, and Ruoxi Jia.
\newblock {Rethinking the Backdoor Attacks' Triggers: {A} Frequency Perspective}.
\newblock In {\em {IEEE International Conference on Computer Vision (ICCV)}}, pages 16453--16461. IEEE, 2021.

\bibitem{ZYWBZ24}
Minxing Zhang, Ning Yu, Rui Wen, Michael Backes, and Yang Zhang.
\newblock {Generated Distributions Are All You Need for Membership Inference Attacks Against Generative Models}.
\newblock In {\em {Winter Conference on Applications of Computer Vision (WACV)}}, pages 4827--4837. IEEE, 2024.

\bibitem{ZJPWLS20}
Yuheng Zhang, Ruoxi Jia, Hengzhi Pei, Wenxiao Wang, Bo~Li, and Dawn Song.
\newblock {The Secret Revealer: Generative Model-Inversion Attacks Against Deep Neural Networks}.
\newblock In {\em {IEEE Conference on Computer Vision and Pattern Recognition (CVPR)}}, pages 250--258. IEEE, 2020.

\bibitem{ZQZ22}
Nan Zhong, Zhenxing Qian, and Xinpeng Zhang.
\newblock {Imperceptible Backdoor Attack: From Input Space to Feature Representation}.
\newblock In {\em {International Joint Conferences on Artifical Intelligence (IJCAI)}}, pages 1736--1742. IJCAI, 2022.

\bibitem{ZHLTSG19}
Chen Zhu, W~Ronny Huang, Hengduo Li, Gavin Taylor, Christoph Studer, and Tom Goldstein.
\newblock {Transferable Clean-label Poisoning Attacks on Deep Neural Nets}.
\newblock In {\em {International Conference on Machine Learning (ICML)}}, pages 7614--7623. JMLR, 2019.

\end{thebibliography}
\appendix

\section{CelebA Attribute Selection}
\label{append:attributeselection}

The CelebA dataset contains 40 binary attributes, which is not suitable for multi-class classification. 
Therefore, we follow previous works~\cite{NT21,SWBMZ22,ZQZ22,LWHSZBCFZ22} that select the three most balanced attributes (Heavy Makeup, Mouth Slightly Open, and Smiling) to create an 8-class ($2^3$) classification task. 

\begin{figure*}[h]
\centering
\begin{subfigure}{0.472\columnwidth}
\includegraphics[width=\columnwidth]{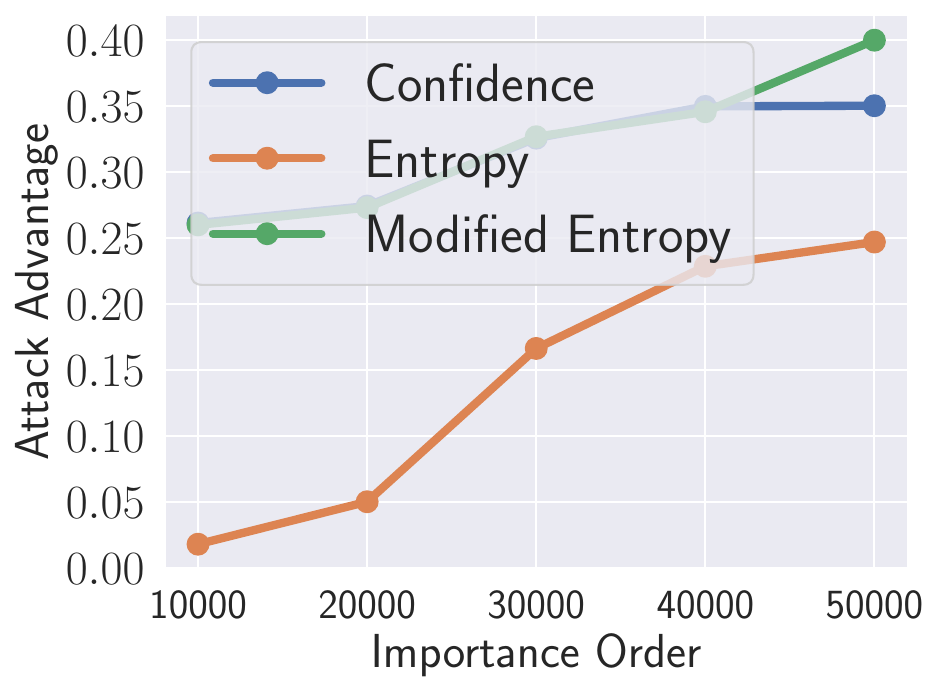}
\caption{MIA (metric-based)}
\label{fig:diffattr_mia_metric}
\end{subfigure}
\begin{subfigure}{0.472\columnwidth}
\includegraphics[width=\columnwidth]{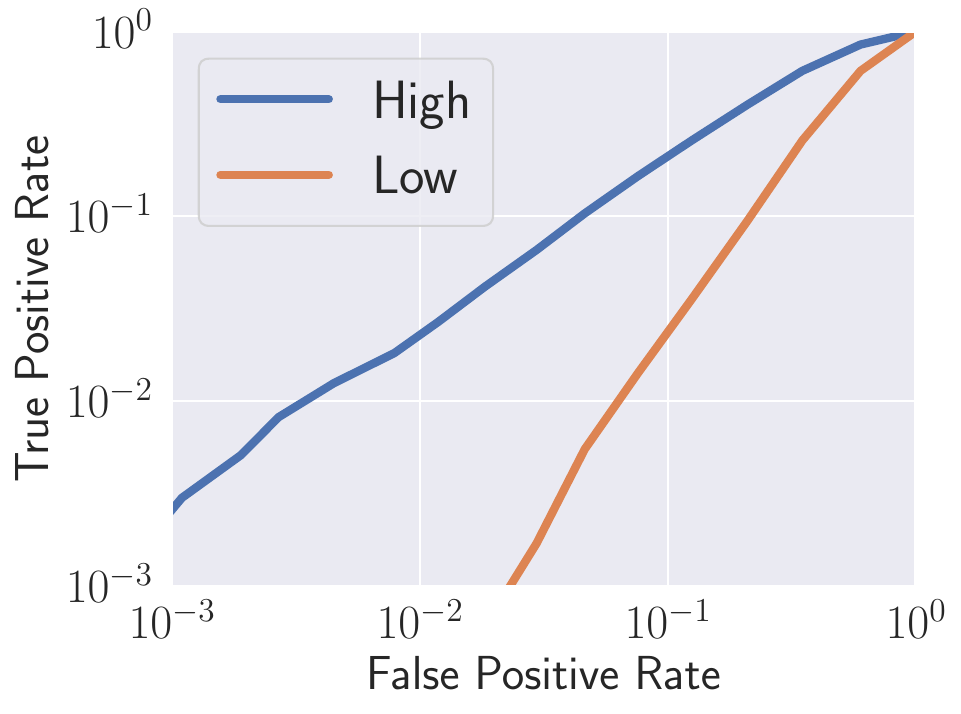}
\caption{MIA (log-scale)}
\label{fig:diffattr_mia_logscale}
\end{subfigure}
\begin{subfigure}{0.452\columnwidth}
\includegraphics[width=\columnwidth]{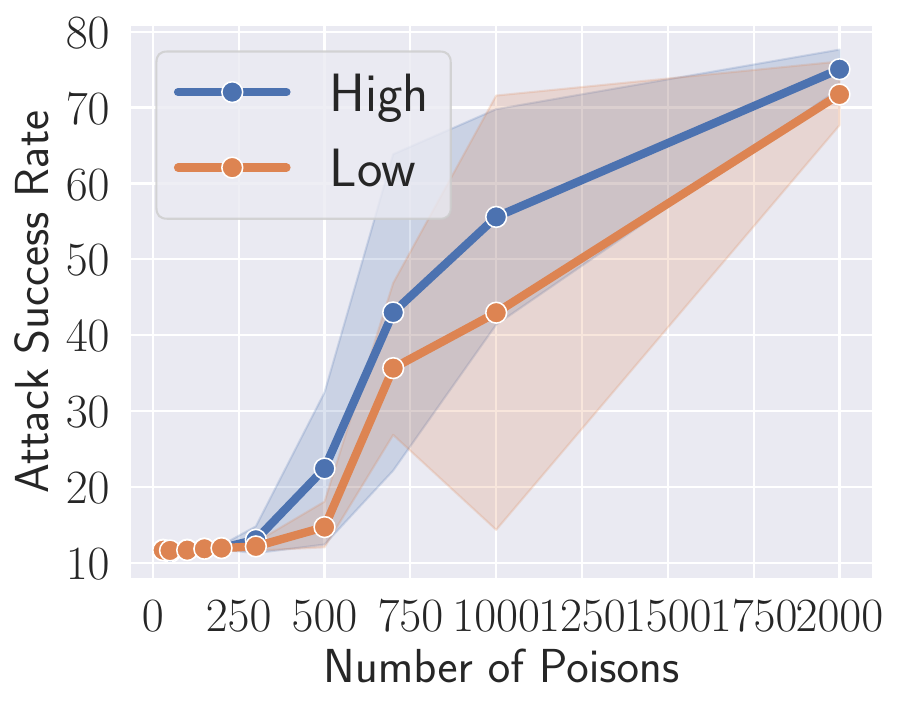}
\caption{Backdoor ASR}
\label{fig:diffattr_backdoorasr}
\end{subfigure}
\begin{subfigure}{0.452\columnwidth}
\includegraphics[width=\columnwidth]{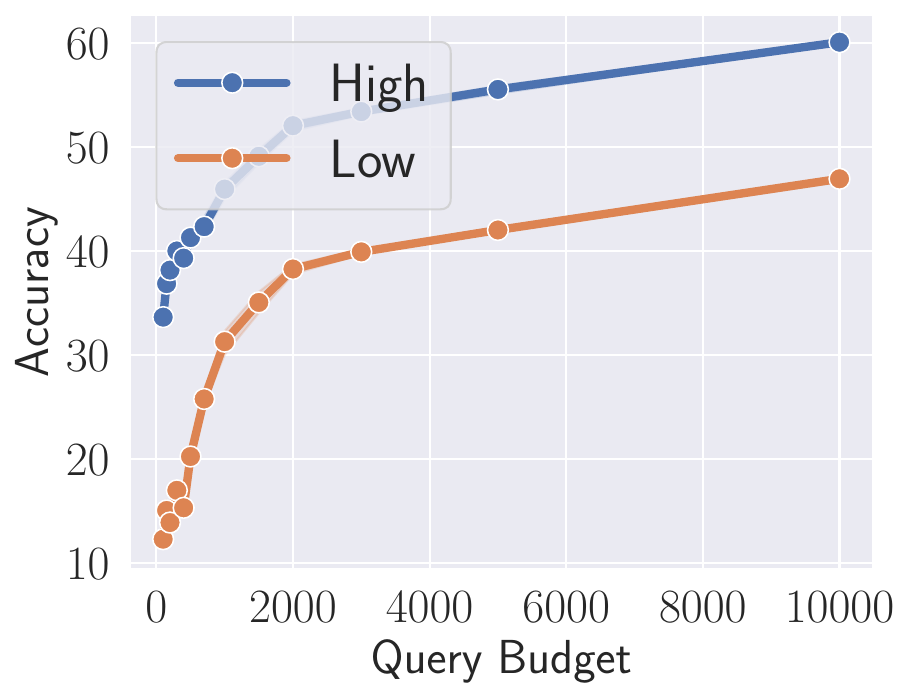}
\caption{Model Stealing}
\label{fig:diffattr_modelsteal}
\end{subfigure}
\caption{Attack performance on samples with high and low importance. 
The results demonstrate that our conclusions are consistent across different sets of selected attributes.}
\label{figure:diffattr}
\end{figure*}

To validate that our findings are not dependent on this specific attribute selection, we conducted the same experiments using another randomly selected set of attributes (High Cheekbones, Arched Eyebrows, and Wearing Lipstick). 
We evaluated the performance on membership inference attacks, model stealing, and backdoor attacks. 
The results are depicted in~\autoref{figure:diffattr}, confirming that our findings are consistent across these different attribute sets. 
For example,~\autoref{fig:diffattr_mia_metric} demonstrate that samples with higher importance are more vulnerable to membership inference attack; it also reflects on the worst-case evaluation as illustrated in~\autoref{fig:diffattr_mia_logscale}. 
The conclusion holds for backdoor and model stealing attack, specifically, with a 1500 query budget, high importance samples can steal a surrogate model with 48\% higher accuracy than that stolen by low importance samples. 

\section{Measurement Hyperparameter}
\label{append:measureselect}

In implementing the $K$NN-Shapley method, we set the hyperparameter $k=6$, following the suggestion in the original paper~\cite{JDWHGLZSS19}. 
Further experimentation with $k=7$ and $k=8$ indicated that performance remains largely consistent across these settings. 
Specifically, the correlation between importance values calculated with $k=6$ and $k=7$ is 0.9988, and between $k=6$ and $k=8$, it's 0.9972, demonstrating the robustness of our results with respect to this hyperparameter.

\begin{figure}[ht]
\centering
\begin{subfigure}{0.49\columnwidth}
\includegraphics[width=\columnwidth]{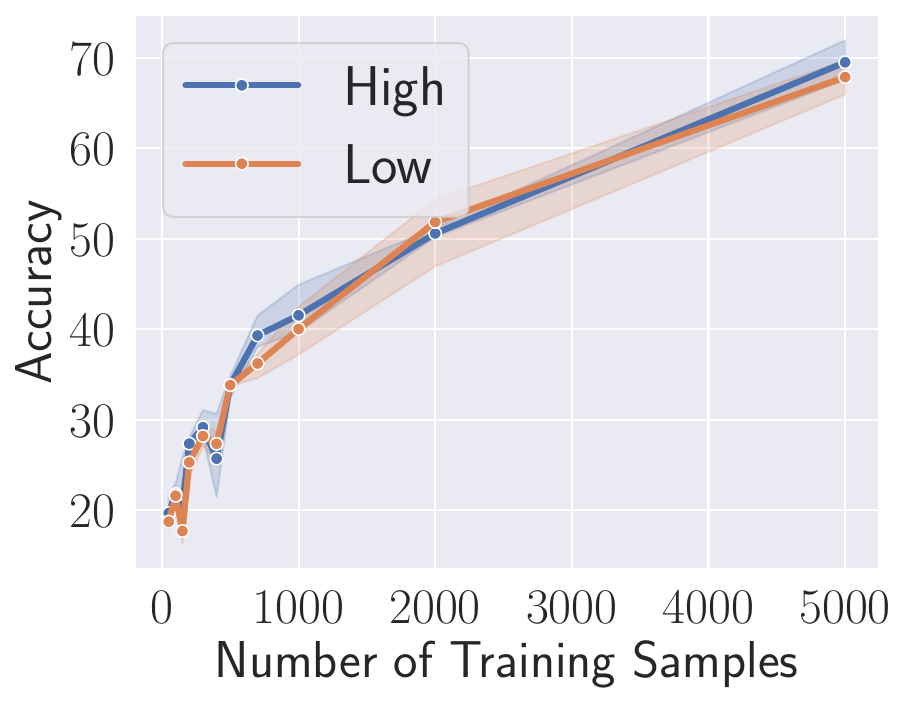}
\caption{LOO}
\label{fig:append_loo}
\end{subfigure}
\begin{subfigure}{0.49\columnwidth}
\includegraphics[width=\columnwidth]{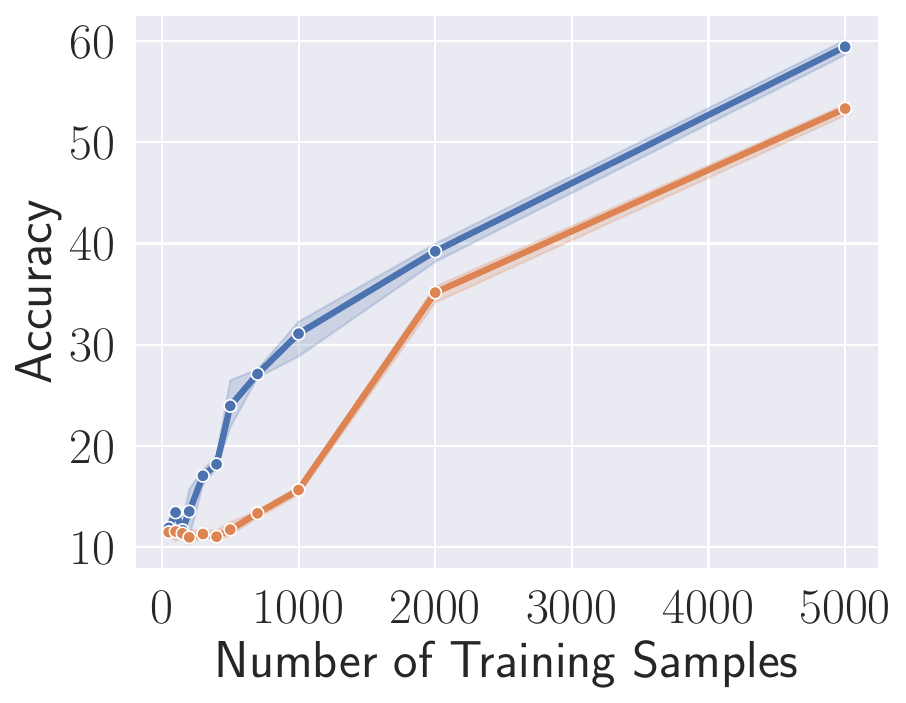}
\caption{Trak}
\label{fig:append_trak}
\end{subfigure}
\caption{LOO and Trak fail to capture the complex interactions between subsets of data, resulting in suboptimal sample importance identification performance.}
\label{figure:append_othermeasure}
\end{figure}

\begin{figure}[t!]
\begin{center}
\centering
\includegraphics[trim=0 0 0 2.2cm,clip,width=0.7\columnwidth]{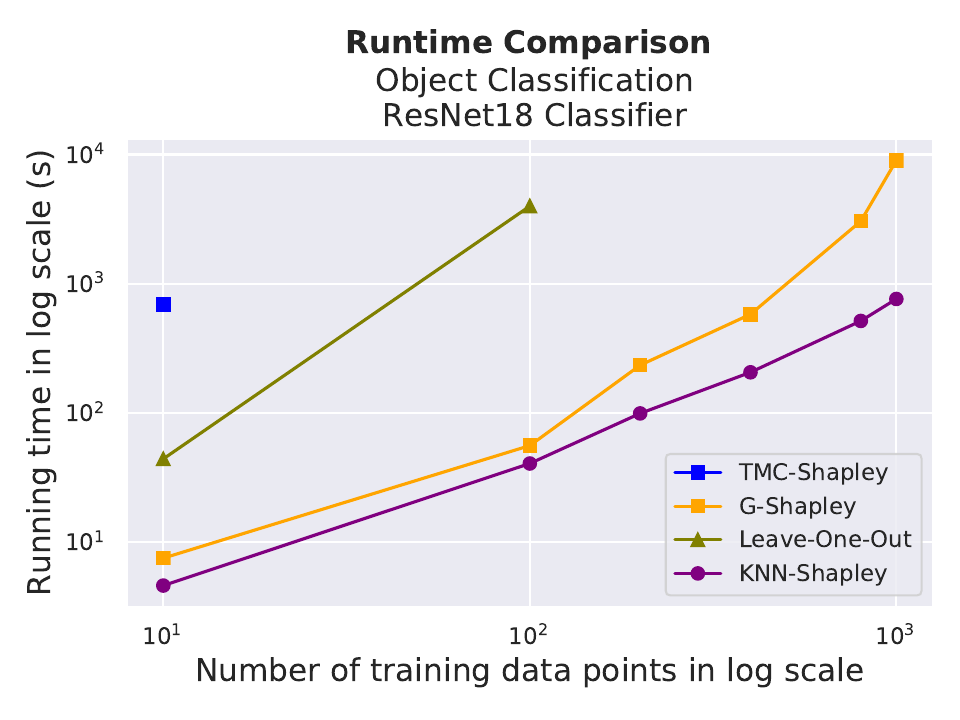}
\caption{A runtime comparison of current measurement methods reveals their significant computational inefficiency. 
(Figure adapted from Figure 1 in~\cite{JWSXDKZLS21})}
\label{fig:runtime}
\end{center}
\end{figure}

\section{Importance Distribution}
\label{append:importance_dis}

\begin{figure*}[ht]
\centering
\begin{subfigure}{0.56\columnwidth}
\includegraphics[width=\columnwidth]{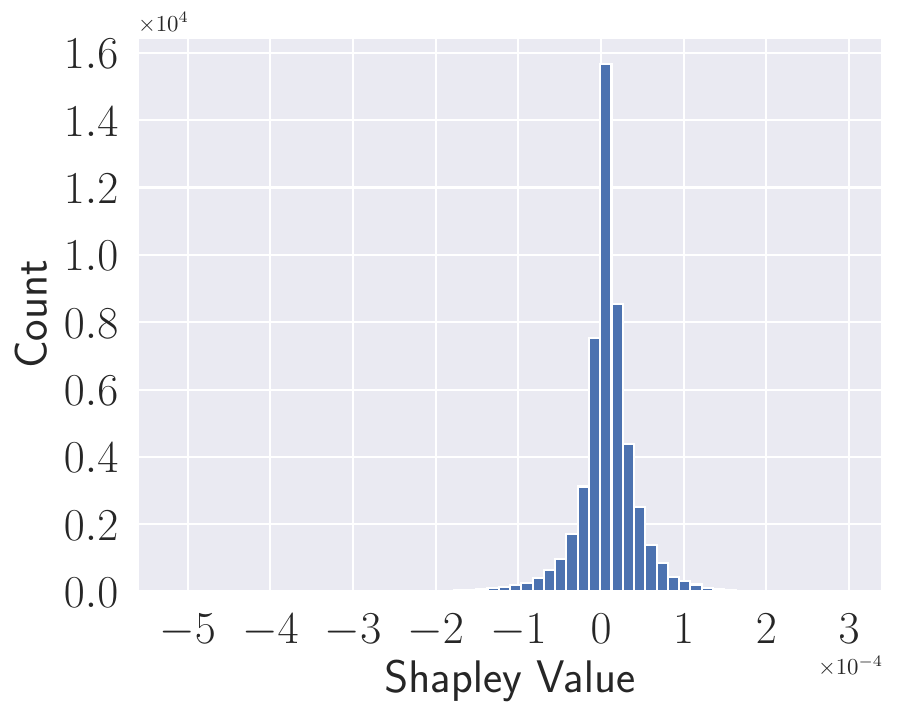}
\caption{CelebA}
\label{fig:append_charac_celeba}
\end{subfigure}
\begin{subfigure}{0.56\columnwidth}
\includegraphics[width=\columnwidth]{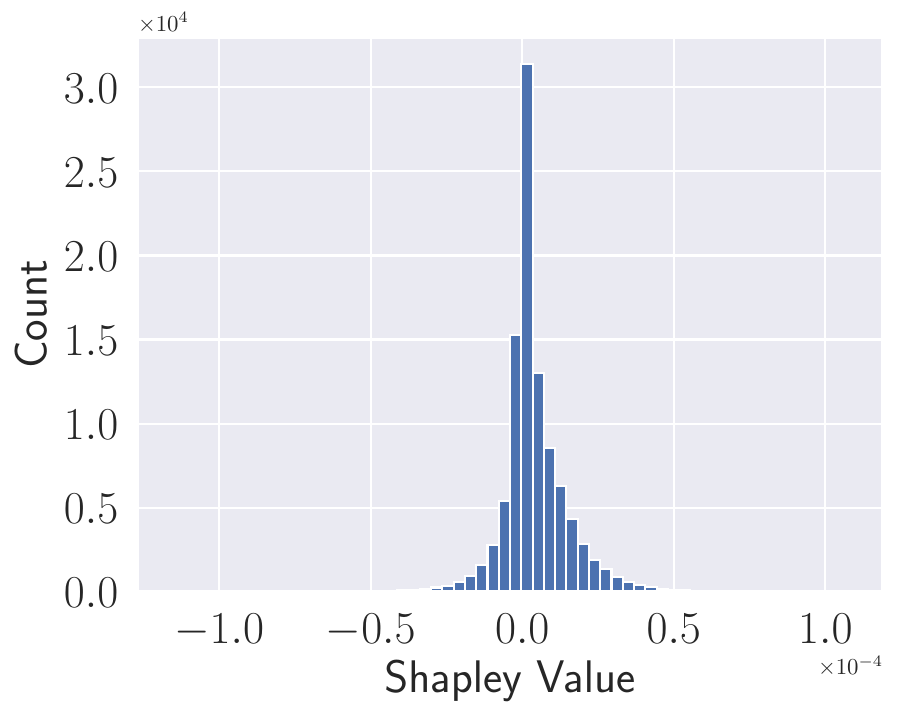}
\caption{TinyImagenet}
\label{fig:append_charac_tiny}
\end{subfigure}
\begin{subfigure}{0.535\columnwidth}
\includegraphics[width=\columnwidth]{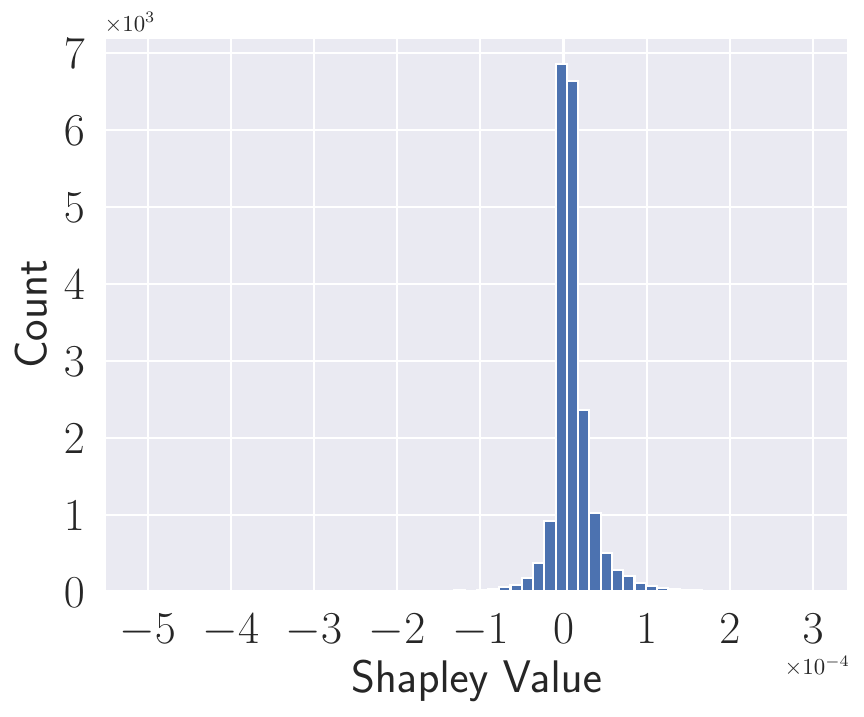}
\caption{Purchase}
\label{fig:append_charac_purchase}
\end{subfigure}
\caption{Importance distribution for CelebA, TinyImagenet, Purchase.}
\label{figure:append_charac}
\end{figure*}

\autoref{figure:append_charac} shows the importance distributions for the CelebA and TinyImageNet dataset.

\begin{figure*}[!t]
\centering
\begin{subfigure}{0.49\columnwidth}
\includegraphics[width=\columnwidth]{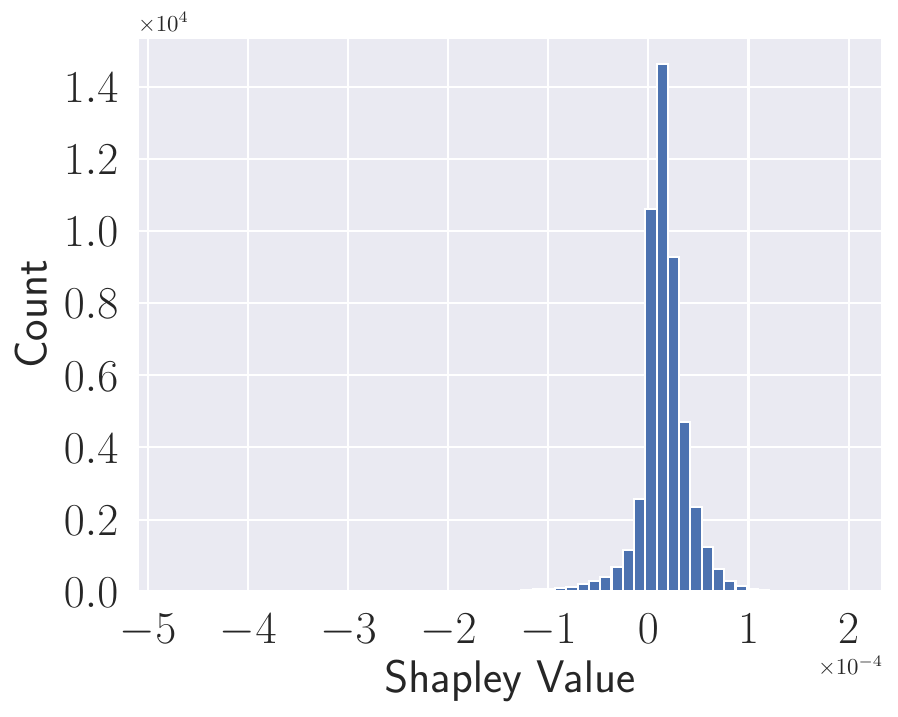}
\caption{Shapley Distribution}
\label{fig:shapley_cifar}
\end{subfigure}
\begin{subfigure}{0.49\columnwidth}
\includegraphics[width=\columnwidth]{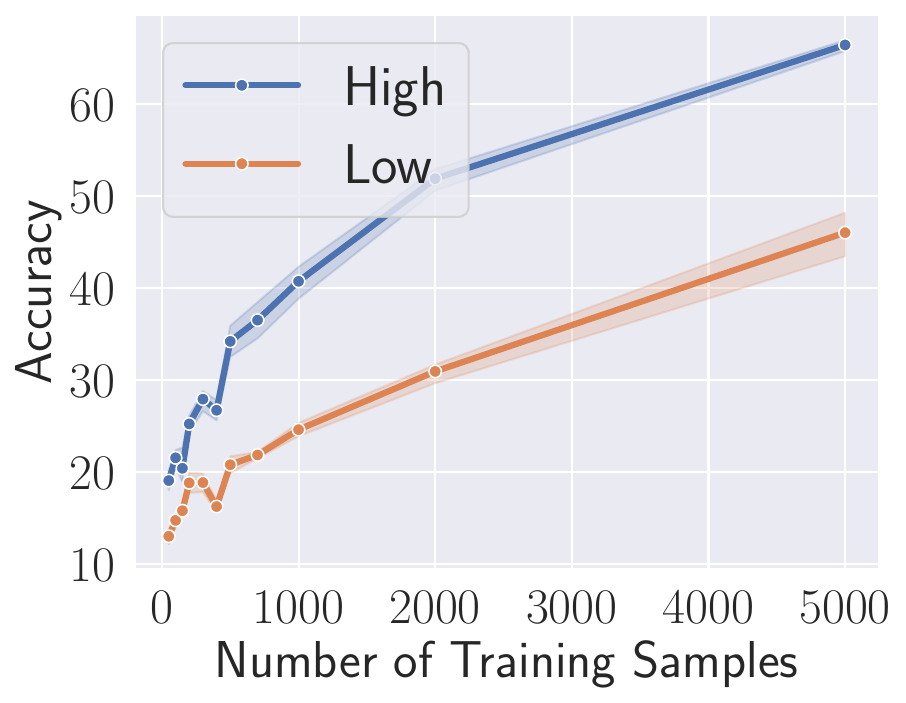}
\caption{CIFAR10}
\label{fig:charac_cifar}
\end{subfigure}
\begin{subfigure}{0.49\columnwidth}
\includegraphics[width=\columnwidth]{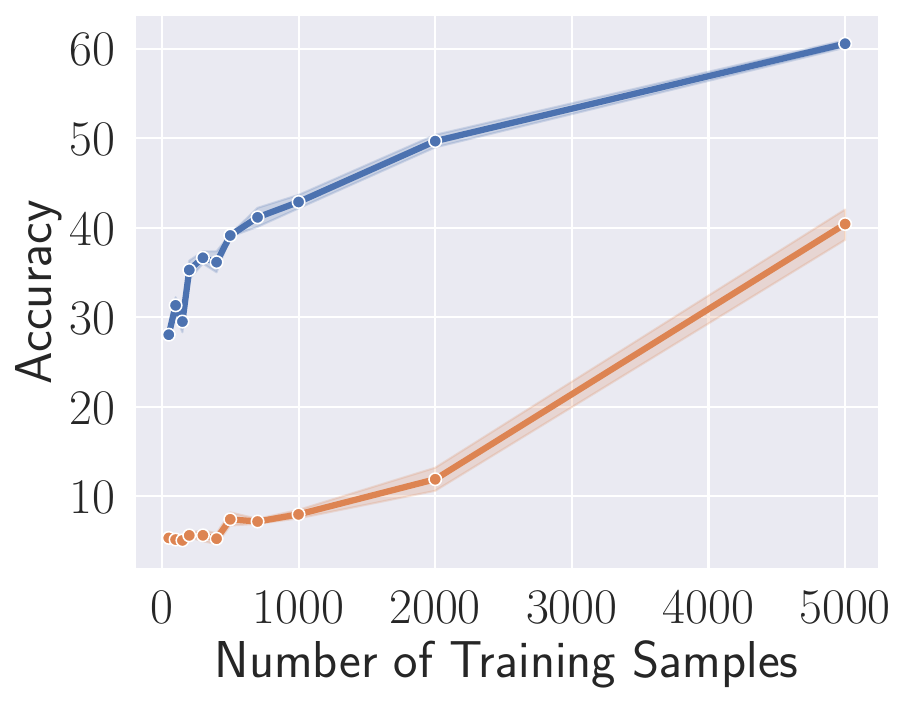}
\caption{CelebA}
\label{fig:charac_celeba}
\end{subfigure}
\begin{subfigure}{0.49\columnwidth}
\includegraphics[width=\columnwidth]{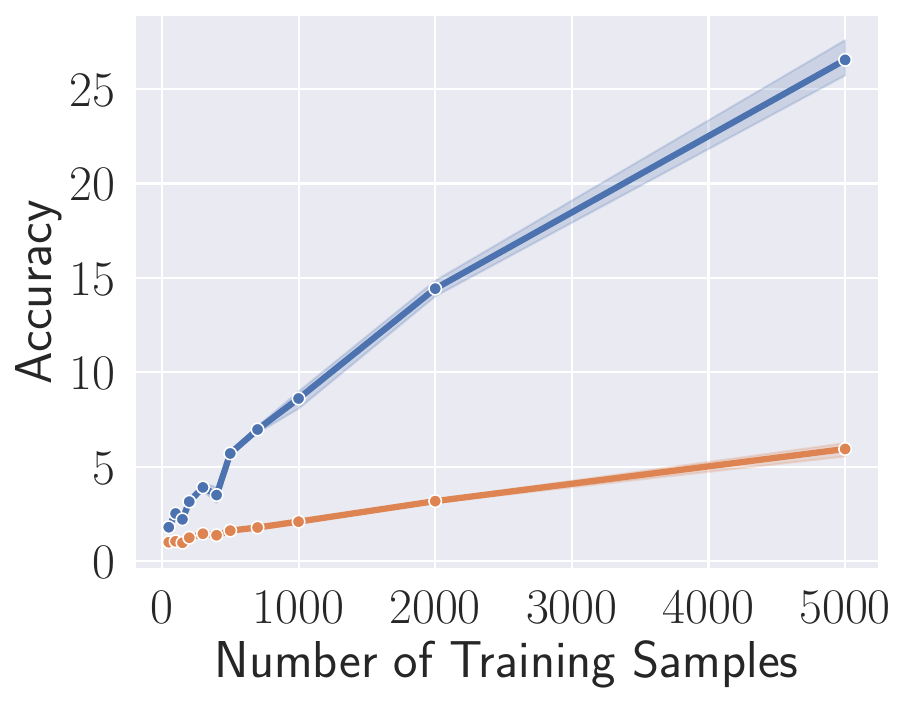}
\caption{TinyImageNet}
\label{fig:charac_tinyimagenet}
\end{subfigure}
\caption{Distribution of importance value and learning characteristics for data with different importance.
High importance samples contribute to better model utility when the dataset has the same size. 
Importance distribution for CelebA and TinyImageNet can be found in~\refapp{append:importance_dis}.}
\label{figure:shapley_character}
\end{figure*}

\section{Effectiveness of \texorpdfstring{$K$NN-Shapley}{kNN-Shapley}}
\label{append:effective_knn}

We validate the efficacy of the $K$NN-Shapley method to ensure its accurate assignment of importance value to individual samples.

We apply the $K$NN-Shapley approach to three distinct datasets and compute the importance value for each sample. 
To gain insight into the contribution of different samples to the model's utility, we visualize the distribution of importance values for the CIFAR10 dataset in~\autoref{fig:shapley_cifar}.

The observed distribution aligns with our expectations, as most samples exhibit similar contributions, while certain samples significantly influence the model's behavior. 
We corroborate this finding with the other two datasets, and include the corresponding visualizations in~\refapp{append:importance_dis}.

Subsequently, we empirically validate whether samples with high importance values and those with low importance values demonstrate distinct training performance. 
To achieve this, we sort the samples based on their importance values and form two sets: one comprising samples with the highest values and the other consisting of samples with the lowest values. 
We employ these two sets to train two separate models and evaluate their performance on the testing dataset. 
We vary the size of these two sets from 50 to 5000 and plot the corresponding testing accuracy in~\autoref{fig:charac_cifar}, ~\autoref{fig:charac_celeba}, and \autoref{fig:charac_tinyimagenet}.

The figures clearly demonstrate that samples with varying importance values exhibit significant differences in training performance. 
Specifically, considering the CIFAR10 dataset trained with 2000 samples, the model trained with high importance samples achieves a testing accuracy that is $1.6\times$ higher compared to the model trained with low importance samples. 
Moreover, in the case of TinyImageNet, the disparity is even more pronounced. 
When the training set comprises 5000 samples, the model trained with valuable data attains a testing accuracy that is $4.4\times$ higher than that of the model trained with low importance samples. 
These experimental findings provide strong evidence supporting the effectiveness of $K$NN-Shapley.

\section{Membership Inference Attack}
\label{appendix_mia}

\begin{figure*}[ht]
\centering
\begin{subfigure}{0.56\columnwidth}
\includegraphics[width=\columnwidth]{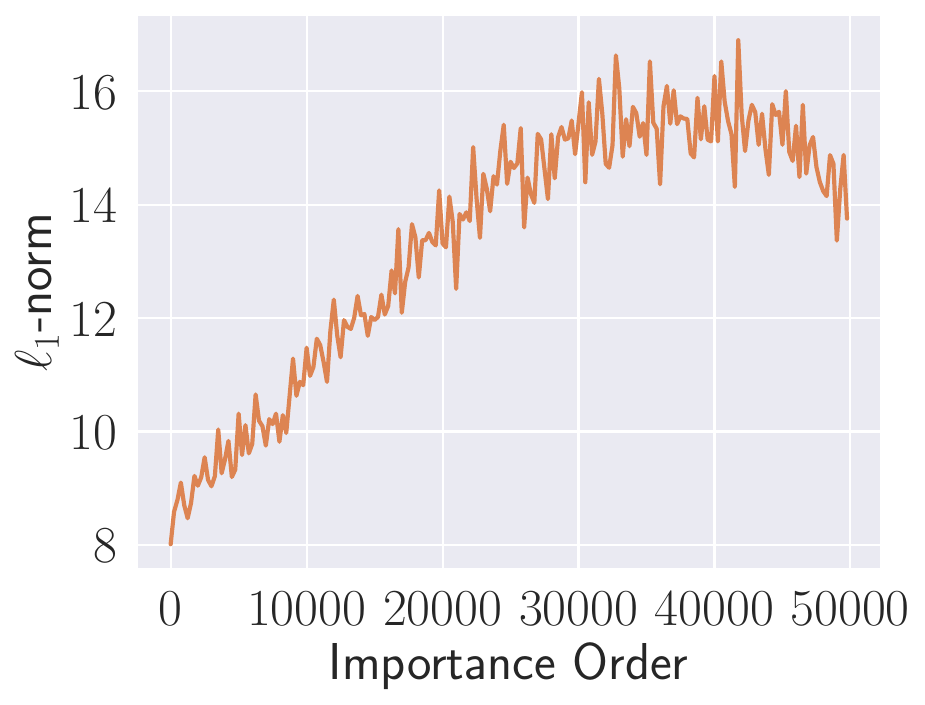}
\caption{CIFAR10}
\label{fig:distance_cifar_l1}
\end{subfigure}
\begin{subfigure}{0.56\columnwidth}
\includegraphics[width=\columnwidth]{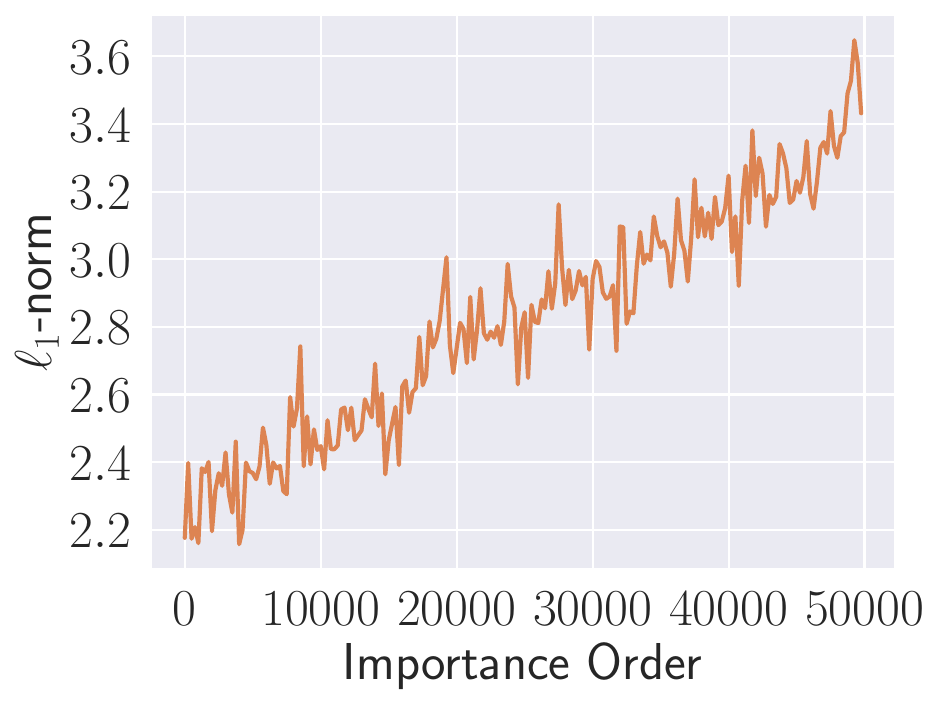}
\caption{CelebA}
\label{fig:distance_celeba_l1}
\end{subfigure}
\begin{subfigure}{0.56\columnwidth}
\includegraphics[width=\columnwidth]{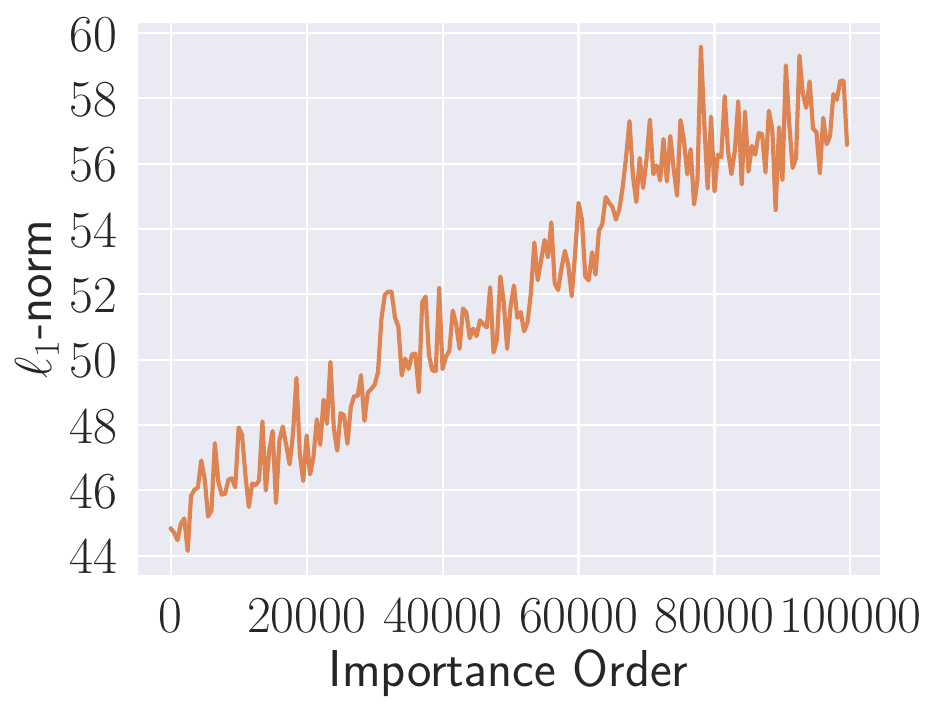}
\caption{TinyImageNet}
\label{fig:distance_tinyimagenet_l1}
\end{subfigure}
\begin{subfigure}{0.56\columnwidth}
\includegraphics[width=\columnwidth]{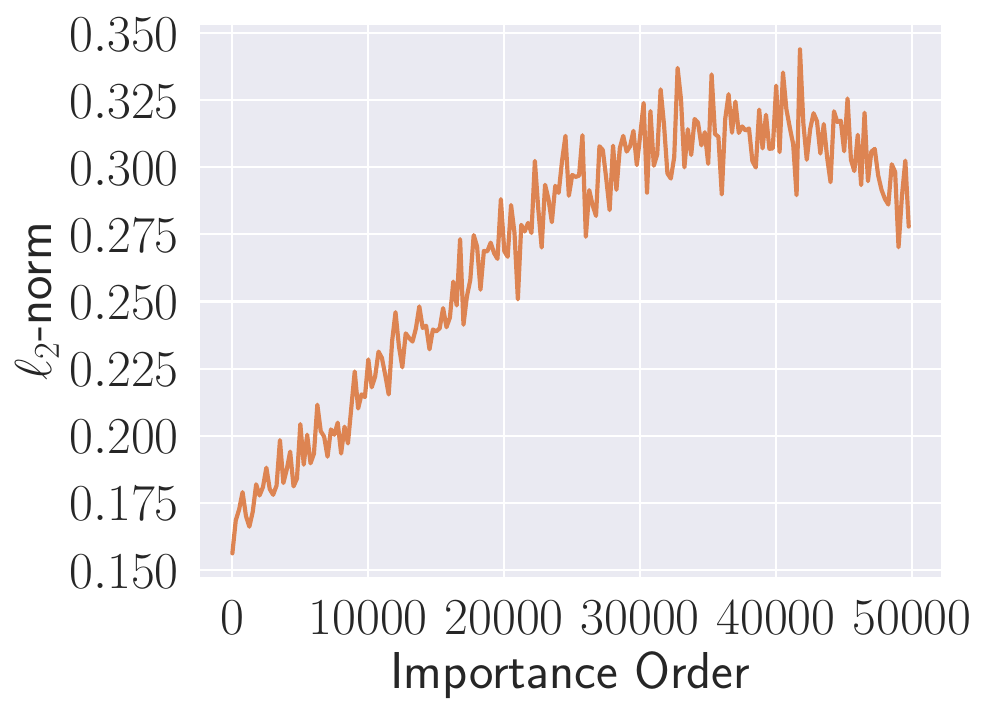}
\caption{CIFAR10}
\label{fig:distance_cifar_l2}
\end{subfigure}
\begin{subfigure}{0.56\columnwidth}
\includegraphics[width=\columnwidth]{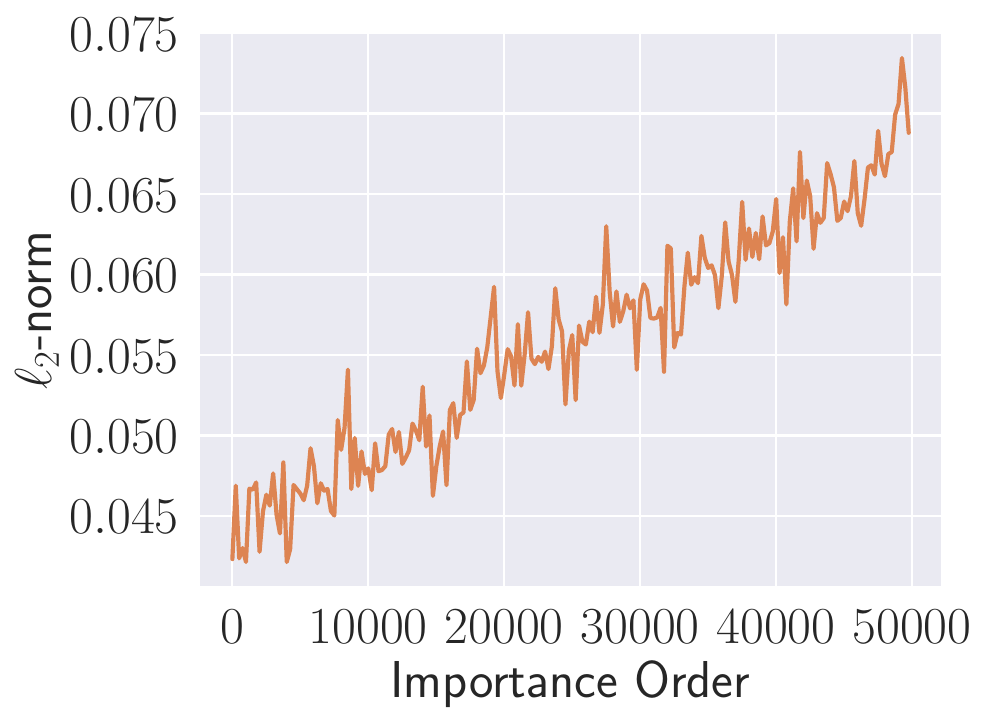}
\caption{CelebA}
\label{fig:distance_celeba_l2}
\end{subfigure}
\begin{subfigure}{0.56\columnwidth}
\includegraphics[width=\columnwidth]{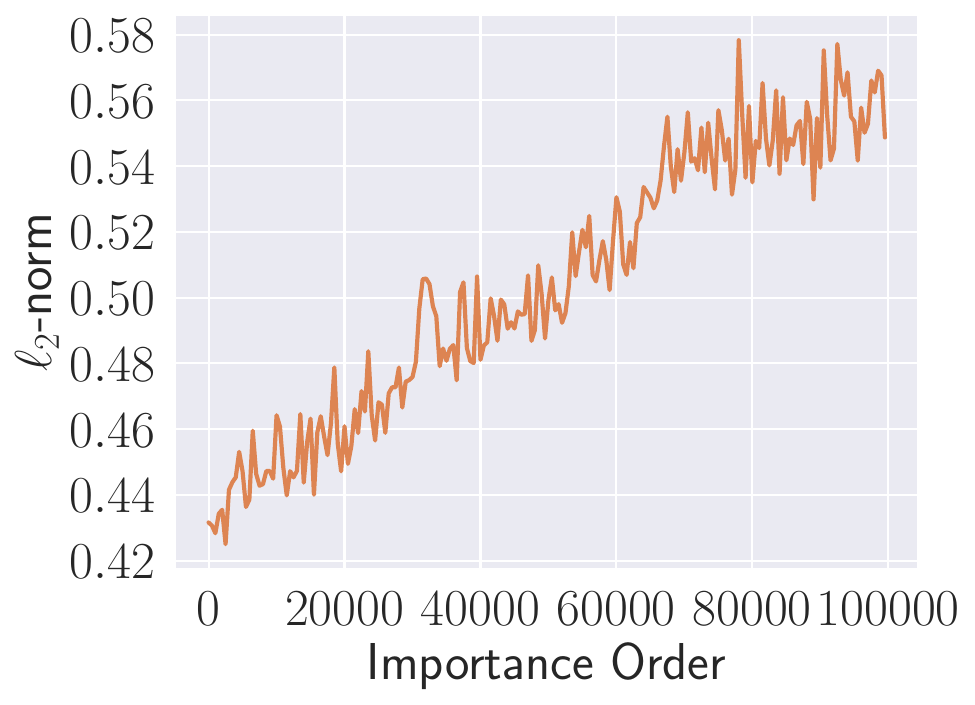}
\caption{TinyImageNet}
\label{fig:distance_tinyimagenet_l2}
\end{subfigure}
\caption{Relationship between distance to the decision boundary and importance value.}
\label{figure:distance_diff_norm}
\end{figure*}

\begin{figure*}[ht]
\centering
\begin{subfigure}{0.56\columnwidth}
\includegraphics[width=\columnwidth]{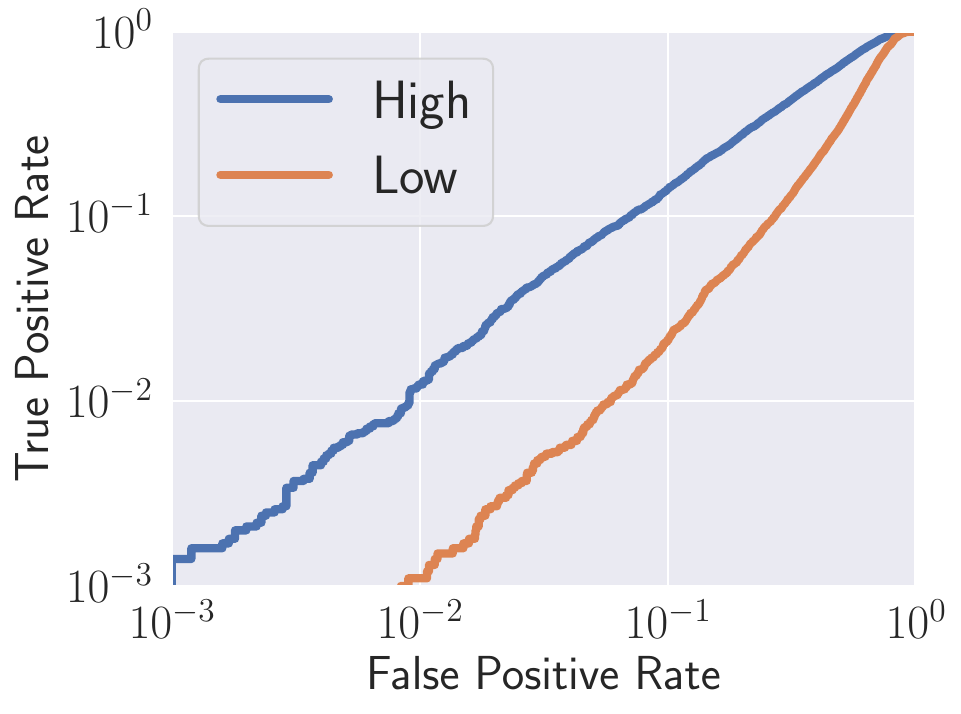}
\caption{CIFAR10}
\label{fig:distance_mia_cifar_l1}
\end{subfigure}
\begin{subfigure}{0.56\columnwidth}
\includegraphics[width=\columnwidth]{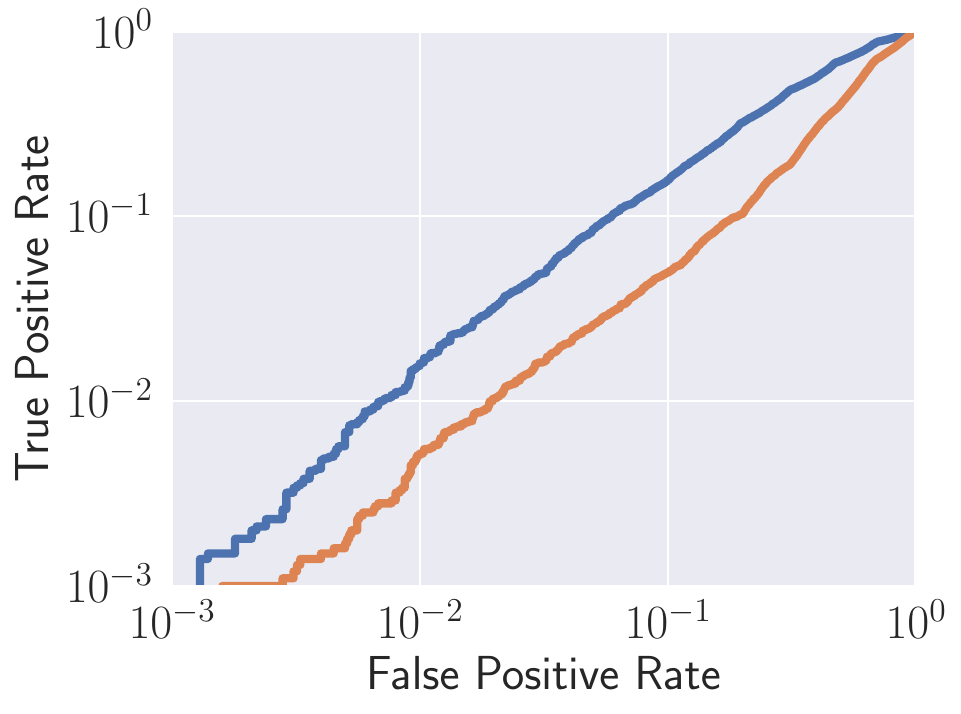}
\caption{CelebA}
\label{fig:distance_mia_celeba_l1}
\end{subfigure}
\begin{subfigure}{0.56\columnwidth}
\includegraphics[width=\columnwidth]{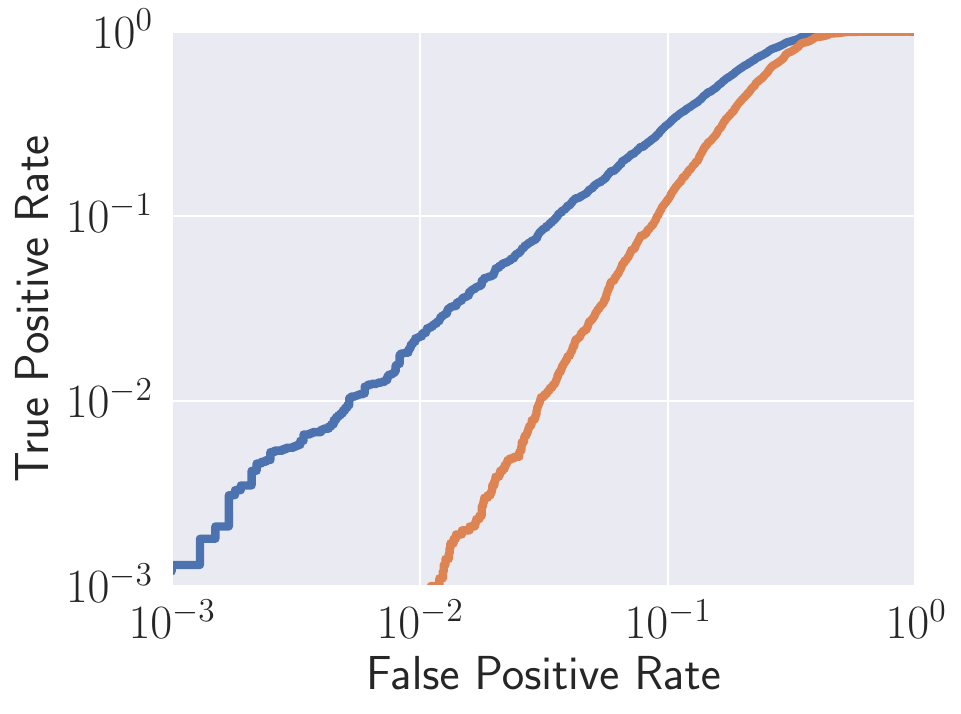}
\caption{TinyImageNet}
\label{fig:distance_mia_tinyimagenet_l1}
\end{subfigure}

\begin{subfigure}{0.56\columnwidth}
\includegraphics[width=\columnwidth]{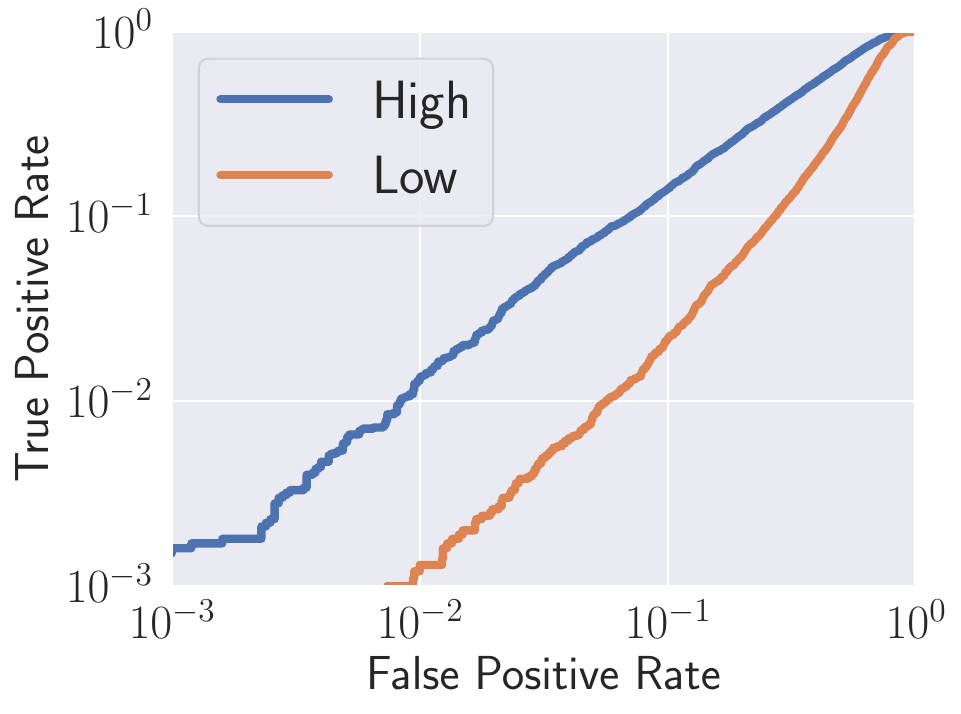}
\caption{CIFAR10}
\label{fig:distance_mia_cifar_l2}
\end{subfigure}
\begin{subfigure}{0.56\columnwidth}
\includegraphics[width=\columnwidth]{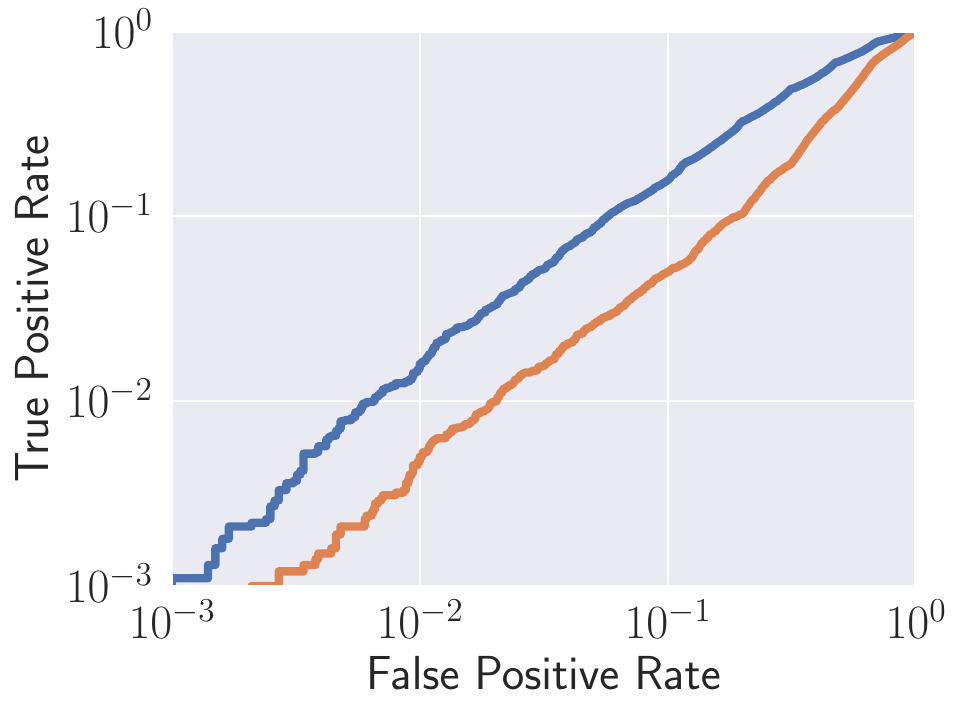}
\caption{CelebA}
\label{fig:distance_mia_celeba_l2}
\end{subfigure}
\begin{subfigure}{0.56\columnwidth}
\includegraphics[width=\columnwidth]{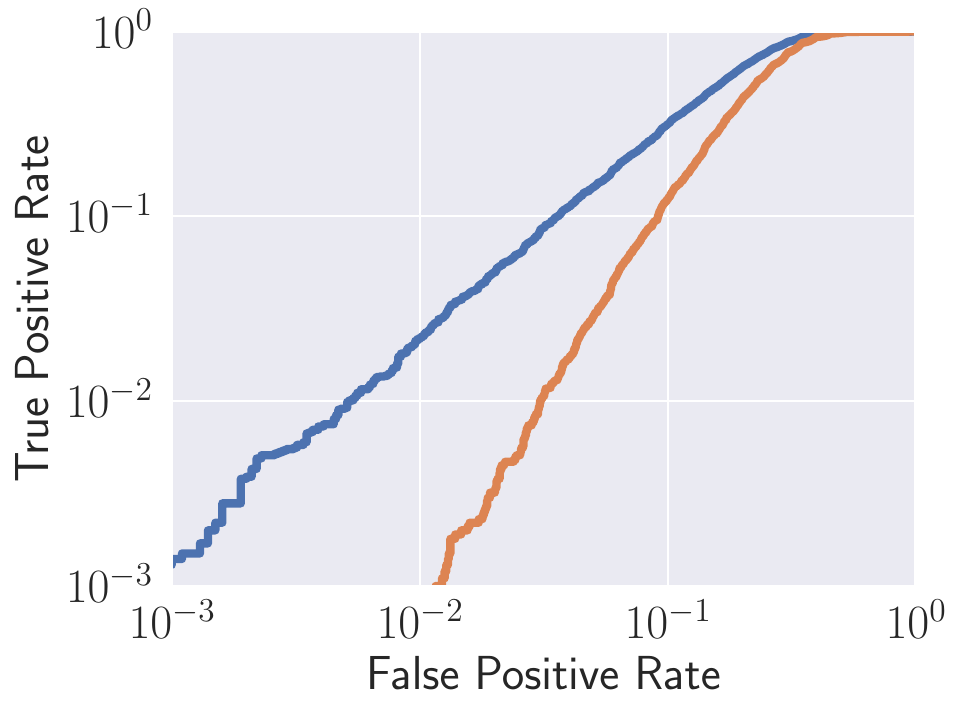}
\caption{TinyImageNet}
\label{fig:distance_mia_tinyimagenet_l2}
\end{subfigure}
\caption{Log-scale ROC curve: membership inference attack based on the distance to the decision boundary. 
The first row is results generated using $\ell_1$ norm while the second row is using $\ell_2$ norm.}
\label{figure:distance_mia_diff_norm}
\end{figure*}

\autoref{figure:distance_diff_norm} depicts the distance to the boundary for samples with different importance values, measured by two different norms.

\autoref{figure:distance_mia_diff_norm} represents the log-scale ROC curves for attacks conducted based on the distance to the boundary.

\section{Correlation Between Different Attributes}
\label{app:attr_corr}

\begin{figure}[ht]
\centering
\includegraphics[width=0.95\columnwidth]{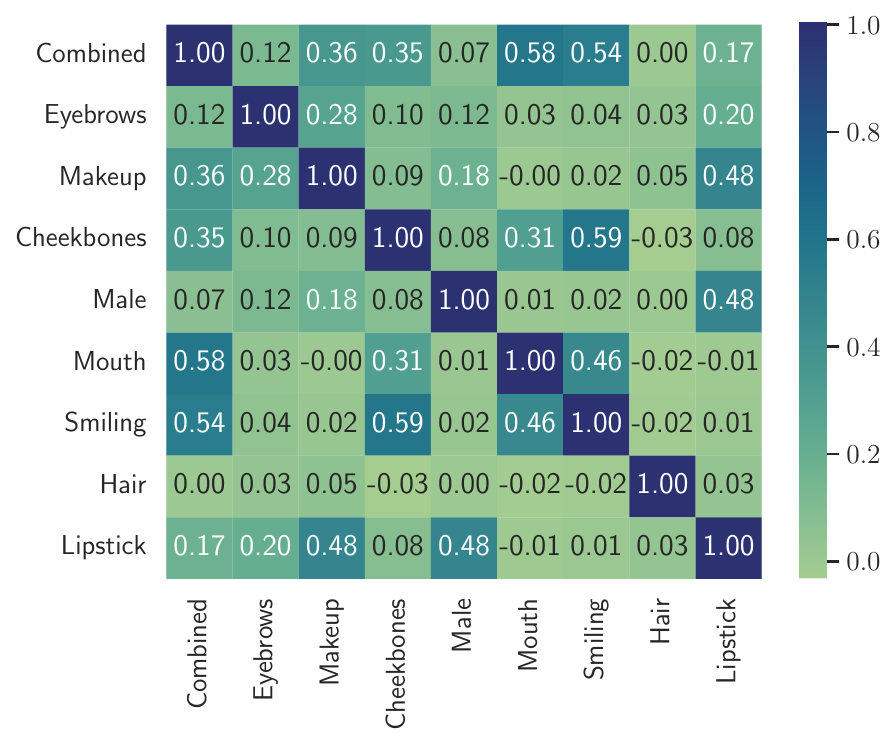}
\caption{The heatmap depicts the correlation among importance values assigned to different attributes. 
It indicates that a sample's elevated importance on one attribute may not align with its importance on another attribute.}
\label{fig:corr_attr}
\end{figure}

We visualize the correlation among importance values assigned to different attributes in~\autoref{fig:corr_attr}.

\section{Hyperparameters for Backdoor Attacks}
\label{append:bd_hyperparameter}

For the three datasets evaluated in our study—CIFAR10, CelebA, and TinyImagenet—a consistent modification was applied to each image: a black square was positioned at the bottom left corner. 

The dimension of this black square, or backdoor trigger, varied according to the image sizes of the respective datasets to maintain proportional consistency. 
Specifically, for the CIFAR10 dataset, with an image resolution of \(32 \times 32\), the trigger was sized at \(2 \times 2\). 
In the case of CelebA, which features larger images with dimensions of \(178 \times 218\), the trigger's size was increased to \(8 \times 8\). 
Lastly, for images from TinyImagenet, which are \(64 \times 64\) pixels, a \(5 \times 5\) square was used as the trigger. 

\section{Clean Accuracy Performance}
\label{append:bd_acc}

The effect of poisoning high and low importance samples on clean accuracy is depicted in ~\autoref{fig:corr_attr}.

\begin{figure*}[!t]
\centering
\begin{subfigure}{0.56\columnwidth}
\includegraphics[width=\columnwidth]{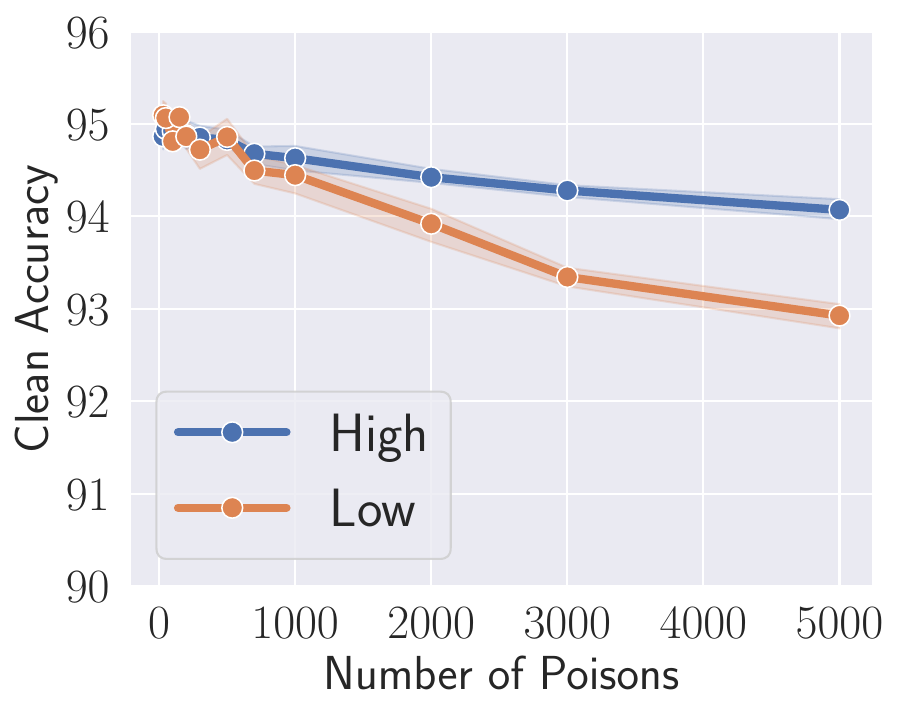}
\caption{CIFAR10}
\label{fig:backdoor_acc_cifar}
\end{subfigure}
\begin{subfigure}{0.57\columnwidth}
\includegraphics[width=\columnwidth]{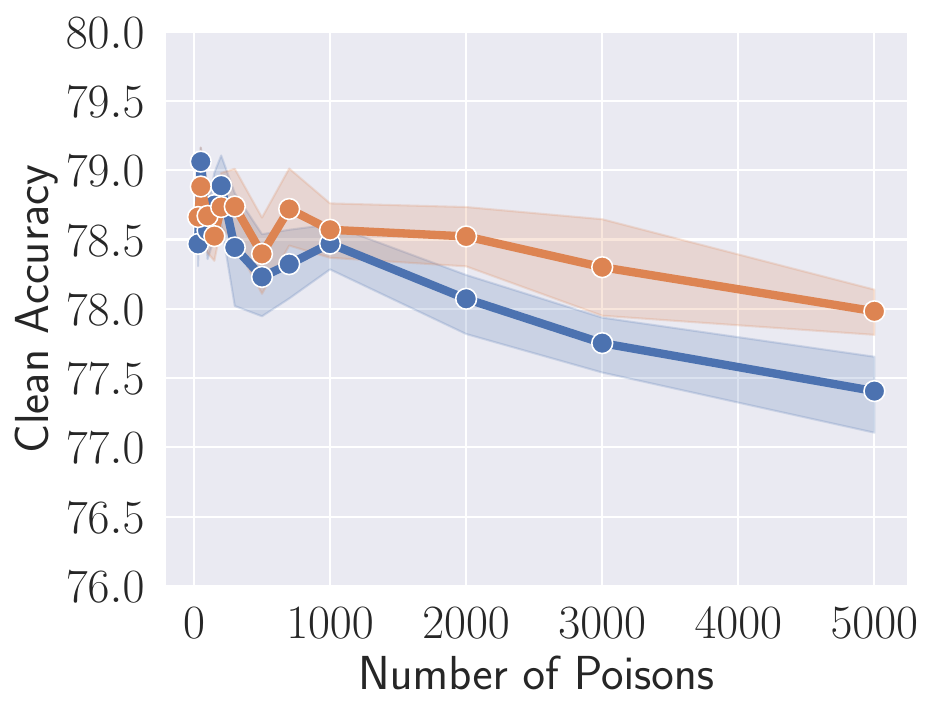}
\caption{CelebA}
\label{fig:backdoor_acc_celeba}
\end{subfigure}
\begin{subfigure}{0.56\columnwidth}
\includegraphics[width=\columnwidth]{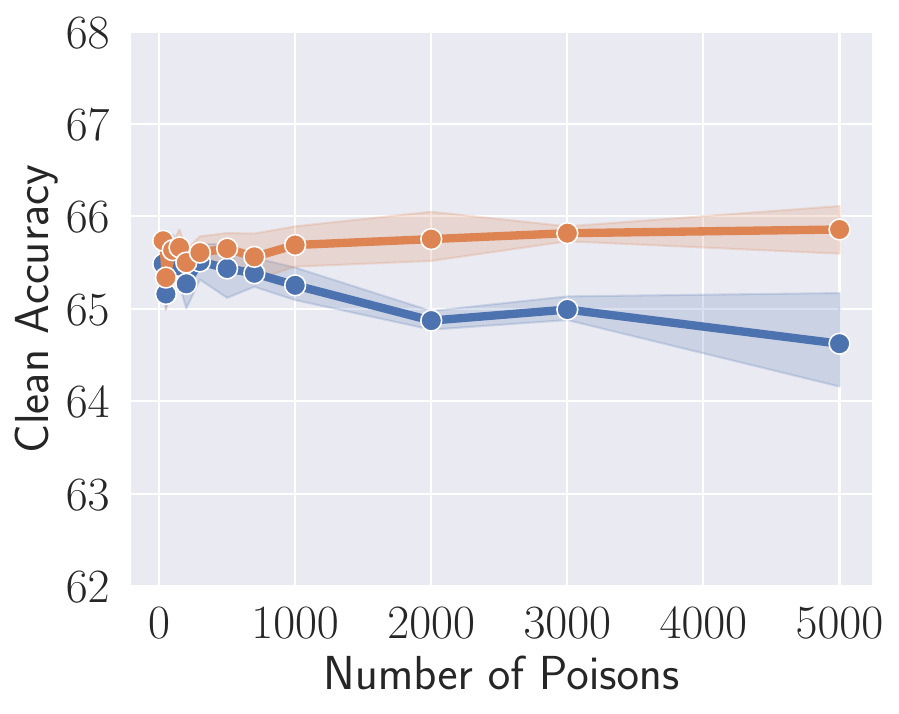}
\caption{TinyImageNet}
\label{fig:backdoor_acc_tinyimagenet}
\end{subfigure}
\caption{Relationship between accuracy and the poisoning rate, poisoning samples with different importance does not have a significant difference.}
\label{figure:backdoor_acc}
\end{figure*}

\section{More Backdoor Attacks}
\label{append:more_backdoor}

To assess whether the observation that poisoning high importance data samples enhances backdoor attack effectiveness is applicable to various trigger patterns, we broadened our study to include three additional backdoor methods. 
These methods comprise Blend~\cite{CLLLS17}, which incorporates triggers covering the entirety of the input; SSBA~\cite{LLWLHL21}, characterized by sample-specific and invisible triggers; LF~\cite{ZPMJ21}, which utilizes triggers of low frequency; SIG~\cite{BKT19}, a method without label poisoning; and CTRL~\cite{LPXDJYW23}, which targets contrastive learning.

We utilized the BackdoorBench tool~\cite{WCZZWYS22} to conduct our experiments, adhering to all default implementation settings with the sole modification being the selection process for poisoning samples.

\begin{figure*}[!t]
\centering
\begin{subfigure}{0.37\columnwidth}
\includegraphics[width=\columnwidth]{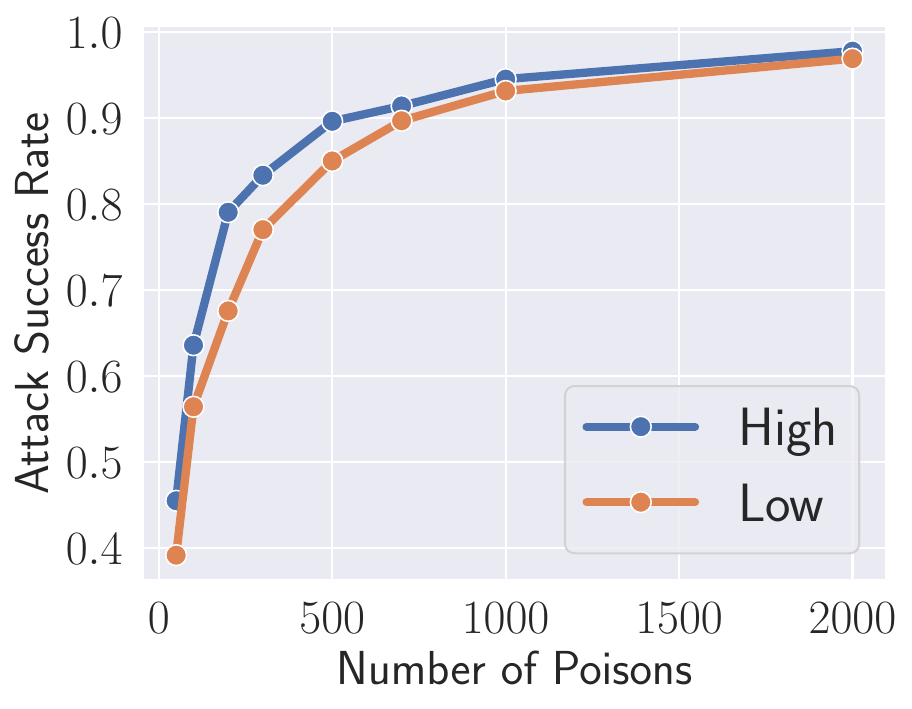}
\caption{Blend~\cite{CLLLS17}}
\label{fig:backdoor_asr_blend}
\end{subfigure}
\begin{subfigure}{0.37\columnwidth}
\includegraphics[width=\columnwidth]{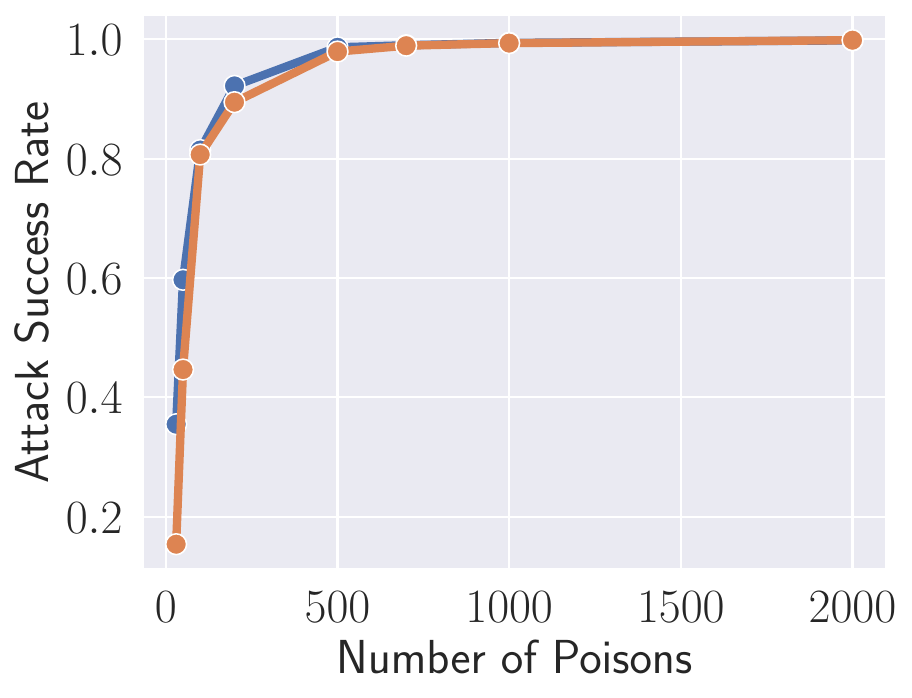}
\caption{SSBA~\cite{LLWLHL21}}
\label{fig:backdoor_asr_ssba}
\end{subfigure}
\begin{subfigure}{0.37\columnwidth}
\includegraphics[width=\columnwidth]{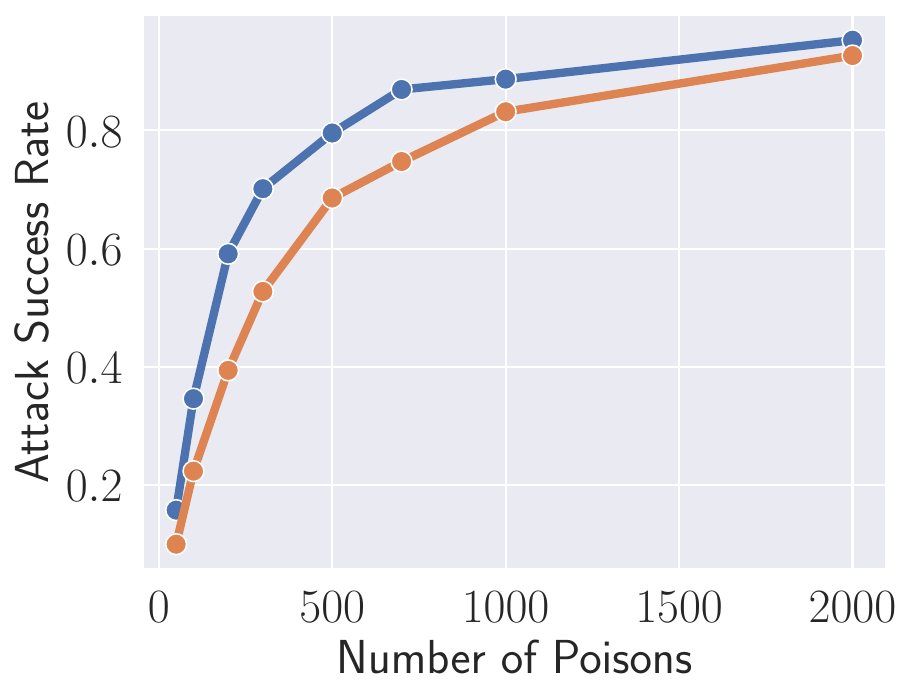}
\caption{LF~\cite{ZPMJ21}}
\label{fig:backdoor_asr_lf}
\end{subfigure}
\begin{subfigure}{0.37\columnwidth}
\includegraphics[width=\columnwidth]{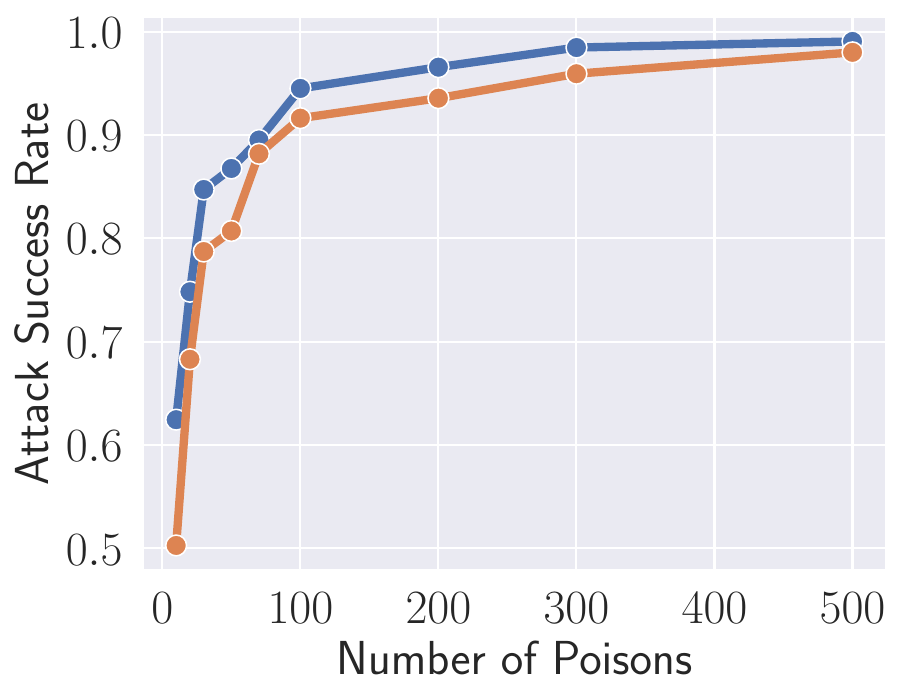}
\caption{SIG~\cite{BKT19}}
\label{fig:bd_sig}
\end{subfigure}
\begin{subfigure}{0.37\columnwidth}
\includegraphics[width=\columnwidth]{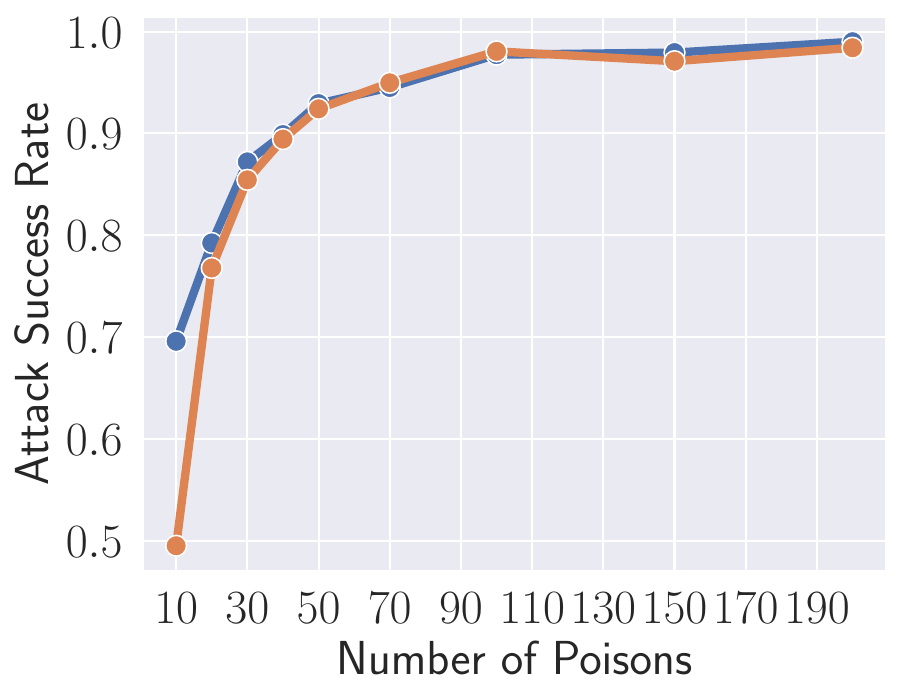}
\caption{CTRL~\cite{LPXDJYW23}}
\label{fig:bd_ctrl}
\end{subfigure}
\caption{Relationship between attack success rate and the poisoning rate on different backdoor attacks, the conclusion that high importance samples enhance the efficiency of the poisoning process holds for other backdoor attacks with different backdoor patterns and learning paradigms.}
\label{figure:more_backdoor_asr}
\end{figure*}

The results, depicted in~\autoref{figure:more_backdoor_asr}, consistently demonstrate that the poisoning of high importance samples significantly improves the efficacy of the backdoor attacks across more complex trigger patterns, thus underscoring the robust generalizability of our conclusions.

\section{Transferability Study}
\label{append:transfer}

\begin{figure*}[!t]
\centering
\begin{subfigure}{0.49\columnwidth}
\includegraphics[width=\columnwidth]{modelsteal/cifar10_steal_cifar10_resnet18_full_l1.pdf}
\caption{ResNet18 (CIFAR10)}
\label{fig:trans_ms_res18}
\end{subfigure}
\begin{subfigure}{0.49\columnwidth}
\includegraphics[width=\columnwidth]{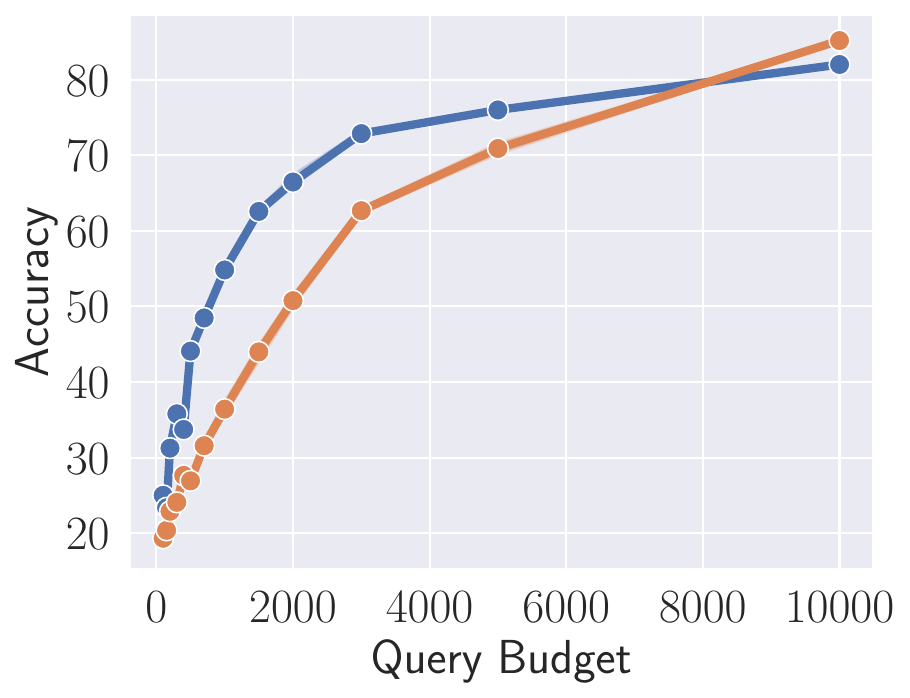}
\caption{MobileNetV2 (CIFAR10)}
\label{fig:trans_ms_mobilev2}
\end{subfigure}
\begin{subfigure}{0.49\columnwidth}
\includegraphics[width=\columnwidth]{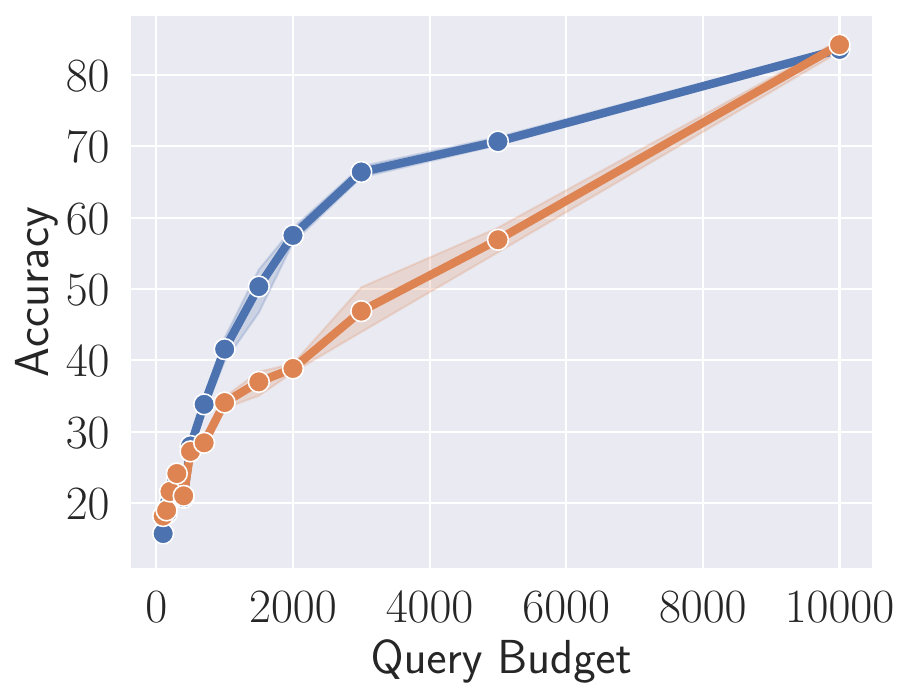}
\caption{ResNet50 (CIFAR10)}
\label{fig:trans_ms_res50}
\end{subfigure}
\begin{subfigure}{0.49\columnwidth}
\includegraphics[width=\columnwidth]{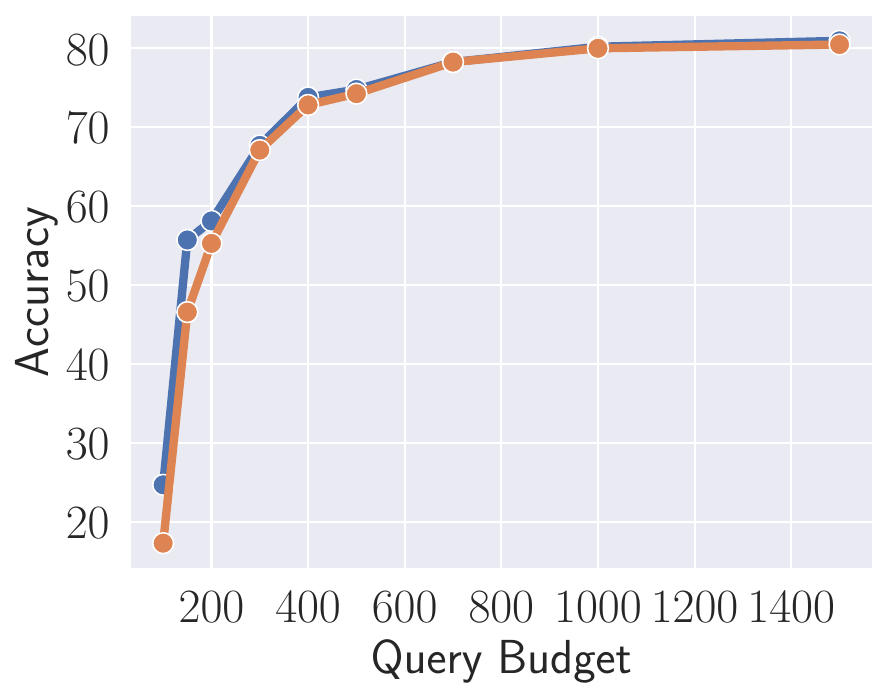}
\caption{MLP (Purchase)}
\label{fig:trans_ms_purchase}
\end{subfigure}
\caption{Relationship between model stealing accuracy and the query budget.}
\label{figure:trans_ms}
\end{figure*}

\begin{figure*}[!t]
\centering
\begin{subfigure}{0.49\columnwidth}
\includegraphics[width=\columnwidth]{backdoor/cifar10_resnet18_ASR_size2.pdf}
\caption{ResNet18 (CIFAR10)}
\label{fig:trans_bd_res18}
\end{subfigure}
\begin{subfigure}{0.49\columnwidth}
\includegraphics[width=\columnwidth]{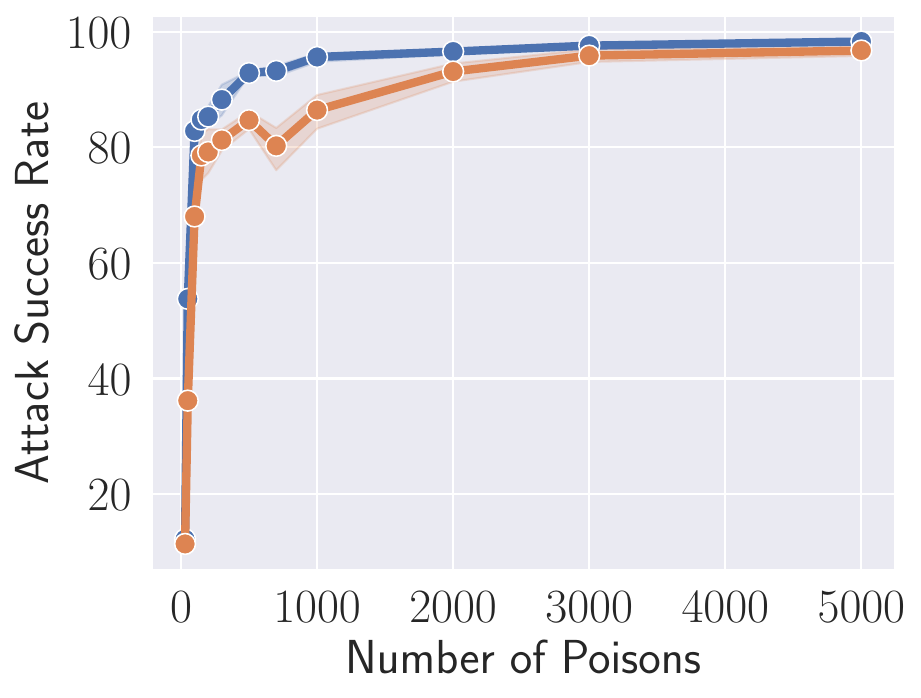}
\caption{MobileNetV2 (CIFAR10)}
\label{fig:trans_bd_mobilev2}
\end{subfigure}
\begin{subfigure}{0.49\columnwidth}
\includegraphics[width=\columnwidth]{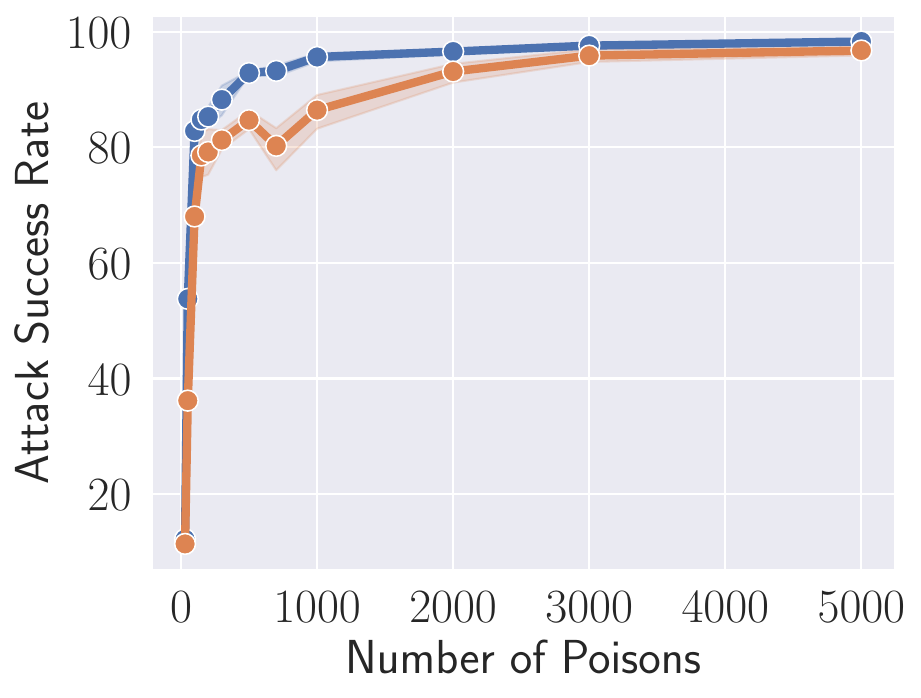}
\caption{ResNet50 (CIFAR10)}
\label{fig:trans_bd_res50}
\end{subfigure}
\begin{subfigure}{0.49\columnwidth}
\includegraphics[width=\columnwidth]{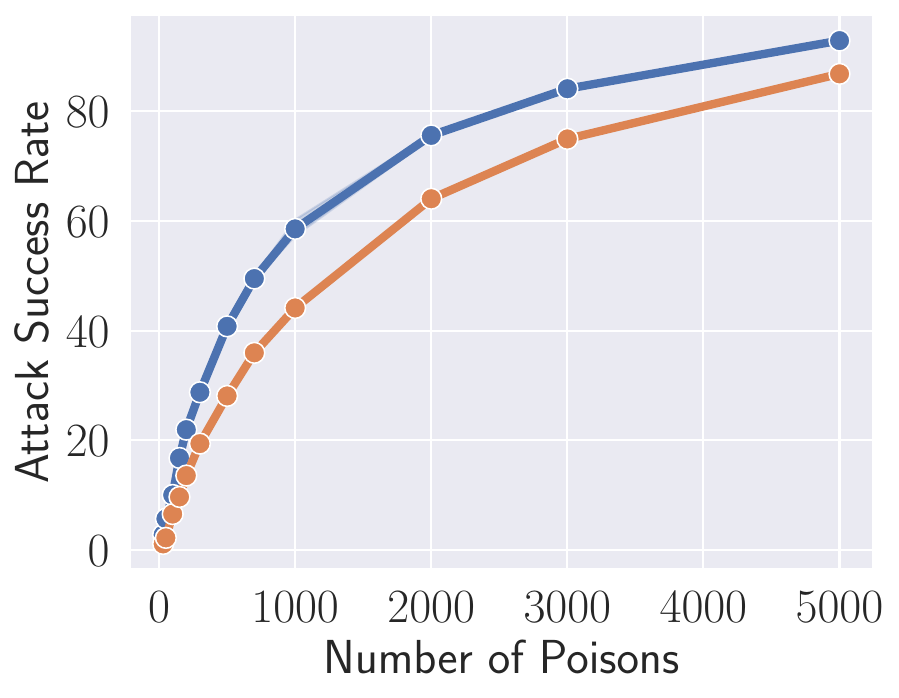}
\caption{MLP (Purchase)}
\label{fig:trans_bd_purchase}
\end{subfigure}
\caption{Relationship between backdoor attack success rate and the poisoning rate. }
\label{figure:trans_bd}
\end{figure*}

\begin{figure*}[ht]
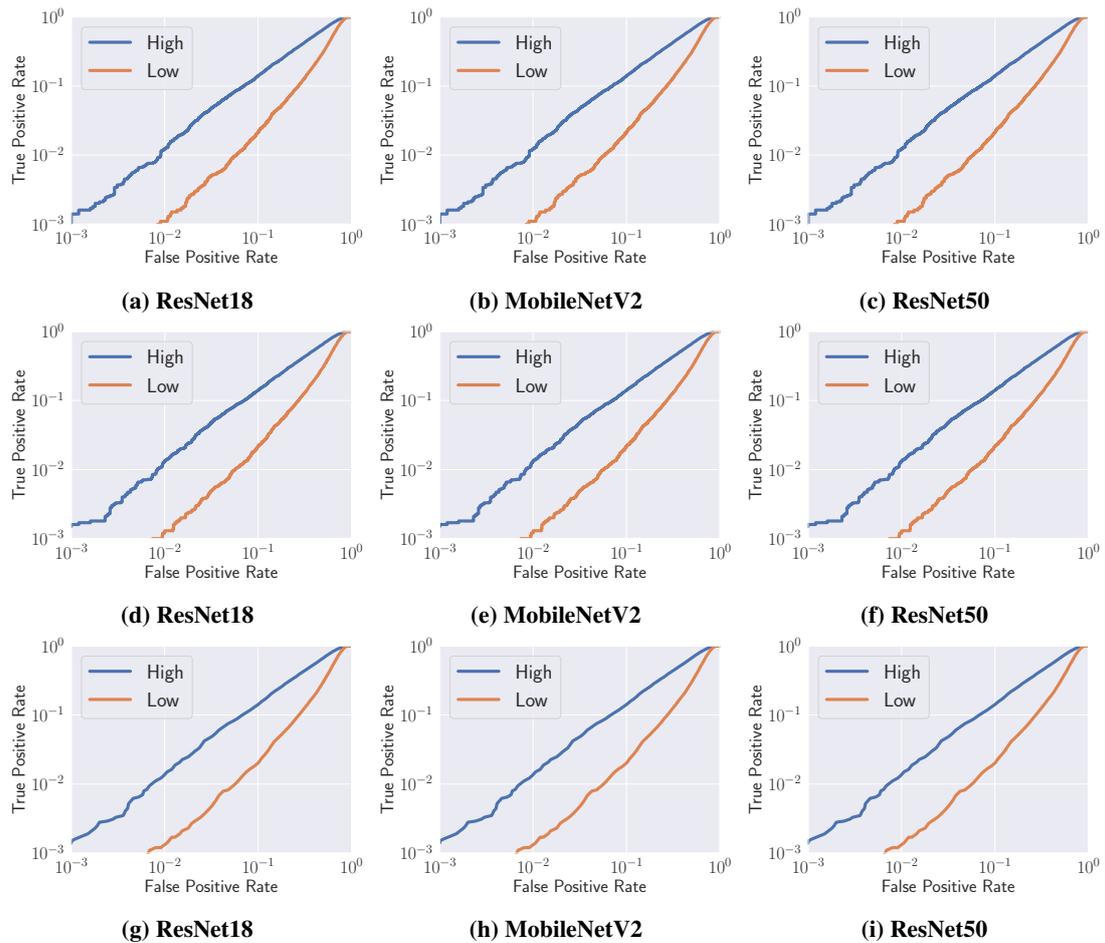

\centering
\begin{subfigure}{0.56\columnwidth}
\includegraphics[width=\columnwidth]{distance_mia/cifar10_resnet18_l1_auc_curve.pdf}
\caption{ResNet18}
\label{fig:trans_mia_res18_l1}
\end{subfigure}
\begin{subfigure}{0.56\columnwidth}
\includegraphics[width=\columnwidth]{distance_mia/cifar10_mobilenetv2_l1_auc_curve.pdf}
\caption{MobileNetV2}
\label{fig:trans_mia_mobilev2_l1}
\end{subfigure}
\begin{subfigure}{0.56\columnwidth}
\includegraphics[width=\columnwidth]{distance_mia/cifar10_resnet50_l1_auc_curve.pdf}
\caption{ResNet50}
\label{fig:trans_mia_res50_l1}
\end{subfigure}
\begin{subfigure}{0.56\columnwidth}
\includegraphics[width=\columnwidth]{distance_mia/cifar10_resnet18_l2_auc_curve.pdf}
\caption{ResNet18}
\label{fig:trans_mia_res18_l2}
\end{subfigure}
\begin{subfigure}{0.56\columnwidth}
\includegraphics[width=\columnwidth]{distance_mia/cifar10_mobilenetv2_l2_auc_curve.pdf}
\caption{MobileNetV2}
\label{fig:trans_mia_mobilev2_l2}
\end{subfigure}
\begin{subfigure}{0.56\columnwidth}
\includegraphics[width=\columnwidth]{distance_mia/cifar10_resnet50_l2_auc_curve.pdf}
\caption{ResNet50}
\label{fig:trans_mia_res50_l2}
\end{subfigure}
\begin{subfigure}{0.56\columnwidth}
\includegraphics[width=\columnwidth]{distance_mia/cifar10_resnet18_inf_auc_curve.pdf}
\caption{ResNet18}
\label{fig:trans_mia_res18_inf}
\end{subfigure}
\begin{subfigure}{0.56\columnwidth}
\includegraphics[width=\columnwidth]{distance_mia/cifar10_mobilenetv2_inf_auc_curve.pdf}
\caption{MobileNetV2}
\label{fig:trans_mia_mobilev2_inf}
\end{subfigure}
\begin{subfigure}{0.56\columnwidth}
\includegraphics[width=\columnwidth]{distance_mia/cifar10_resnet50_inf_auc_curve.pdf}
\caption{ResNet50}
\label{fig:trans_mia_res50_inf}
\end{subfigure}
\caption{Log-scale ROC curve: membership inference attack based on the distance to the decision boundary. 
The first row is results generated using $\ell_1$ norm, the second row is using $\ell_2$ norm, and the third row is using $\ell_{\infty}$ norm.}
\label{figure:trans_mia_append}
\end{figure*}

\autoref{figure:trans_ms} and~\autoref{figure:trans_bd} demonstrate the transferability on model stealing and backdoor attack. 

\autoref{figure:trans_mia_append} showcases Log-scale ROC curves for three distinct architectures utilizing three different norms. 
The experiments are conducted on the CIFAR10 target dataset.

\section{Related Work}
\label{sec:related}

In addition to the attack approach investigated in our work, several methods exist for privacy and security attacks against machine learning models. 
In the following section, we provide a brief overview of these approaches.

\subsection{Membership Inference Attack}
\label{sec:related_mia}

Membership Inference Attacks (MIA)~\cite{SSSS17,SZHBFB19,LZ21,WLBZ24,HWWBSZ21,ZYWBZ24} have emerged as a significant threat to privacy in the context of machine learning models. 
These attacks aim to reveal the membership status of a target sample, i.e., whether the sample was part of the training dataset or not, thereby directly breaching privacy.

The seminal work by Shokri et al.\cite{SSSS17} introduced MIA against machine learning models, wherein multiple shadow models were trained to mimic the behavior of the target model. 
This attack originally required access to data from the same distribution as the training dataset. 
However, Salem et al.\cite{SZHBFB19} relaxed this assumption by demonstrating the effectiveness of using only a single shadow model, substantially reducing the computational cost involved.

Subsequent research~\cite{CTCP21,LZ21} has explored more challenging settings for MIA. 
In these scenarios, the adversary only has access to hard-label predictions from the target model. 
Li and Zhang~\cite{LZ21} proposed a method that approximates the distance between the target sample and its decision boundary using adversarial examples, enabling the attacker to make decisions based on this distance.

Recent advancements in MIA have focused on enhancing attack performance. 
Carlini et al.\cite{CCNSTT22} leveraged the discrepancy between models trained with and without the target sample to improve attack effectiveness. 
Liu et al.\cite{LZBZ22} demonstrated the utility of loss trajectory analysis in MIA. 
Furthermore, Tram{\`e}r et al.~\cite{TSJLJHC22} highlighted the potential of data poisoning, showing that even with access to a small fraction of the training dataset, the attacker can significantly boost the performance of membership inference attacks.

\subsection{Model Stealing Attack}
\label{sec:related_ms}

Model stealing attacks~\cite{TZJRR16,OSF19,KPQ21,TMWP21,LWBZ24,WWBZS23} aim to extract information from a victim model and construct a local surrogate model. 
This attack was initially proposed by Tramèr et al.\cite{TZJRR16}, assuming that the adversary has access to a surrogate dataset for stealing the model. 
Orekondy et al. further advanced this approach by developing a reinforcement learning-based framework that optimizes query time and effectiveness~\cite{OSF19}.

Recent research has focused on the more stringent data-free setting, where adversaries lack access to any data. 
In this context, Kariyappa et al.~\cite{KPQ21} propose MAZE, which employs a generative model to generate synthetic data samples for launching the attack. 
The generator is trained to maximize disagreement between the victim model and the clone model, requiring the gradients from the victim model. 
To approximate these gradients with only black-box access, zeroth-order gradient estimation techniques are adopted.

Truong et al.~\cite{TMWP21} present a similar approach, where they replace the loss function from Kullback-Leibler (KL) divergence to $\ell_1$ norm loss for training the student model.
In contrast to the previous attacks that generate ``hard'' queries that differ in predictions between the victim and clone models, Sanyal et al.~\cite{SAB22} adopt a different strategy by generating "diverse" queries to increase predictions belonging to different classes.

\subsection{Backdoor Attack}
\label{sec:related_poi}

Backdoor attacks~\cite{GDG17,LMALZWZ18,SWBMZ22,BS21,LPXDJYW23} are training-time attacks that introduce malicious behavior into the model, making it behave like a benign model for normal inputs, while intentionally misclassifying the input to a predetermined class when the trigger appears. 

The seminal work by Gu et al.\cite{GDG17} introduced the concept of the backdoor attack on machine learning models. 
Building upon this, Liu et al.\cite{LMALZWZ18} proposed an advanced backdooring technique that incorporates enhanced triggers and relies on fewer assumptions. 
However, these attacks were limited to injecting static triggers, making them susceptible to detection.

Salem et al.\cite{SWBMZ22} integrated generative models to perform dynamic backdoor attacks, where the trigger is not fixed thus increasing the difficulty of detection. 
Nguyen and Tran\cite{NT20} further extended this concept to design an input-aware attack.
Most existing attacks in this domain are based on poisoning attacks~\cite{WZLBWZ23,MBDPWLR17,SHNSSDG18,ZHLTSG19}, which involve poisoning the training dataset. 
In contrast, Bagdasaryan and Shmatikov~\cite{BS21} propose a distinct attack target in the scenario where the learning algorithm itself is poisoned, presenting an alternative approach in this field of study.

\subsection{Data Reconstruction Attack}
\label{sec:related_ds}

Data reconstruction attacks~\cite{FJR15,YZCL19,HVYSI22,SBBFZ20} aim to recover the target dataset with limited access to the target model, with the aid of additional knowledge possessed by the adversary.

In the realm of data reconstruction attacks, existing approaches can be broadly classified into three categories: optimization-based attacks, training-based attacks, and analysis-based attacks.

Optimization-based attacks, first introduced by Fredrikson et al.\cite{FJR15}, represent the majority of existing reconstruction attacks. 
These attacks employ an iterative optimization process to reconstruct the training dataset, with the objective of obtaining a high likelihood score for the desired class. 
Notably, the integration of generative models by Zhang et al.\cite{ZJPWLS20} has contributed to improving the quality of reconstruction. 
Building on this line of research, several studies have explored diverse architectural choices~\cite{CKJQ21,WFLKZM21} and loss functions~\cite{SHCAK22} to further enhance reconstruction performance.

Conversely, training-based attacks~\cite{YZCL19} regard the target model as an encoder and train a corresponding decoder network to reconstruct inputs based on the model's outputs. 
Recently, Haim et al.~\cite{HVYSI22} presented a theoretical demonstration that, under specific assumptions, the training data can be completely recovered, leading to a new attack approach.

\end{document}